\documentclass[final,3p,times]{elsarticle}

\biboptions{sort&compress}
\journal{Comput. Methods Appl. Mech. Engrg. [Submitted: 14th Dec. 2018]}

\usepackage{graphicx}
\usepackage{amsfonts}
\usepackage{amsmath}
\usepackage{amssymb}
\usepackage{amsthm}
\usepackage{bm}
\usepackage{fancyhdr}
\usepackage{indentfirst}
\usepackage{booktabs}
\usepackage{float}
\usepackage{tikz}
\usepackage{subcaption}
\usepackage{xcolor}
\usepackage{enumerate}
\usepackage{multirow}
\usepackage{mathtools}
\usepackage{lipsum}
\usepackage{tabularx}

\usepackage[]{bibentry}
\nobibliography*

\newtheorem{remark}{Remark}

\newcommand{\rd}{\mathrm{d}}

\newcommand{\I}{{(i)}}
\newcommand{\J}{{(j)}}
\renewcommand{\IJ}{{(ij)}}
\renewcommand{\ij}{{ij}}
\newcommand{\one}{{(1)}}
\newcommand{\two}{{(2)}}

\newcommand{\bx}{x}

\newcommand{\Kn}{\mathrm{Kn}}

\usepackage[normalem]{ulem}

\usepackage{url}
\usetikzlibrary{patterns}

\begin{document}

\begin{frontmatter}
\title{A discontinuous Galerkin fast spectral method for the multi-species Boltzmann equation}
\author[labelAero]{Shashank Jaiswal}
\ead{jaiswal0@purdue.edu}
\author[labelAero]{Alina A. Alexeenko}
\ead{alexeenk@purdue.edu}
\author[labelMath]{Jingwei Hu\corref{cor1}}
\ead{jingweihu@purdue.edu}
\cortext[cor1]{Corresponding author.}
\address[labelAero]{School of Aeronautics and Astronautics, Purdue University, West Lafayette, IN 47907, USA}
\address[labelMath]{Department of Mathematics, Purdue University, West Lafayette, IN 47907, USA}
\begin{abstract}
We introduce a fast Fourier spectral method for the multi-species Boltzmann collision operator. The method retains the riveting properties of the single-species fast spectral method (\citeauthor{GHHH17} (\citeyear{GHHH17}) \cite{GHHH17}) including: (a) spectral accuracy, (b) reduced computational complexity compared to direct spectral method, (c) reduced memory requirement in the precomputation, and (d) applicability to general collision kernels. The fast collision algorithm is then coupled with discontinuous Galerkin discretization in the physical space (\citeauthor{JAH19} (\citeyear{JAH19}) \cite{JAH19}) to result in a highly accurate deterministic method (DGFS) for the full Boltzmann equation of gas mixtures. A series of numerical tests is performed to illustrate the efficiency and accuracy of the proposed method. Various benchmarks highlighting different collision kernels, different mass ratios, momentum transfer, heat transfer, and in particular the diffusive transport have been studied. The results are directly compared with the direct simulation Monte Carlo (DSMC) method.
\end{abstract}
\end{frontmatter}

\section{Introduction}

The Boltzmann equation is an integro-differential equation describing the evolution of the distribution function in six-dimensional phase space. It governs the dilute gas behavior at the molecular level and its solution is required to accurately describe a wide range of non-continuum flow  phenomena such as shocks, expansions into vacuum \cite{Muntz1989} as well as velocity and thermal slip at gas-solid interfaces \cite{sharipov2003velocity,sharipov2004velocity}. Most rarefied flows of technological interest involve gas mixtures with species diffusion playing a decisive role in turbulent, chemically reacting flows, and evaporation/condensation processes \cite{takata2007half}. This paper focuses on the development and verification of a deterministic numerical solution to the full Boltzmann equation for gas mixtures. 

The physics of Boltzmann equation is now most often simulated computationally using the direct simulation Monte Carlo (DSMC) method \cite{Bird}. Based on the kinetic theory of gases, DSMC models the binary interactions between particles stochastically. The DSMC method can be rigorously derived as the Monte Carlo solution of the $N$-particle master kinetic equation \cite{DSMC2016}. Under the assumption that molecular interactions are Markov processes, in the limit of infinite number of particles $N \rightarrow \infty$, Wagner established the convergence of Bird's DSMC method to the Boltzmann equation \cite{wagner1992convergence}. DSMC is widely used for simulating high-speed phenomena, whereas low-speed and unsteady flows are less tractable by stochastic simulations due to the inherent statistical noise. 

To avoid the complexity of solving the full Boltzmann equation, many simplified multi-species kinetic models have been proposed and this is a very active research direction in the mathematical and engineering communities, see for instance some early works \cite{GK56, sirovich1962kinetic, mccormack1973construction}, and more recently \cite{AAP02, Brull15, HHM17, BGSP18}, and references therein. These simplified models perform better at low Knudsen numbers for flows in the slip and early transition regimes. Yet they often fail to capture the physics at high Knudsen numbers and for \textit{diffusion} dominated flows at low Knudsen numbers (see \cite{gallis2014direct,gallis2006normal}). Consequently,  in this work, rather than searching for a simple kinetic model to mimic some properties of the Boltzmann equation, we propose a deterministic evaluation of the full multi-species Boltzmann equation with an intention of correctly reproducing the mass, momentum, and energy transport in gas mixtures.

The main difficulty of numerically solving the full Boltzmann equation lies in its complicated collision term. Over the past years, the deterministic methods that approximate the Boltzmann collision operator/equation have undergone considerable development. This includes the discrete velocity methods, spectral methods, etc. The readers are referred to \cite{Mieussens14, pareschi} for a comprehensive review. In particular, the Fourier spectral method has been applied to solve the multi-species Boltzmann equation in the past. In \cite{MHGM14}, a spectral-Lagrangian Boltzmann solver was proposed for a multi-energy level gas for elastic/inelastic interactions with a Lagrangian based post-processing procedure to guarantee the conservation of macroscopic quantities. However, the method was implemented in a straightforward manner without any acceleration strategy and is therefore very expensive. In \cite{wu2015fast}, a fast spectral method was introduced for the multi-species Boltzmann equation along with a strategy to treat large mass ratios. The method is based on the so-called Carleman representation which in its original form can only treat hard sphere molecules \cite{MP06}. Extension to general collision kernels requires additional assumption on the kernel and parameter fitting/recalibration. This could be a reason that all the numerical tests were restricted to hard spheres in \cite{wu2015fast}. Recently, a fast Fourier spectral method for the single-species Boltzmann collision operator was introduced in \cite{GHHH17}. The complexity for a single evaluation of the collision operator is reduced from $O(N^6)$ (direct calculation) to $O(MN^4 \log N)$, where $N$ is the number of discretization points in each velocity dimension, and $M \ll N^2$ is the number of discretization points on the sphere. Moreover, the method does not employ any assumptions or parametric fitting on the collision kernel, and is directly applicable for general molecular interactions. Based on \cite{GHHH17}, a discontinuous Galerkin fast spectral (DGFS) method was proposed in \cite{JAH19} for solving the full single-species Boltzmann equation. DGFS can produce high order spatially and temporally accurate solutions for low-speed and unsteady flows in micro-systems, and is amenable to excellent nearly-linear scaling characteristics on massively parallel architectures \cite{JHA18}. 

Along similar lines, we develop in this work the DGFS method for the multi-species Boltzmann equation and validate it on various benchmark tests including species diffusion which is of paramount importance in engineering applications. Specifically, we first generalize the method in \cite{GHHH17} to derive a fast Fourier spectral method for the multi-species collision operator. The proposed method retains the riveting properties of the single-species fast spectral method including: (a) spectral accuracy in the velocity space, (b) reduced computational complexity compared to direct spectral method, (c) reduced memory requirement in the precomputation, and (d) applicability to general collision kernels. Next, we couple the fast collision algorithm with the discontinuous Galerkin discretization \cite{JAH19} in the physical space to result in a highly accurate deterministic method for the full Boltzmann equation of gas mixtures. 

The rest of this paper is organized as follows. In section~\ref{sec_theEquation}, we give an overview of the multi-species Boltzmann equation, the self/cross collision integrals, H-theorem, and the phenomenological collision kernels used in practical engineering applications. The nondimensionalization of the equation is performed in section~\ref{sec_nondimensionalization}. Section~\ref{sec_numericalMethod} introduces the fast Fourier spectral method for multi-species Boltzmann collision operator. The discontinuous Galerkin method for the full Boltzmann equation is described in section~\ref{sec_DG}. Results of numerical experiments for the Krook-Wu solution, normal shock, Fourier flow, oscillatory Couette flow, Couette flow, and Fick's diffusion are presented in section~\ref{sec_numericalExperiments}. Concluding remarks are given in section~\ref{sec_conclusions}.

\section{The multi-species Boltzmann equation}
\label{sec_theEquation}

In this section, we give a brief description of the multi-species Boltzmann equation along with its basic mathematical properties.

Suppose we consider a many-particle system comprised of a mixture of $s$ species ($s\geq 2$). Each species is represented by a distribution function $f^{(i)}(t,x,v)$, where $t$ is time, $x$ is position, and $v$ is particle velocity. $f^{(i)}\,\rd{x}\,\rd{v}$ gives the number of particles of species $i$ to be found in an infinitesimal volume $\rd{x}\,\rd{v}$ centered at the point $(x,v)$ of the phase space. The multi-species Boltzmann equation describing the time evolution of $f^{(i)}$ is written as (cf. \cite{Cercignani, Harris})
\begin{equation} \label{MBE}
\partial_t f^{(i)}+v\cdot \nabla_x f^{(i)}=\sum_{j=1}^{s}\mathcal{Q}^{(ij)}(f^{(i)},f^{(j)}), \quad t>0, \quad x\in\Omega\subset \mathbb{R}^3,\quad v\in \mathbb{R}^3,\quad i=1,2,\dots,s.
\end{equation}
Here $\mathcal{Q}^{(ij)}$ is the collision operator that models binary collisions between species $i$ and $j$, and acts only in the velocity space:
\begin{equation} 
\mathcal{Q}^{(ij)}(f^{(i)},f^{(j)})(v)=\int_{\mathbb{R}^3}\int_{S^2}\mathcal{B}_{ij}(v-v_*,\sigma)\left[f^{(i)}(v')f^{(j)}(v_*')-f^{(i)}(v)f^{(j)}(v_*)\right]\rd{\sigma}\,\rd{v_*},
\end{equation}
where $(v, v_*)$ and $(v', v'_*)$ denote the pre- and post- collision velocity pairs. During collisions, the momentum and energy are conserved:
\begin{equation} 
m_iv+m_jv_*=m_iv'+m_jv_*', \quad m_i|v|^2+m_j|v_*|^2=m_i|v'|^2+m_j|v_*'|^2,
\end{equation}
where $m_i$, $m_j$ denote the mass of particles of species $i$ and $j$ respectively. Hence one can parameterize $v'$ and $v_*'$ as follows
\begin{align}
\left\{
\begin{array}{l}
\displaystyle v'=\frac{v+v_*}{2}+\frac{(m_i-m_j)}{2(m_i+m_j)}(v-v_*)+\frac{m_j}{(m_i+m_j)}|v-v_*|\sigma, \\[12pt]
\displaystyle v_*'=\frac{v+v_*}{2}+\frac{(m_i-m_j)}{2(m_i+m_j)}(v-v_*)-\frac{m_i}{(m_i+m_j)}|v-v_*|\sigma,
\end{array}\right.
\end{align}
with $\sigma$ being a vector varying on the unit sphere $S^2$. Finally $\mathcal{B}_{ij}=\mathcal{B}_{ji}(\geq 0)$ is the collision kernel characterizing the interaction mechanism between particles. It can be shown that
\begin{equation}
    \mathcal{B}_{ij} = B_{ij}(|v - v_*|, \cos\chi), \quad \cos \chi = \frac{\sigma \cdot (v - v_*)}{| v - v_*|},
\end{equation}
where $\chi$ is the deviation angle between $v-v_*$ and $v'-v'_*$.

Given the interaction potential between particles, the specific form of $B_{ij}$ can be determined using the classical scattering theory:
\begin{equation}
B_{ij}(|v - v_*|, \cos\chi)=|v-v_*|\,\Sigma_{ij}(|v-v_*|,\chi),
\label{eq_dim_B1}
\end{equation}
where $\Sigma_{ij}$ is the differential cross-section given by
\begin{equation}
\Sigma_{ij}(|v-v_*|,\chi)=\frac{b_{ij}}{\sin \chi}\left | \frac{\rd{b_{ij}}}{\rd{ \chi}} \right|,
\label{eq_differentialCrossSection}
\end{equation}
with $b_{ij}$ being the impact parameter. With a few exceptions, e.g. Hard Sphere (HS) model, the explicit form of $\Sigma_{ij}$ can be hard to obtain since $b_{ij}$ is related to $\chi$ implicitly. To avoid this complexity, phenomenological collision kernels are often used in practice with the aim to reproduce the correct transport coefficients. Koura et al. \cite{koura1991variable} introduced the so-called Variable Soft Sphere (VSS) model by assuming 
\begin{equation}
    \chi = 2 \cos^{-1} \{(b_{ij}/d_{ij})^{1/\alpha_{ij}}\},
\end{equation}
where $\alpha_{ij}$ is the scattering parameter, and $d_{ij}$ is the diameter borrowed from Bird's Variable Hard Sphere (VHS) model (cf. eqn.~(4.79) in \cite{Bird}):
\begin{equation}
    d_{ij} = d_{\mathrm{ref},ij} \Bigg[ \Bigg(\frac{2 k_B T_{\mathrm{ref},ij}}{\mu_{ij}|v-v_*|^2}\Bigg)^{\omega_{ij}-0.5} \frac{1}{\Gamma(2.5-\omega_{ij})} \Bigg]^{1/2},
    \label{eq_dVSS}
\end{equation}
with $\Gamma$ being the Gamma function, $\mu_{ij}=\frac{m_im_j}{m_i+m_j}$ the reduced mass, $d_{\mathrm{ref},ij}$, $T_{\mathrm{ref},ij}$, and $\omega_{ij}$, respectively, the reference diameter, reference temperature, and viscosity index. Substituting the eqns.~(\ref{eq_differentialCrossSection})-(\ref{eq_dVSS}) into (\ref{eq_dim_B1}), one can obtain $B_{ij}$ as
\begin{equation} \label{VSS}
    B_{ij} = b_{\omega_{ij},\,\alpha_{ij}} \, |v - v_*|^{2(1 - \omega_{ij})} \,(1 + \cos \chi)^{\alpha_{ij}-1},
\end{equation}
where $b_{\omega_{ij},\,\alpha_{ij}}$ is a constant given by
\begin{equation}
    b_{\omega_{ij},\,\alpha_{ij}} = \frac{d_{\mathrm{ref},ij}^2}{4} \Bigg(\frac{2 k_B T_{\mathrm{ref},ij}}{\mu_{ij}}\Bigg)^{\omega_{ij}-0.5} \frac{1}{\Gamma(2.5 - \omega_{ij})} \,\frac{\alpha_{ij}}{2^{\alpha_{ij}-1}}.
\end{equation}
In particular, the VHS kernel is obtained when $\alpha_{ij} = 1$ and $0.5\leq \omega_{ij} \leq 1$ ($\omega_{ij}=1$: Maxwell molecules; $\omega_{ij}=0.5$: HS); and the VSS kernel is obtained when $1< \alpha_{ij} \leq 2$ and $0.5\leq \omega_{ij} \leq 1$.

Given the distribution function $f^{(i)}$, the number density, mass density, velocity, and temperature of species $i$ are defined as
\begin{equation}
n^{(i)}=\int_{\mathbb{R}^3}f^{(i)}\,\rd{v}, \quad \rho^{(i)}=m_in^{(i)}, \quad u^{(i)}=\frac{1}{n^{(i)}}\int_{\mathbb{R}^3}vf^{(i)}\,\rd{v}, \quad T^{(i)}=\frac{m_i}{3n^{(i)}k_B}\int_{\mathbb{R}^3}(v-u^{(i)})^2f^{(i)}\,\rd{v}.
\end{equation}
The total number density, mass density, and velocity are given by
\begin{equation}
n=\sum_{i=1}^s n^{(i)}, \quad \rho=\sum_{i=1}^s \rho^{(i)}, \quad u=\frac{1}{\rho}\sum_{i=1}^s\rho^{(i)}u^{(i)}.
\end{equation}
Further, the diffusion velocity, stress tensor, and heat flux vector of species $i$ are defined as
\begin{equation}
v^{(i)}_D=\frac{1}{n^{(i)}}\int_{\mathbb{R}^3}cf^{(i)}\,\rd{v}=u^{(i)}-u, \quad \mathbb{P}^{(i)}=\int_{\mathbb{R}^3}m_ic\otimes c f^{(i)}\,\rd{v}, \quad q^{(i)}=\int_{\mathbb{R}^3}\frac{1}{2}m_ic|c|^2 f^{(i)}\,\rd{v},
\end{equation}
where $c=v-u$ is the peculiar velocity. Finally, the total stress, heat flux, pressure, and temperature are given by
\begin{equation}
\mathbb{P}=\sum_{i=1}^s\mathbb{P}^{(i)}, \quad q=\sum_{i=1}^s q^{(i)}, \quad p= n k_B T=\frac{1}{3}\text{tr}(\mathbb{P}).
\end{equation}

It can be shown that the collision operator $\mathcal{Q}^{(ij)}$ satisfies the following weak forms:
\begin{equation}
\begin{split}
&\int_{\mathbb{R}^3}\mathcal{Q}^{(ij)}(f^{(i)},f^{(j)})(v)\varphi(v)\,\rd{v}=\int_{\mathbb{R}^3}\int_{\mathbb{R}^3}\int_{S^2}\mathcal{B}_{ij}(v-v_*,\sigma)\left[f^{(i)}(v')f^{(j)}(v_*')-f^{(i)}(v)f^{(j)}(v_*)\right]\\
&\hspace{2in} \cdot \frac{\varphi(v)+\varphi(v_*)-\varphi(v')-\varphi(v_*')}{4}\,\rd{\sigma}\,\rd{v}\,\rd{v_*},\\
&\int_{\mathbb{R}^3}\mathcal{Q}^{(ij)}(f^{(i)},f^{(j)})(v)\varphi(v)\,\rd{v}+\int_{\mathbb{R}^3}\mathcal{Q}^{(ji)}(f^{(j)},f^{(i)})(v)\phi(v)\,\rd{v}\\
=&\int_{\mathbb{R}^3}\int_{\mathbb{R}^3}\int_{S^2}\mathcal{B}_{ij}(v-v_*,\sigma)\left[f^{(i)}(v')f^{(j)}(v_*')-f^{(i)}(v)f^{(j)}(v_*)\right]\frac{\varphi(v)+\phi(v_*)-\varphi(v')-\phi(v_*')}{2}\,\rd{\sigma}\,\rd{v}\,\rd{v_*}.
\end{split}
\end{equation}
Using these weak forms, it is easy to derive
\begin{equation} \label{consv}
\begin{split}
&\int_{\mathbb{R}^3}\mathcal{Q}^{(ij)}(f^{(i)},f^{(j)})\,\rd{v}=0,\\
&\int_{\mathbb{R}^3}\mathcal{Q}^{(ij)}(f^{(i)},f^{(j)})m_iv\,\rd{v}+\int_{\mathbb{R}^3}\mathcal{Q}^{(ji)}(f^{(j)},f^{(i)})m_jv\,\rd{v}=0,\\
&\int_{\mathbb{R}^3}\mathcal{Q}^{(ij)}(f^{(i)},f^{(j)})m_i|v|^2\,\rd{v}+\int_{\mathbb{R}^3}\mathcal{Q}^{(ji)}(f^{(j)},f^{(i)})m_j|v|^2\,\rd{v}=0,
\end{split}
\end{equation}
and the well-known Boltzmann's H-theorem
\begin{equation} \label{Hthm}
\sum_{i,j=1}^s \int_{\mathbb{R}^3}\mathcal{Q}^{(ij)}(f^{(i)},f^{(j)})\ln f^{(i)}\,\rd{v} \leq 0.
\end{equation}

(\ref{Hthm}) implies that the total entropy of the system decays with time:
\begin{equation}
 \sum_{i=1}^{s} \left\{ \partial_t \int_{\mathbb{R}^3}  f^\I \ln f^{(i)}\, \rd{v} + \nabla_x \cdot \int_{\mathbb{R}^3} v f^\I \ln f^{(i)}\, \rd{v} \right\} \leq 0,
 \end{equation}
 and the equality holds if and only if $f^{(i)}$ attains the local equilibrium
\begin{align} \label{Max}
f^{(i)}=\frac{n^\I}{(2\pi R_iT)^{3/2}}\exp\Bigl(-\frac{(v-u)^2}{2R_iT}\Bigr):=\mathcal{M}^\I,
\end{align}
where $R_i=k_B/m_i$ is the specific gas constant.

On the other hand, using (\ref{consv}), one can take the moments of eqn.~(\ref{MBE}) to obtain the following local conservation laws:
\begin{equation}
\begin{split}
&  \partial_t \int_{\mathbb{R}^3} f^\I \rd{v} + \nabla_x \cdot \int_{\mathbb{R}^3} v f^\I \rd{v} = 0,\\
& \sum_{i=1}^{s} \left\{ \partial_t \int_{\mathbb{R}^3} m_i v f^\I \rd{v} + \nabla_x \cdot \int_{\mathbb{R}^3} m_i v \otimes v f^\I \rd{v} \right\} = 0,\\
& \sum_{i=1}^{s} \left\{ \partial_t \int_{\mathbb{R}^3} \frac{1}{2} m_i |v|^2 f^\I \rd{v} + \nabla_x \cdot \int_{\mathbb{R}^3} \frac{1}{2} m_i v |v|^2 f^\I \rd{v} \right\} = 0,
\end{split}
\end{equation}
which, using the previously defined macroscopic quantities, can be recast as
\begin{equation}
\begin{split}
& \partial_t n^{(i)} + \nabla_x \cdot \left( n^{(i)} u^{(i)} \right) = 0\quad  \Longrightarrow \quad \partial_t \rho + \nabla_x \cdot \left( \rho u \right) = 0,\\
& \partial_t(\rho u) + \nabla_x \cdot ( \rho u \otimes u+ \mathbb{P}) = 0,\\
& \partial_tE + \nabla_x \cdot \left(Eu + \mathbb{P} u+q\right)=0,
\end{split}
\label{eqn_macroscopicBoltzmann}
\end{equation}
where $E= 3 n k_B T/2 + \rho u^2/2$ is the total energy. Note that this system is not closed. However, replacing $f^{(i)}$ by $\mathcal{M}^\I$ in (\ref{eqn_macroscopicBoltzmann}) yields a closed system, i.e., the compressible Euler equations. With more involved calculations (so-called Chapman-Enskog expansion), one can derive the Navier-Stokes equations. We omit the detail but mention that the heat flux term will contain the diffusion velocity $v_D^{(i)}$, a property unique to the mixtures (see for instance \cite{Harris}).

For the eqn.~(\ref{MBE}), one can consider the in-flow equilibrium boundary condition:
\begin{equation} \label{eq_bc_inlet}
	f^\I(t,\bx,v)=  \frac{n_\mathrm{in}^{(i)}}{(2\pi R_i T_\mathrm{in})^{3/2}} \exp \Big( - \frac{(v - u_\mathrm{in})^2}{2 R_i T_\mathrm{in}} \Big), \quad x \in \partial \Omega, \quad v\cdot \hat{n}<0,
\end{equation}
where $\hat{n}$ is the outward pointing normal at $x$, $n_\mathrm{in}^{(i)}$, $T_\mathrm{in}$ and $u_\mathrm{in}$ are the prescribed density, temperature and velocity. Another commonly used one is the Maxwell boundary condition:
\begin{equation} \label{eq_bc_wall}
    f^\I(t,\bx,v)=(1-\alpha) f^{(i)}(t,x,v-2[(v-u_w)\cdot \hat{n}]\hat{n})+ \alpha \,n^\I_w \exp \Big( - \frac{(v - u_w)^2}{2 R_i T_w} \Big),  \quad x \in \partial \Omega, \quad (v-u_w)\cdot \hat{n}<0,
\end{equation}
where $T_w$ and $u_w$ are the temperature and velocity of the wall, $n^\I_w$ is determined from conservation of mass as
\begin{equation} \label{eq_bc_nwall}
    n^\I_w = - \frac{\int_{(v - u_w)\cdot { \hat{n}} \geq 0}(v - u_w)\cdot { \hat{n}} \, f^\I \, \rd{v}}
    { \int_{(v - u_w)\cdot {\hat{n}} <0} (v - u_w)\cdot { \hat{n}}  \, \exp \Big( - \frac{(v - u_w)^2}{2 R_i T_w} \Big) \, \rd{v}},
\end{equation}
and $\alpha$ is the accommodation coefficient, with $\alpha=1$ corresponds to purely diffusive boundary and $\alpha=0$ to purely reflective boundary.

\section{Nondimensionalization}
\label{sec_nondimensionalization}

For easier manipulation, we perform a nondimensionalization of the eqn.~(\ref{MBE}). We first choose the characteristic length $H_0$, temperature $T_0$, number density $n_0$, and mass $m_0$, and then define the characteristic velocity $u_0 = \sqrt{2 k_B T_0/m_0}$ and time $t_0=H_0/u_0$. We rescale $t$, $x$, $v$, $m_i$, and $f^{(i)}$ as follows:
\begin{equation}
\hat{t}= \frac{t}{t_0}, \quad 
\hat{x}= \frac{x}{H_0}, \quad 
\hat{v}= \frac{v}{u_0}, \quad 
\hat{m}_i=\frac{m_i}{m_0}, \quad
\hat{f}^{(i)}= \frac{f^{(i)}}{n_0/u^3_0},
\end{equation}
and rescale the collision kernel as
\begin{equation}
	\hat{\mathcal{B}}_{ij} = \frac{\mathcal{B}_{ij}}{B_{0,ij}},
\end{equation}
where 
\begin{align}
B_{0,ij} &= u_0 \sqrt{1 +m_i/m_j}\, \pi\, d^2_{\mathrm{ref}, ij}\, (T_{\mathrm{ref},ij}/T_0)^{\omega_{ij}-0.5}.
\end{align}
Then the eqn.~(\ref{MBE}) becomes (dropping $\hat{~}$ for simplicity)
\begin{equation}
\begin{split}
\partial_{t} f^{(i)} +v\cdot \nabla_{x} f^{(i)} =\sum_{j=1}^s \frac{n_0H_0}{u_0} B_{0,ij}   \int_{\mathbb{R}^3}\int_{S^2}\mathcal{B}_{ij}\left[f^{(i)}(v')f^{(j)}(v_*')-f^{(i)}(v)f^{(j)}(v_*)\right]\rd{\sigma}\,\rd{v_*}.
\end{split}
\end{equation}
The factor 
\begin{equation}
\frac{u_0}{n_0\,H_0\,B_{0,ij}} = \frac{\frac{u_0}{n_0 B_{0,ij}}}{H_0} = \mathrm{Kn}_{ij}
\end{equation}
is the Knudsen number defined as the ratio of the mean free path and characteristic length scale, hence
\begin{align} \label{eq_Kn}
\mathrm{Kn}_{ij} &= \frac{1}{\sqrt{1 +m_i/m_j}\, \pi\, n_0\, d^2_{\mathrm{ref}, ij}\, (T_{\mathrm{ref},ij}/T_0)^{\omega_{ij}-0.5}\, H_0}.
\end{align}
One can also define the ``average" Knudsen number for each species $i$ as 
\begin{equation}
\mathrm{Kn}_i=\left(\sum_{j=1}^s\frac{1}{\mathrm{Kn}_{ij}}\right)^{-1}.
\end{equation}
This is consistent with eqn.~(4.76) in \cite{Bird}.

Therefore, the dimensionless Boltzmann equation for the VSS kernel (\ref{VSS}) reads as
\begin{equation} \label{NMBE}
\partial_t f^{(i)}+v\cdot \nabla_x f^{(i)}= \sum_{j=1}^s \frac{1}{\mathrm{Kn}_{ij}}\mathcal{Q}^{(ij)}(f^{(i)},f^{(j)})(v),
\end{equation}
with
\begin{equation} \label{NMCO}
\mathcal{Q}^{(ij)}(f^{(i)},f^{(j)})(v)=\int_{\mathbb{R}^3}\int_{S^2}B_{ij}(|v-v_*|,\cos \chi)\left[f^{(i)}(v')f^{(j)}(v_*')-f^{(i)}(v)f^{(j)}(v_*)\right]\rd{\sigma}\,\rd{v_*},
\end{equation}
\begin{align} \label{NVSS}
B_{ij} &= \frac{\alpha_{ij}}{ \sqrt{1+m_i/m_j} \, \mu_{ij}^{\omega_{ij}-0.5} 2^{1+\alpha_{ij}}\,\Gamma(2.5-\omega_{ij})\pi}|v - v_*|^{2(1 - \omega_{ij})} \, (1 + \cos \chi)^{\alpha_{ij}-1}.
\end{align}

\begin{remark}
We adopt the VSS kernel in this paper for easy comparison with DSMC solutions. The fast algorithm for the collision operator does not rely on the specific form (\ref{NVSS}) (see Section~\ref{sec_numericalMethod}).
\end{remark}

In addition, we rescale the macroscopic quantities as
\begin{align}
\hat{n}^\I = \frac{n^\I}{n_0}, \quad
\hat{\rho}^\I = \frac{\rho^\I}{m_0 n_0}, \quad
\hat{u}^\I = \frac{u^\I}{u_0}, \quad
\hat{T}^\I = \frac{T^\I}{T_0}, \quad
\hat{\mathbb{P}}^\I = \frac{\mathbb{P}^\I}{\frac{1}{2} m_0 n_0 u_0^2}, \quad
\hat{q}^\I = \frac{q^\I}{\frac{1}{2} m_0 n_0 u_0^3},
\end{align}
then in rescaled variables (again dropping $\hat{~}$ for simplicity)
\begin{align}
&n^{(i)} = \int_{\mathbb{R}^3}f^{(i)}\,\rd{v}, \quad 
\rho^{(i)} = m_i n^{(i)}, \quad 
u^{(i)} = \frac{1}{n^{(i)}} \int_{\mathbb{R}^3}vf^{(i)}\,\rd{v}, \quad T^{(i)}=\frac{2m_i}{3 n^{(i)}}\int_{\mathbb{R}^3}(v-u^{(i)})^2f^{(i)}\,\rd{v}, \nonumber \\
&\mathbb{P}^\I = 2\, m_i \int_{\mathbb{R}^3} (v-u) \otimes (v-u) f^{(i)} \,\rd{v}, \quad
q^\I = m_i \int_{\mathbb{R}^3} (v-u) |v-u|^2 f^{(i)} \,\rd{v},
\end{align}
and the Maxwellian (\ref{Max}) becomes
\begin{equation}
\mathcal{M}^{(i)} = n^{(i)} \left(\frac{m_i}{\pi T}\right)^{3/2} \exp{\left( -\frac{m_i| v-u |^2 }{T}\right)}.
\end{equation}

\begin{remark}

For the normal shock (see section~\ref{sec:normal}), it is often convenient to define the so-called parallel ($T^\I_\parallel$) and perpendicular ($T^\I_\perp$) components of temperature as
\begin{align}
T^{(i)}_\parallel=\frac{2m_i}{n^{(i)}}\int_{\mathbb{R}^3}(v_x-u_x^{(i)})^2f^{(i)}\,\rd{v}, \quad T^{(i)}_\perp=\frac{2m_i}{ n^{(i)}}\int_{\mathbb{R}^3}(v_y-u_y^{(i)})^2f^{(i)}\,\rd{v},
\end{align}
where subscripts $x$ and $y$ denote the first and second components of respective vector fields.

\end{remark}

\section{A fast Fourier spectral method for the multi-species Boltzmann collision operator}
\label{sec_numericalMethod}

The main difficulty of numerically solving the multi-species Boltzmann equation (\ref{NMBE}) lies in the collision operator (\ref{NMCO}). In this section, we introduce a fast Fourier spectral method (in the velocity space) to approximate this operator. Discussion for the spatially inhomogeneous equation will be given in the next section.

We first perform a change of variables $v_*$ to $g=v-v_*$ in (\ref{NMCO}) to obtain
\begin{equation} 
\mathcal{Q}^{(ij)}(f^{(i)},f^{(j)})(v)=\int_{\mathbb{R}^3}\int_{S^2}B_{ij}(|g|,\sigma\cdot \hat{g})\left[f^{(i)}(v')f^{(j)}(v_*')-f^{(i)}(v)f^{(j)}(v_*)\right]\rd{\sigma}\,\rd{g},
\end{equation}
where $\hat{g}$ is the unit vector along $g$ and
\begin{align}
v'=v-\frac{m_j}{m_i+m_j}g+\frac{m_j}{m_i+m_j}|g|\sigma, \quad v_*'=v-\frac{m_j}{m_i+m_j}g-\frac{m_i}{m_i+m_j}|g|\sigma.
\end{align}

Next we need to choose a finite computational domain $\mathcal{D}_L=[-L,L]^3$. This is based on the following criterion (similar discussion for the single-species case can be found in \cite{PR00}).

Assume the support of functions $f^{(i)}$, $f^{(j)}$ can be approximated by a ball with radius $S$: $\text{Supp}(f^{(i)}(v),f^{(j)}(v)) \subset \mathcal{B}_S$, then one has
\begin{enumerate}

\item $\text{Supp}(\mathcal{Q}^{(ij)}(f^{(i)},f^{(j)})(v))\subset \mathcal{B}_{\sqrt{1+m_j/m_i}S}$.

This is because if $|v|>\sqrt{1+m_j/m_i}S$, then $f^{(i)}(v)=0$; also $m_i|v'|^2+m_j|v_*'|^2\geq m_i|v|^2>(m_i+m_j)S^2$, then either $|v'|>S$ or $|v_*'|>S$, so $f^{(i)}(v')=0$ or $f^{(j)}(v_*')=0$; either way $\mathcal{Q}^{(ij)}(f^{(i)},f^{(j)})(v)=0$.

\item It is enough to truncate $g$ to a ball $\mathcal{B}_R$ with $R=2S$:
\begin{equation}
\mathcal{Q}^{(ij)}(f^{(i)},f^{(j)})(v)=\int_{\mathcal{B}_R}\int_{S^2}B_{ij}(|g|,\sigma\cdot \hat{g})\left[f^{(i)}(v')f^{(j)}(v_*')-f^{(i)}(v)f^{(j)}(v_*)\right]\rd{\sigma}\,\rd{g}.
\end{equation}

This is because if $2S<|g|=|v-v_*|\leq |v|+|v_*|$, then $|v|>S$ or $|v_*|>S$, so $f^{(i)}(v)=0$ or $f^{(j)}(v_*)=0$; also $2S<|g|=|v-v_*|=|v'-v_*'|\leq |v'|+|v_*'|$, then $|v'|>S$ or $|v_*'|>S$, so $f^{(i)}(v')=0$ or $f^{(j)}(v_*')=0$; either way $\mathcal{Q}^{(ij)}(f^{(i)},f^{(j)})(v)=0$.

\item Since $|v|\leq \sqrt{1+m_j/m_i}S$ and $|g|\leq 2S$ in $\mathcal{Q}^{(ij)}(f^{(i)},f^{(j)})(v)$, we have

$|v_*|=|v-g|\leq |v|+|g|\leq(2+\sqrt{1+m_j/m_i})S$;

$|v'|=\left|v-\frac{m_j}{m_i+m_j}g+\frac{m_j}{m_i+m_j}|g|\sigma\right|\leq |v|+\frac{2m_j}{m_i+m_j}|g|\leq(4m_j/(m_i+m_j)+\sqrt{1+m_j/m_i})S$;

$|v_*'|=\left|v-\frac{m_j}{m_i+m_j}g-\frac{m_i}{m_i+m_j}|g|\sigma\right|\leq |v|+|g|\leq(2+\sqrt{1+m_j/m_i})S$.

\item To avoid aliasing, need

\begin{equation} \label{domain}
2L\geq \left(\max(4m_j/(m_i+m_j),2)+\sqrt{1+m_j/m_i}\right)S+S.
\end{equation}
\end{enumerate}

\begin{remark}
From (\ref{domain}), it can be seen that the computational domain needs to be very large for large mass ratios $m_j/m_i \gg 1$. This is a common issue appearing in multi-species problems. Possible remedies include adaptive mesh in velocity space (cf. \cite{TCS16}), using an asymptotic model valid for large mass ratios (cf. \cite{DL96}), or introducing independent velocity grid for each species wherein different collision types for every ($i,j$) pair are treated independently (cf. \cite{clarke2014discrete, wu2015fast}). In this paper, we only consider moderate mass ratios and postpone these studies to a future work.
\end{remark}

Now we approximate $f^{(i)}$ (similarly for $f^{(j)}$) by a truncated Fourier series on $\mathcal{D}_L$:
\begin{equation}
f^{(i)}(v)\approx\sum_{k=-\frac{N}{2}}^{\frac{N}{2}-1}\hat{f}_k^{(i)} e^{i\frac{\pi}{L}k\cdot v}, \quad \hat{f}_k^{(i)}=\frac{1}{(2L)^3}\int_{\mathcal{D}_L}f^{(i)}(v) e^{-i\frac{\pi}{L}k\cdot v}\,\rd{v},
\end{equation}
note here an abuse of notation: the summation over the 3D index $k$ means $-N/2\leq k_i \leq N/2-1$, where $k_i$ is each component of $k$. Upon substitution of $f^{(i)}$, $f^{(j)}$ into $\mathcal{Q}^{(ij)}(f^{(i)},f^{(j)})$ and a Galerkin projection to the same Fourier space, we obtain the $k$-th Fourier mode of the collision operator as
\begin{equation} \label{sum}
\hat{\mathcal{Q}}^{(ij)}_k=\sum_{\substack{l,m=-\frac{N}{2}\\l+m=k}}^{\frac{N}{2}-1}G^{(ij)}(l,m)\hat{f}^{(i)}_l \hat{f}^{(j)}_m,
\end{equation} 
with the weight
\begin{equation*} 
G^{(ij)}(l,m)=\int_{\mathcal{B}_R}\int_{S^2}B_{ij}(|g|,\sigma\cdot \hat{g})\left[e^{-i\frac{\pi}{L}\frac{m_j}{m_i+m_j}(l+m)\cdot g+i\frac{\pi}{L}|g|\left(\frac{m_j}{m_i+m_j}l-\frac{m_i}{m_i+m_j}m\right)\cdot \sigma}-e^{-i\frac{\pi}{L}m \cdot g}\right]\rd{\sigma}\,\rd{g}.
\end{equation*}

Without special treatment, the summation (\ref{sum}) has to be evaluated directly, resulting in a computational cost of $O(N^6)$. Furthermore, the weight $G^{(ij)}(l,m)$ needs to be precomputed and the storage requirement is $O(N^6)$. This can quickly become a bottleneck even for moderate $N$. Motivated by our previous work for the single-species Boltzmann equation \cite{GHHH17}, we propose the following strategy to accelerate the direct summation as well as alleviate its memory bottleneck.

For the gain term (positive part) of $G^{(ij)}(l,m)$, we decompose it as
\begin{align} \label{GG}
G^{(ij)+}(l,m)&=\int_0^R\int_{S^2}F^{(ij)}(l+m,\rho,\sigma)e^{i\frac{\pi}{L}\rho\left(\frac{m_j}{m_i+m_j}l-\frac{m_i}{m_i+m_j}m\right)\cdot \sigma}\rd{\sigma}\,\rd{\rho},
\end{align}
where $\rho=|g|$ is the radial of $g$ and 
\begin{align}
F^{(ij)}(l+m,\rho,\sigma)=\rho^2\int_{S^2}B_{ij}(\rho,\sigma\cdot \hat{g})e^{-i\frac{\pi}{L}\rho\frac{m_j}{m_i+m_j}(l+m)\cdot \hat{g}}\,\rd{\hat{g}},
\end{align}
while for the loss term (negative part) of $G^{(ij)}(l,m)$, 
\begin{align}
G^{(ij)-}(m)&=\int_0^R\int_{S^2}\int_{S^2}\rho^2B_{ij}(\rho,\sigma\cdot \hat{g})e^{-i\frac{\pi}{L}\rho\, m \cdot \hat{g}}\rd{\sigma}\,\rd{\hat{g}}\,\rd{\rho}.
\end{align}

The idea is to precompute $F^{(ij)}(l+m,\rho,\sigma)$ and $G^{(ij)-}(m)$ up to a high accuracy, and approximate the integral in (\ref{GG}) on the fly using a quadrature rule:
\begin{equation}
G^{(ij)+}(l,m)\approx \sum_{\rho, \sigma} w_{\rho}w_{\sigma} F^{(ij)}(l+m,\rho,\sigma)e^{i\frac{\pi}{L}\rho\left(\frac{m_j}{m_i+m_j}l-\frac{m_i}{m_i+m_j}m\right)\cdot \sigma},
\end{equation}
where for the radial direction, we use the Gauss-Legendre quadrature with $N_{\rho}=O(N)$ points (since the integral oscillates roughly on $O(N)$); for the integral over the sphere, we use the $M$-point spherical design quadrature \cite{Womersley,womersley2018efficient} (usually $M\ll N^2$).

Therefore, the gain term of the collision operator can be approximated as
\begin{align}
\hat{\mathcal{Q}}^{(ij)+}_k\approx \sum_{\rho,\sigma}w_{\rho}w_{\sigma}F^{(ij)}(k,\rho,\sigma)\sum_{\substack{l,m=-\frac{N}{2}\\l+m=k}}^{\frac{N}{2}-1} \left(e^{i\frac{\pi}{L}\rho \frac{m_j}{m_i+m_j}l\cdot \sigma}\hat{f}^{(i)}_l  \right)\left(e^{-i\frac{\pi}{L}\rho \frac{m_i}{m_i+m_j}m\cdot \sigma}\hat{f}^{(j)}_m  \right).
\end{align}
Written in the above form, we see that the inner sum is a convolution of two functions so that it can be evaluated efficiently in $O(N^3\log N)$ operations via the fast Fourier transform (FFT). Together with the outer sum, the total complexity of evaluating $\hat{\mathcal{Q}}^{(ij)+}_k$ (for all $k$) is $O(MN^4\log N)$ (recall the total number of quadrature points needed for $\rho$ and $\sigma$ is $O(MN)$).

On the other hand, the loss term of the collision operator can be written as
\begin{align}
\hat{\mathcal{Q}}^{(ij)-}_k=\sum_{\substack{l,m=-\frac{N}{2}\\l+m=k}}^{\frac{N}{2}-1} \hat{f}^{(i)}_l \left(G^{(ij)-}(m) \hat{f}^{(j)}_m  \right),
\end{align}
which is readily a convolution, hence can be evaluated in $O(N^3\log N)$.

Putting both pieces together, we have obtained a fast algorithm of complexity $O(MN^4\log N)$ for evaluating the collision operator $\mathcal{Q}^{(ij)}(f^{(i)},f^{(j)})$, where $M\ll N^2$. In addition, the memory requirement to store the weight $F^{(ij)}(l+m,\rho,\sigma)$ and $G^{(ij)-}(m)$ is $O(MN^4)$.

\section{The discontinuous Galerkin method for the spatial discretization}
\label{sec_DG}

The previously introduced fast spectral method allows us to compute the collision operator efficiently. To solve the full spatially inhomogeneous equation (\ref{NMBE}), we also need an accurate and efficient spatial and time discretization. Here we adopt the RKDG (Runge-Kutta discontinuous Galerkin) method \cite{CS01} widely used for hyperbolic type equations. Since the transport term is linear in the Boltzmann equation, the application of DG method is straightforward. We give a brief description below for completeness.

We first decompose the physical domain $\Omega$ into $N_e$ variable-sized disjoint elements $D^e_\bx$:
\begin{align}
	\Omega\approx \bigcup\limits_{e=1}^{N_e} D^e_\bx, \quad D^e_\bx \cap D^{e'}_\bx = \emptyset, \quad \forall \; e \neq e', \quad 1\leq e,e' \leq N_e.
\end{align}
In each element $D_\bx^e$, we approximate the distribution function $f^\I(t,\bx,v)$ for each species by a polynomial of order $N_p$:
\begin{equation}
	\bx \in D_\bx^{e}: \quad f^\I_e(t,\bx,v) = \sum_{l = 1}^{K} \mathcal{F}^\I_{e,\,l}(t,v) \,\phi_l^e(x), \quad 1\leq i \leq s,
	\label{eq_DG_local_approx}
\end{equation}
where $\phi_l^e(x)$ is the basis function supported in $D_\bx^{e}$, $K$ is the total number of terms in the local expansion, and $\mathcal{F}_{e,l}^\I(t,v)$ is the elemental degree of freedom. 

We form the residual by substituting the expansion (\ref{eq_DG_local_approx}) into the eqn.~(\ref{NMBE}):
\begin{align}
\mathcal{R}^\I_e=	\sum_{l=1}^{K} \phi_l^e\, \partial_t\mathcal{F}^\I_{e,\,l}+ \sum_{l=1}^{K} \mathcal{F}^\I_{e,\,l} \, v \cdot \nabla_{\bx} \phi_l^e - \sum_{j=1}^{s} \frac{1}{\Kn_\ij}\sum_{l_1,l_2=1}^{K}  \mathcal{Q}^\IJ \left(\mathcal{F}^\I_{e,\,l_1}, \mathcal{F}^\J_{e,\,l_2}\right) \phi_{l_1}^e \phi_{l_2}^e, \quad 1\leq i \leq s,
\label{eq_bze_weak_multiply}
\end{align}
where we used the quadratic property of the collision operator. We then require that the residual is orthogonal to all test functions. In the Galerkin formulation, the test function is the same as the basis function, thus
\begin{equation}
	\int_{D_\bx^e} \mathcal{R}^\I_e \, \phi_m^e \, \rd{\bx} = 0, \quad 1\leq m \leq K, \quad 1\leq i \leq s.
	\label{eq_innerProduct}
\end{equation}

Substituting (\ref{eq_bze_weak_multiply}) into (\ref{eq_innerProduct}) and applying the divergence theorem, we obtain
\begin{align}
	&\sum_{l=1}^{K} \left(\int_{D_x^e} \phi_m^e\, \phi_l^e \,\rd{\bx}\right)\partial_t\mathcal{F}^\I_{e,\,l} -\sum_{l=1}^K \mathcal{F}^\I_{e,\,l}\, v\cdot \int_{D_\bx^e}  \phi_l^e \,\nabla_{\bx}  \phi_m^e\, \rd{\bx} \nonumber\\ 
	=&- \int_{\partial D_\bx^e}  \phi_m^e  \left( F^\I_*\cdot \hat{n}^e \right)\rd{\bx} + \sum_{j=1}^{s}\frac{1}{\Kn_\ij}\sum_{l_1,l_2=1}^{K} \mathcal{Q}^\IJ (\mathcal{F}^\I_{e,\,l_1}, \mathcal{F}^\J_{e,l_2}) \left(\int_{D_\bx^e} \phi_m^e \,\phi_{l_1}^e\, \phi_{l_2}^e\, \rd{\bx}\right),
	\label{eq_dg_innerProd}
\end{align}
where $\hat{n}^e$ is the local outward pointing normal and $F^\I_*$ denotes the numerical flux. Specifically, the surface integral in the above equation is defined as follows
\begin{equation} \label{eq_flux}
\int_{\partial D_\bx^e}  \phi_m^e  \left( {F}^\I_*\cdot \hat{n}^e \right)\rd{\bx} =\sum_{E\,\in\,\partial D_\bx^e} \int_E  \phi_m^e  \left( {F}^\I_{*,\,E}\cdot \hat{n}^e_E \right) \rd{\bx},
\end{equation}
with $\hat{n}^e_E$ and ${F}^\I_{*,\,E}$ being the outward normal and numerical flux along the face $E$. In our implementation, we choose the upwind flux:
\begin{equation}
{F}^\I_{*,\,E} = 
\begin{cases}
v \,f_e^\I(t, \bx_{E,\,int(D_\bx^e)},v), \quad v \cdot \hat{n}_E^e \geq 0
\\
v \, f_e^\I(t, \bx_{E,\,ext(D_\bx^e)},v), \quad v \cdot \hat{n}_E^e < 0
\end{cases}
\end{equation}
where \textit{int} and \textit{ext} denote interior and exterior of the face $e$ respectively.

Finally, define the mass matrix $\mathcal{M}_{ml}$, stiffness matrix $\mathcal{S}_{ml}$, and the tensor $\mathcal{H}_{m l_1 l_2}$ as
\begin{equation}
\begin{split}
&\mathcal{M}_{ml}^e = \int_{D^e_x} \phi_m^{e}(x) \, \phi_l^{e}(x) \,\rd{x}, \quad \mathcal{S}_{ml}^e = \int_{D^e_x} \phi_l^{e}(x) \, \nabla_x \phi_m^{e}(x)\, \rd{\bx},\\
&\mathcal{H}_{m\, l_1 l_2}^e = \int_{D^e_x} \phi_m^{e}(x) \, \phi_{l_1}^{e}(x) \,\phi_{l_2}^{e}(x)\, \rd{\bx},
\end{split}
\end{equation}
then (\ref{eq_dg_innerProd}) can be written as
\begin{equation}
\begin{split}
&\sum_{l=1}^{K} \mathcal{M}_{ml}^e \,\partial_t\mathcal{F}^\I_{e,\,l} - \sum_{l=1}^{K} v\cdot \mathcal{S}_{ml}^e\, \mathcal{F}^\I_{e,\,l} 
=- \int_{\partial D_\bx^e}  \phi_m^e  \left( F^\I_*\cdot \hat{n}^e \right)\rd{\bx} + \sum_{j=1}^{s} \frac{1}{\Kn_\ij}\sum_{l_1,\,l_2=1}^{K} \mathcal{H}_{m \, l_1  l_2}^e  \mathcal{Q}^{(ij)} \left(\mathcal{F}^\I_{e,\,l_1},  \mathcal{F}^\J_{e,\,l_2}\right),
\label{eq_weakForm}
\end{split}
\end{equation}
for $1\leq m \leq K$, $1\leq i \leq s$. 

(\ref{eq_weakForm}) is the DG system we are going to solve in each element $D_\bx^e$ of the physical domain. The fast spectral method introduced in the previous section is used to evaluate the term $\mathcal{Q}^{(ij)} \left(\mathcal{F}^\I_{e,\,l_1},  \mathcal{F}^\J_{e,\,l_2}\right)$. The second-order strong-stability-preserving (SSP) RK scheme \cite{gottlieb2001strong} is applied for the time derivative.

\subsection{Structure of $\mathcal{H}_{m\,l_1\,l_2}^e$: a spectral element approach}

Needless to say, the main computational bottleneck when solving the system (\ref{eq_weakForm}) lies in the term $\mathcal{Q}^{(ij)} \left(\mathcal{F}^\I_{e,\,l_1},  \mathcal{F}^\J_{e,\,l_2}\right)$, whose complexity is $O(MN^4\log N)$ for given $i$, $j$, $l_1$, and $l_2$. For general polynomial basis (e.g., the modal DG basis), $\mathcal{H}_{m\,l_1\,l_2}^e$ is a full tensor, hence the total complexity to evaluate the collision part would be $O(s^2K^2MN^4\log N)$ (for all pairs of $(i,j)$ and $(l_1,l_2)$) inside each element $D_\bx^e$. This is still computationally demanding, even though we are equipped with the fast collision solver. Therefore, the sparsity of $\mathcal{H}_{m\,l_1\,l_2}^e$ would potentially save the computational cost since the collision operator only needs to be evaluated for $l_1$, $l_2$ such that $\mathcal{H}_{m\,l_1\,l_2}^e\neq 0$. It is known that in the spectral element method \cite{Patera84}, if the nodal basis (\cite{hesthaven2007nodal}) is used and the interpolation points are chosen the same as the quadrature points, the mass matrix will become diagonal. Here to achieve better efficiency, we propose to use the same approach to treat the tensor $\mathcal{H}_{m\,l_1\,l_2}^e$. We present the 1D case for simplicity.

Suppose $D_\bx^{e}= [x^{e}_{l}, x^{e}_{r}]$, with $x^{e}_{l}$ and $x^{e}_{r}$ being, respectively, the left and right ends of the element $D_\bx^e$. $h^{e} = |x^{e}_{r} - x^{e}_{l}|$ is the element size. The DG convention is to define an element in the standard interval $D^{(st)}=[-1,1]$ and map the standard element $D^{(st)}$ to the local element $D_\bx^{e}$ using an affine mapping 
 \begin{equation}
 x = x_{l}^{e}\frac{1-\xi}{2} + x_{r}^{e}\frac{1+\xi}{2}, \quad \xi \in D^{(st)}.
 \label{eq_init_mapLine}
 \end{equation}
Then
 \begin{align}
 \mathcal{H}_{m\,l_1\,l_2}^{e} &= \int_{D^{(st)}} \phi_m^{(st)}(\xi) \; \phi_{l_1}^{(st)}(\xi) \; \phi_{l_2}^{(st)}(\xi)\,\Big|\frac{\partial x}{\partial \xi}\Big|\;\rd{\xi}=\frac{h^e}{2} \int_{D^{(st)}} \phi_m^{(st)}(\xi) \; \phi_{l_1}^{(st)}(\xi) \; \phi_{l_2}^{(st)}(\xi)\,\rd{\xi} \nonumber \\
 &\approx \frac{h^e}{2} \sum_{q=1}^{N_{q}} w_{q} \phi_m^{(st)}(\xi_q) \; \phi_{l_1}^{(st)}(\xi_q) \; \phi_{l_2}^{(st)}(\xi_q):=\frac{h^{e}}{2} \mathcal{H}_{m\,l_1\,l_2}^{(st)},
 \end{align}
where $\{\xi_q, \,w_q\}_{q=1}^{N_q}$ are the quadrature points and weights. 
 
Consider the Lagrange polynomials as basis functions, i.e.,
 \begin{align}
 \phi_{m}^{(st)}(\xi) := \prod_{\begin{smallmatrix}1\le {n}\le K\\ n\neq m\end{smallmatrix}} \frac{\xi-\xi_n}{\xi_m-\xi_n}, \quad m=1,...,K, 
 \label{eq_nodal1D_Basis}
 \end{align}
 where $\{\xi_m\}_{m=1}^K$ are the Gauss-Lobatto-Legendre (GLL) quadrature points. When $\{\xi_q\}_{q=1}^{N_q}$ are taken the same as $\{\xi_m\}_{m=1}^K$, $\phi_{m}^{(st)}(\xi_q)=\delta_{mq}$, hence the mass matrix becomes diagonal. Similarly,
 \begin{align}
 \mathcal{H}_{m\,l_1 l_2}^{(st)} = \sum_{q=1}^{N_{q}} w_{q} \delta_{mq}\; \delta_{l_1q} \; \delta_{l_2q}= \begin{cases} 
 \displaystyle w_m, \quad &\mathrm{iff} \quad m=l_1=l_2, \\ 
 0, \quad& \mathrm{otherwise}.
 \end{cases}
 \end{align}
For example, $N_q=K=3$ GLL quadrature yields
 \begin{align}
 \mathcal{H}^{(st)}_{1\,l_1\,l_2} = \text{diag} \begin{Bmatrix}
 1/3 \\ 0 \\ 0  
 \end{Bmatrix}, \quad 
 \mathcal{H}^{(e)}_{2\,l_1\,l_2} =\text{diag} 
 \begin{Bmatrix}
 0 \\ 4/3 \\ 0  
 \end{Bmatrix}, \quad
 \mathcal{H}^{(e)}_{3\,l_1\,l_2} = \text{diag} 
 \begin{Bmatrix}
 0 \\ 0 \\ 1/3  
 \end{Bmatrix}.
 \end{align}
Therefore, in this special case, the total complexity to evaluate the collision term $\mathcal{Q}^{(ij)} \left(\mathcal{F}^\I_{e,\,l_1},  \mathcal{F}^\J_{e,\,l_2}\right)$ is reduced to $O(s^2KMN^4\log N)$ (for all pairs of $(i,j)$ and $(l_1,l_2)$ such that $\mathcal{H}_{m\,l_1\,l_2}^e\neq 0$). Of course, this improvement in efficiency comes with an accuracy loss which is quite complicated to analyze. Nevertheless, all numerical results presented in this paper are produced using the above described approach and the bulk properties such as density, temperature, etc. are found to be in good agreement with reference solutions (available finite difference solutions or DSMC solutions). A detailed study of numerical accuracy would be a subject of future work.

\section{Numerical experiments}
\label{sec_numericalExperiments}

\subsection{Spatially homogeneous case: Krook-Wu exact solution}

For constant collision kernel, an exact solution to the spatially homogeneous multi-species Boltzmann equation can be constructed (see \cite{krook1977exact}). We use this solution to verify the accuracy of the proposed fast spectral method for approximating the collision operator. Considering a binary mixture, the equation simplifies to
\begin{align}
\partial_t f^\I = \sum_{j=1}^{2}\int_{\mathbb{R}^3}\int_{S^{2}} B_{ij}\left[f^{(i)}(v')f^{(j)}(v_*')-f^{(i)}(v)f^{(j)}(v_*)\right]\rd{\sigma}\,\rd{v_*},
\label{eq_bkw_pde}
\end{align}
where $B_{ij}=B_{ji}:=\frac{\lambda_{ji}}{4\pi n^{(j)}}$ and $\lambda_{ij}$ is some positive constant. The exact solution is given by
\begin{align}
f^\I(t,v) = n^\I \Bigg( \frac{m_i}{2\pi K} \Bigg)^{3/2} \exp{\Bigg(-\frac{m_i v^2}{2K}\Bigg)} \Bigg((1-3Q_i) + \frac{m_i}{K} Q_i v^2 \Bigg), \quad i=1,2,
\label{eq_bkw_f}
\end{align}
where
\begin{align}
& \mu=\frac{4m_1 m_2}{(m_1+m_2)^2}, \quad p_1 = \lambda_{22} - \lambda_{21} \mu (3-2\mu), \quad p_2 = \lambda_{11} - \lambda_{12} \mu (3-2\mu), \nonumber \\
& A = \frac{1}{6} \Bigg( \lambda_{11} + \lambda_{21} \mu\left(3 -2 \mu \frac{p_2}{p_1}\right) \Bigg), \quad B = \frac{1}{3} \Bigg( \lambda_{11} p_1 + \lambda_{21} \mu (3 -2 \mu) p_2 \Bigg),\nonumber \\
& Q(t) = \frac{A}{A\exp(A t)-B}, \quad Q_i(t) = p_i Q(t),   \nonumber\\
&K(t) = \frac{n^\one + n^\two}{ (n^\one + n^\two) + 2( n^\one p_1 + n^\two p_2) Q(t) }.
\end{align}
Furthermore, the following condition needs to be satisfied
\begin{equation}
(p_1-p_2)\left(2\mu^2\left(\frac{\lambda_{21}}{p_1}-\frac{\lambda_{12}}{p_2}\right)-1\right)=0.
\end{equation}
For simplicity, we choose $n^\one=n^\two=1$, $\lambda_{11}=\lambda_{22}=1$, $\lambda_{12}=\lambda_{21}=1/2$ but vary the mass ratio $m_1/m_2$ in the following tests.

It is also helpful to take the derivative of eqn.~(\ref{eq_bkw_f}), which yields
\begin{align}
\partial_t f^\I&= f^\I \Bigg( -\frac{3}{2K}K' + \frac{m_i\;v^2}{2 K^2} K'\Bigg) + n^\I \Bigg(\frac{m_i}{2\pi K}\Bigg)^{3/2} \exp{\Bigg(-\frac{m_i v^2}{2K}\Bigg)} \Bigg(-3Q'_i + \frac{m_i}{K} Q'_i v^2 - \frac{m_i}{K^2} K' Q_i v^2 \Bigg) \nonumber\\
&:=\sum_{j=1}^2\mathcal{Q}^{(ij)}(f^{(i)},f^{(j)}),
\end{align}
where
\begin{align}
&Q'(t)=-\frac{A^3\exp(At)}{(A\exp(At)-B)^2}, \quad Q_i'(t)=p_iQ'(t), \quad K'(t)=-\frac{2(n^{(1)}+n^{(2)})(n^{(1)}p_1+n^{(2)}p_2)}{[(n^{(1)}+n^{(2)})+2(n^{(1)}p_1+n^{(2)}p_2)Q(t)]^2}Q'(t).
\end{align}
This allows us to check the accuracy of the collision solver without introducing time discretization error.

\begin{figure}[!ht]
\centering
\begin{subfigure}{.5\textwidth}
  \centering
  \includegraphics[width=85mm]{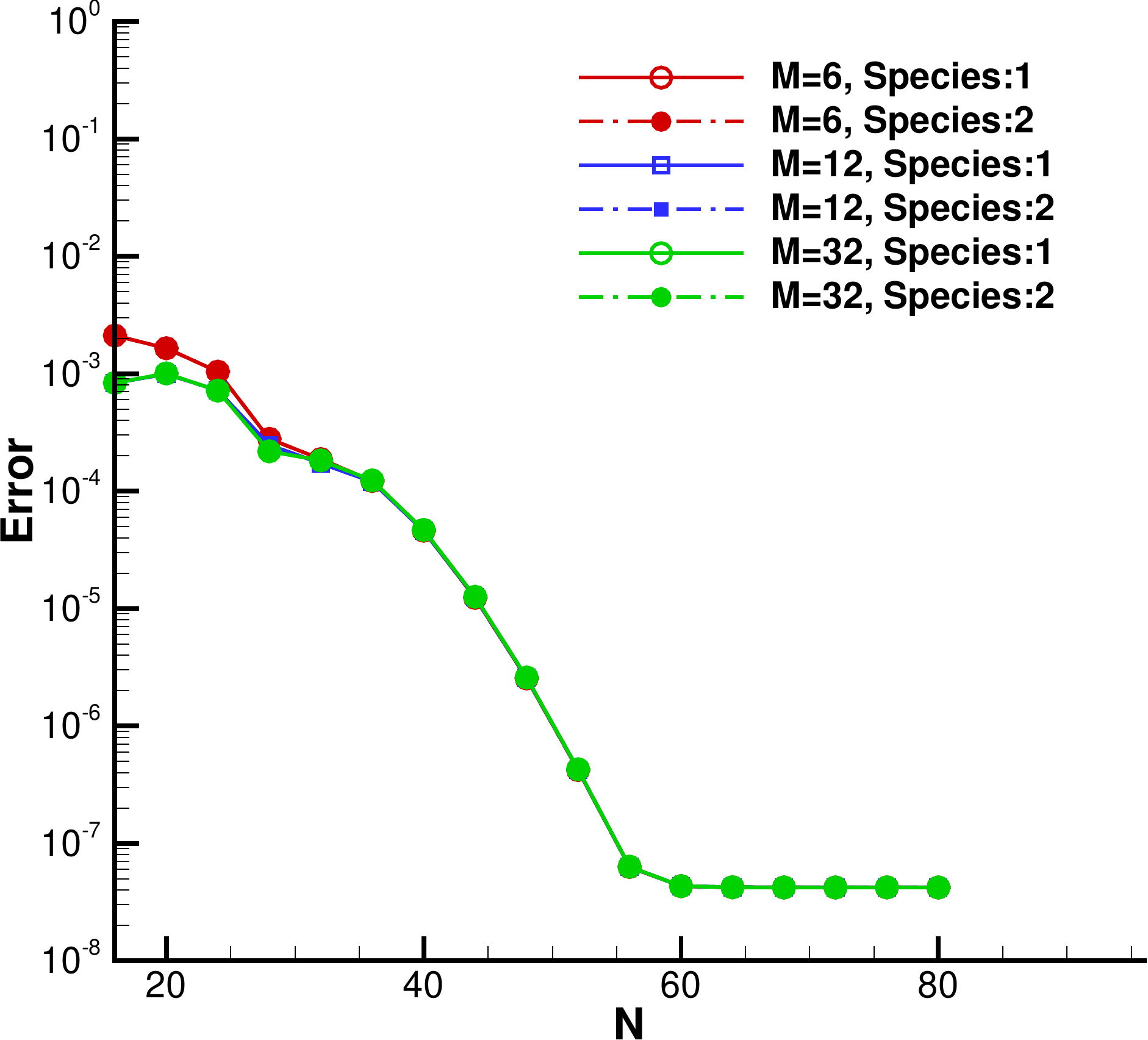}
  \caption{$m_1/m_2=1$}
  \label{fig_bkw_errorVsMassRatio_mr_1}
\end{subfigure}%
\begin{subfigure}{.5\textwidth}
  \centering
  \includegraphics[width=85mm]{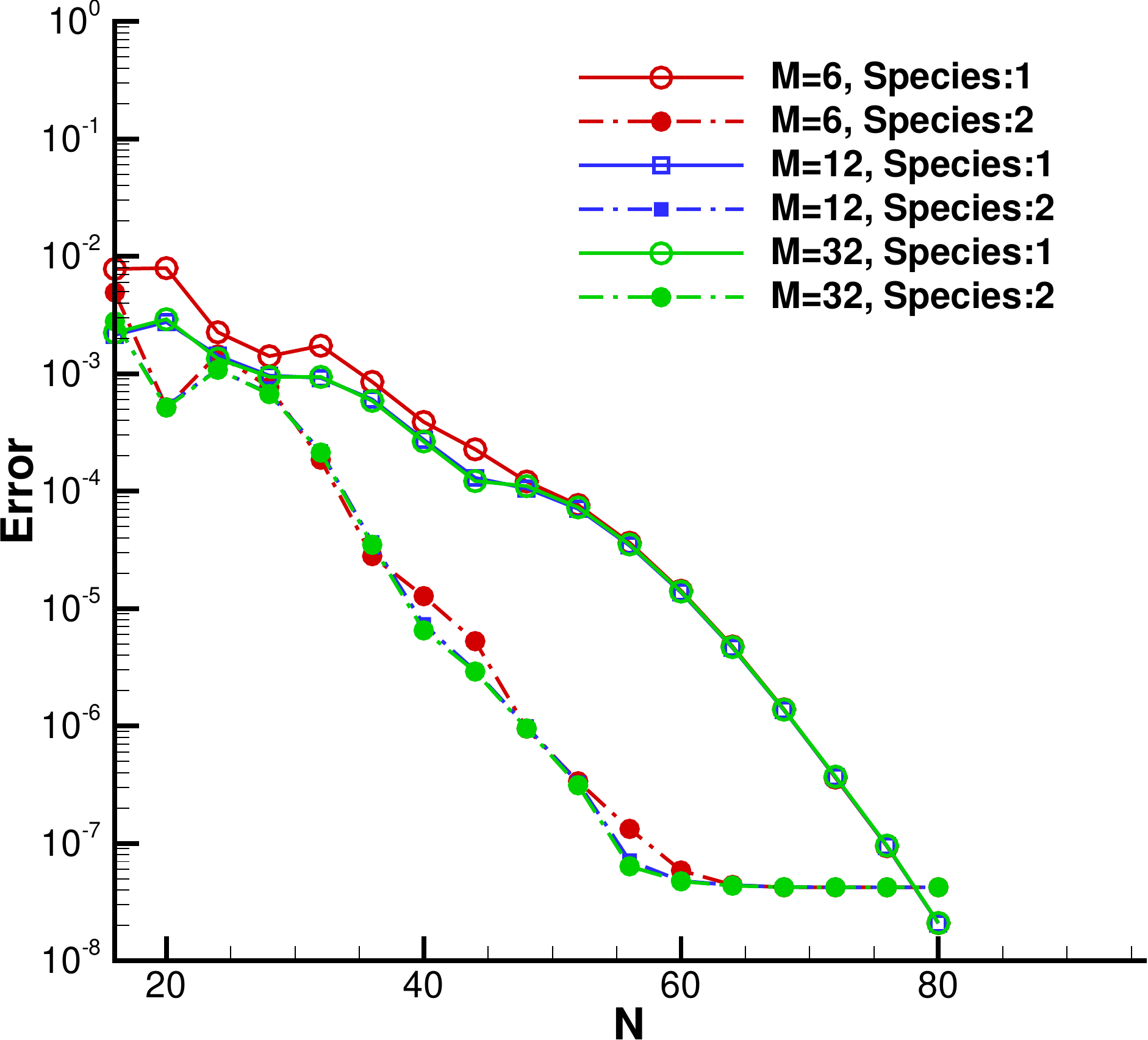}
  \caption{$m_1/m_2=2$}
  \label{fig_bkw_errorVsMassRatio_mr_2}
\end{subfigure}
\begin{subfigure}{.5\textwidth}
  \centering
  \includegraphics[width=85mm]{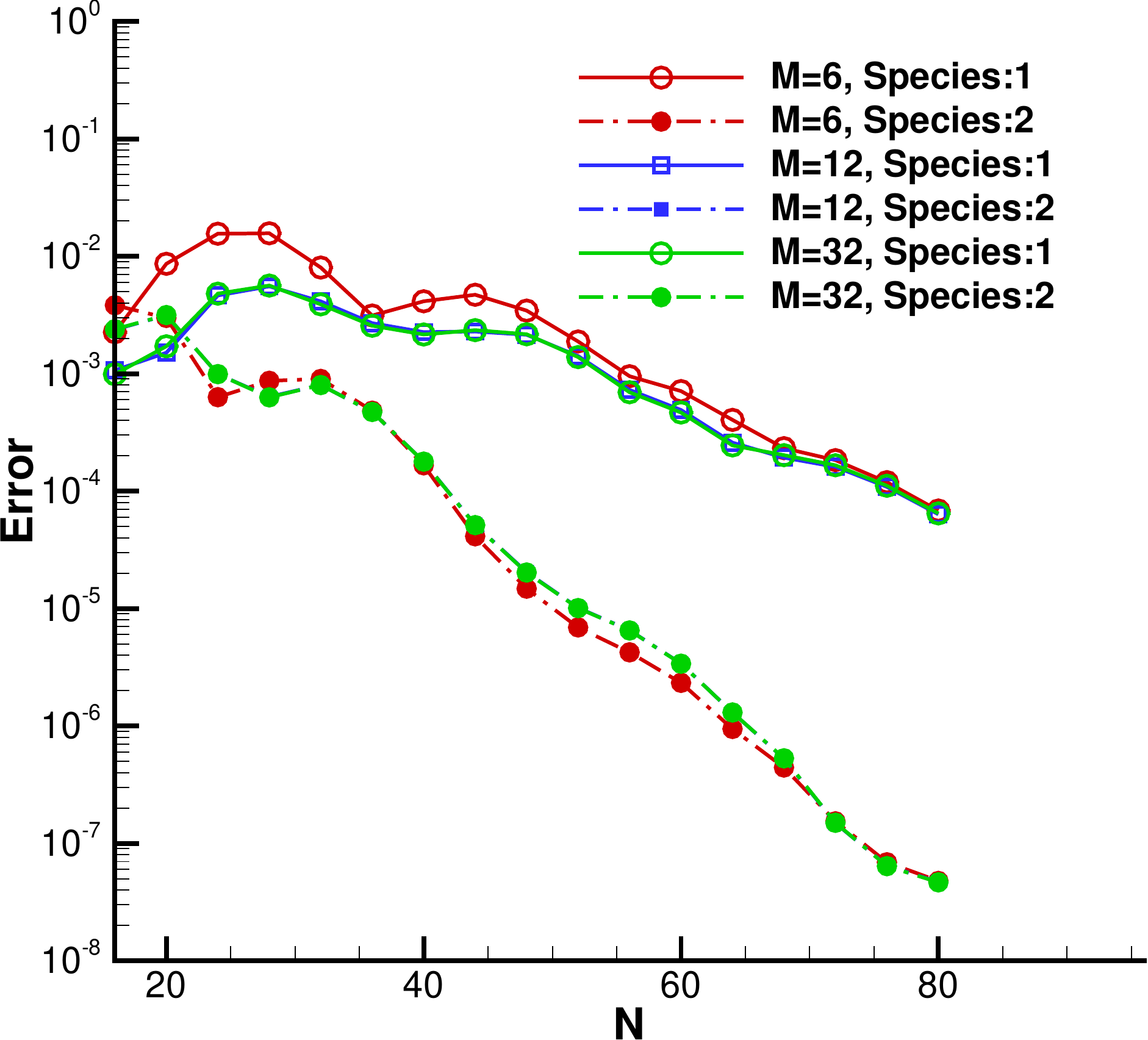}
  \caption{$m_1/m_2=4$}
  \label{fig_bkw_errorVsMassRatio_mr_4}
\end{subfigure}%
\begin{subfigure}{.5\textwidth}
  \centering
  \includegraphics[width=85mm]{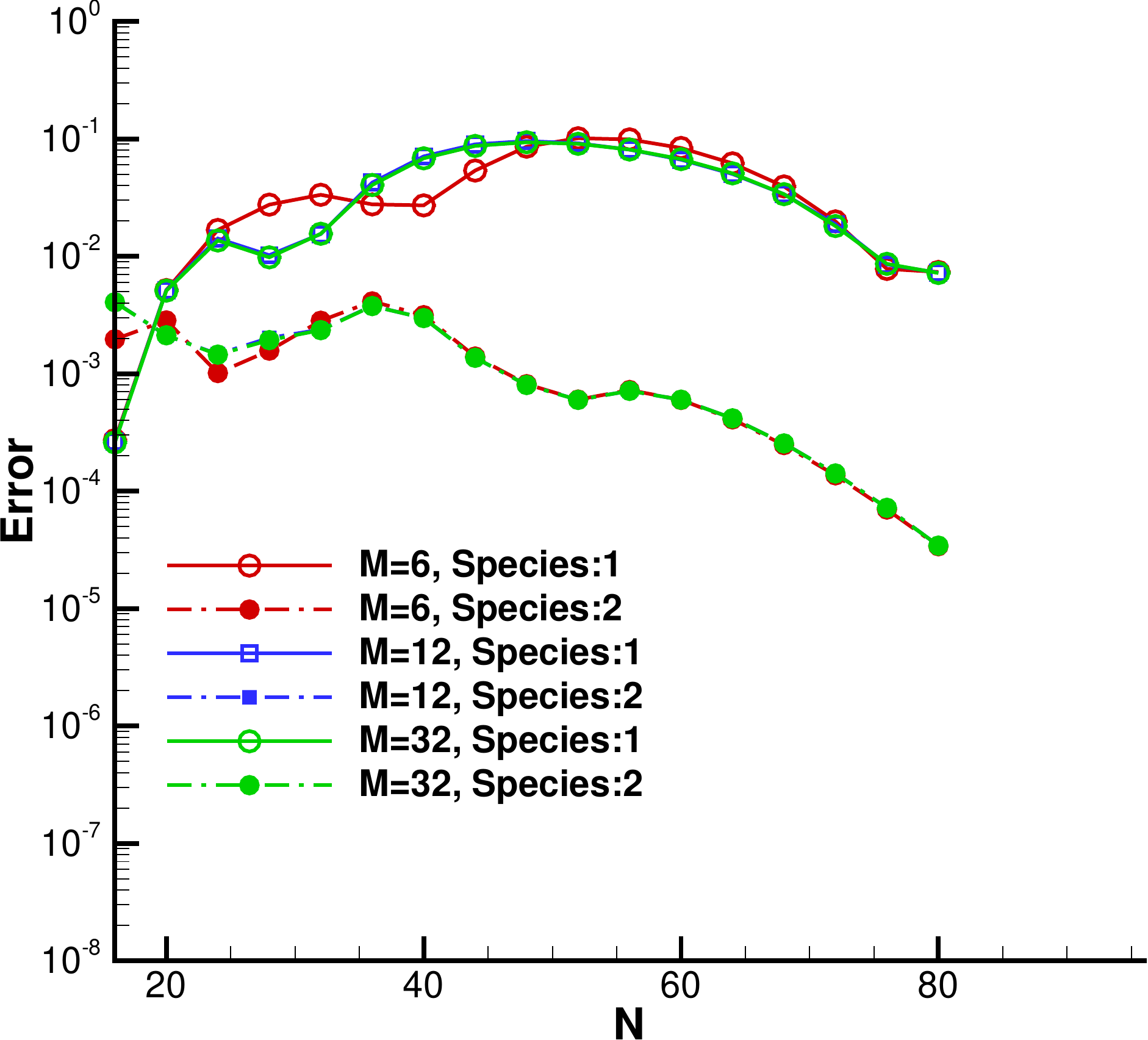}
  \caption{$m_1/m_2=8$}
  \label{fig_bkw_errorVsMassRatio_mr=8}
\end{subfigure}
\caption{Spatially homogeneous Krook-Wu solution. $L^\infty$ error $\mathcal{E}^\I = \| \partial_t f_{exact}^\I - \partial_t f_{numerical}^\I \|_{L^{\infty}},\;i=\{1,2\}$ at $t=4$ for different mass ratios. $N$ is the number of discretization points in each velocity dimension and $M$ is the number of spherical design quadrature points used on the sphere. Number of Gauss-Legendre quadrature points $N_{\rho}$ in the radial direction is fixed to $N$. A fixed velocity domain $[-12,\;12]^3$ has been used for all cases.}
\label{fig_bkw_errorVsMassRatio}
\end{figure}

Figure~\ref{fig_bkw_errorVsMassRatio} depicts the convergence behavior of the proposed fast algorithm with respect to $N$ for different mass ratios. Due to the isotropic nature of the solution, we observe that the errors remain relatively unaffected for different $M$ (number of quadrature points used on the sphere). On the other hand, the method exhibits a spectral convergence as $N$ (number of discretization points in each velocity dimension) increases. It is also clear that the accuracy deteriorates for large mass ratios (to keep the same level of accuracy, larger $N$ is needed). To understand the influence of $N_{\rho}$ (number of quadrature points in the radial direction), we list in Table~\ref{tab_bkw_error} the errors of the method with respect to different $N_{\rho}$. It can be observed that the error is relatively unaffected upon reducing $N_\rho$ from $N$ to $N/2$.

\begin{table}[!ht]
\centering
\setlength{\tabcolsep}{0.3em}
\begin{tabular}{@{}cc|cc|cc|cc|cc|cc@{}}
\toprule
\multirow{3}{*}{$N$} & \multirow{3}{*}{$N_\rho$} & \multicolumn{2}{c|}{$m_1/m_2=1$} & \multicolumn{2}{c|}{$m_1/m_2=2$} & \multicolumn{2}{c|}{$m_1/m_2=4$} & \multicolumn{2}{c|}{$m_1/m_2=6$} & \multicolumn{2}{c}{$m_1/m_2=8$} \\ 
   &   &      $\mathcal{E}^\one$ & $\mathcal{E}^\two$ & $\mathcal{E}^\one$ & $\mathcal{E}^\two$ & $\mathcal{E}^\one$ & $\mathcal{E}^\two$ & $\mathcal{E}^\one$ & $\mathcal{E}^\two$ & $\mathcal{E}^\one$ & $\mathcal{E}^\two$ \\ \midrule
16 & 4 & 1.528e-03 & 1.528e-03 & 6.675e-03 & 4.444e-03 & 2.048e-03 & 3.633e-03 & 1.414e-03 & 4.941e-04 & 2.709e-04 & 2.481e-03 \\
 & 8 & 2.114e-03 & 2.114e-03 & 7.795e-03 & 4.917e-03 & 2.253e-03 & 3.828e-03 & 1.425e-03 & 7.087e-04 & 2.715e-04 & 1.963e-03 \\
 & 16 & 2.114e-03 & 2.114e-03 & 7.795e-03 & 4.917e-03 & 2.253e-03 & 3.828e-03 & 1.425e-03 & 7.086e-04 & 2.715e-04 & 1.963e-03 \\
32 & 8 & 1.526e-04 & 1.526e-04 & 1.671e-03 & 2.237e-04 & 7.249e-03 & 9.770e-04 & 1.692e-02 & 1.701e-03 & 3.029e-02 & 2.928e-03 \\
 & 16 & 1.873e-04 & 1.873e-04 & 1.729e-03 & 1.852e-04 & 8.018e-03 & 9.020e-04 & 1.935e-02 & 1.652e-03 & 3.338e-02 & 2.822e-03 \\
 & 32 & 1.873e-04 & 1.873e-04 & 1.729e-03 & 1.852e-04 & 8.018e-03 & 9.020e-04 & 1.935e-02 & 1.652e-03 & 3.338e-02 & 2.822e-03 \\
64 & 16 & 4.227e-08 & 4.227e-08 & 4.704e-06 & 4.749e-08 & 4.201e-04 & 3.460e-06 & 5.263e-03 & 4.056e-05 & 6.231e-02 & 4.053e-04 \\
 & 32 & 4.227e-08 & 4.227e-08 & 4.754e-06 & 4.422e-08 & 4.043e-04 & 9.441e-07 & 5.153e-03 & 3.381e-05 & 6.186e-02 & 4.082e-04 \\
 & 64 & 4.227e-08 & 4.227e-08 & 4.754e-06 & 4.422e-08 & 4.043e-04 & 9.441e-07 & 5.153e-03 & 3.381e-05 & 6.186e-02 & 4.082e-04 \\
\bottomrule
\end{tabular}
\caption{Spatially homogeneous Krook-Wu solution. $L^\infty$ error $\mathcal{E}^\I = \| \partial_t f_{exact}^\I - \partial_t f_{numerical}^\I \|_{L^{\infty}},\;i=\{1,2\}$ at $t=4$ for different mass ratios. $N$ is the number of discretization points in each velocity dimension and $N_{\rho}$ is the number of Gauss quadrature points used in the radial direction. Number of quadrature points $M$ used on the sphere is fixed to $6$. A fixed velocity domain $[-12,\;12]^3$ has been used for all cases.}
\label{tab_bkw_error}
\end{table}

\begin{figure*}[tbp]
\centering
\begin{subfigure}{.5\textwidth}
  \centering
  \includegraphics[width=70mm]{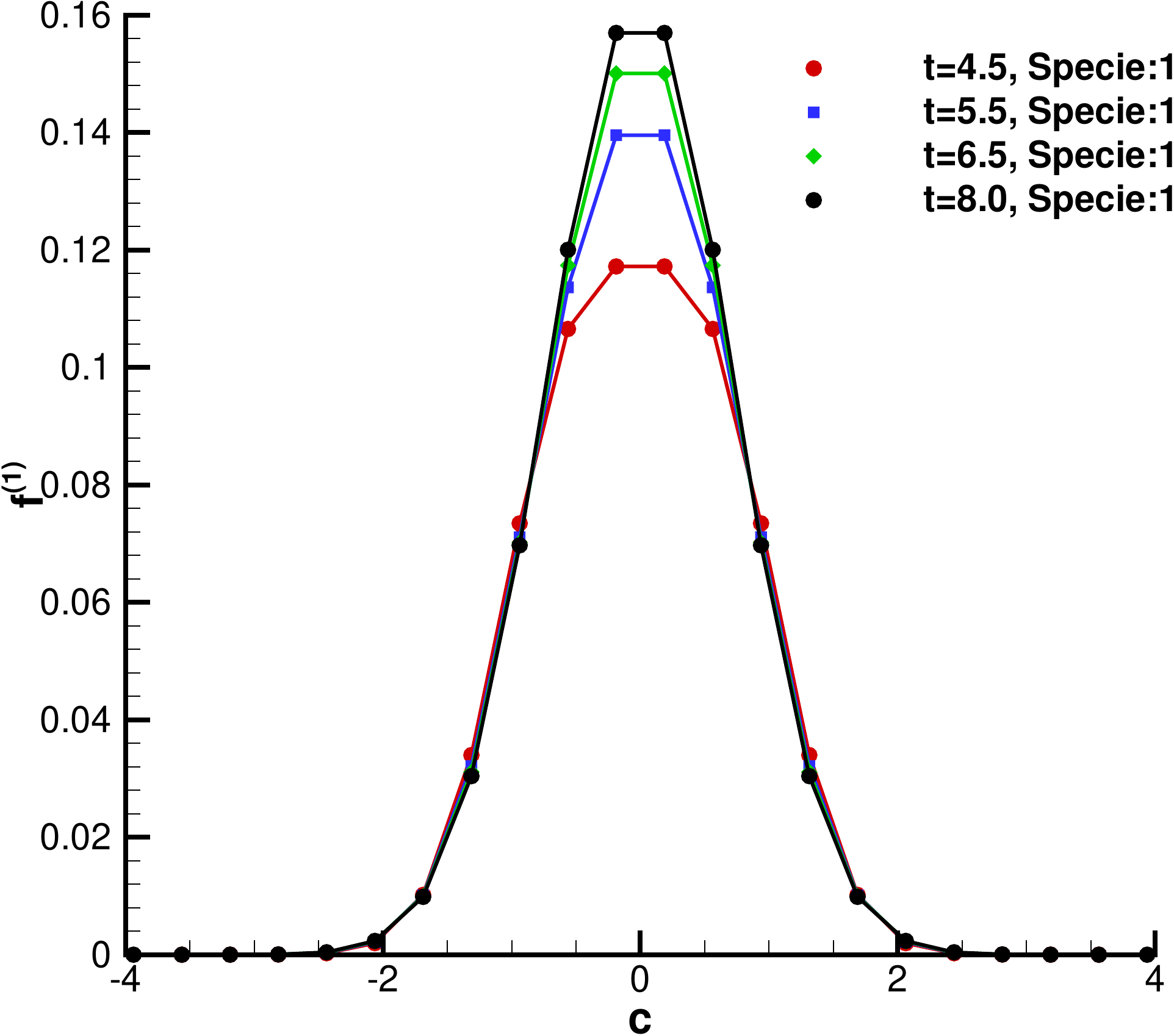}
  \caption{$m_1/m_2=2$, species 1}
  \label{fig_bkw_variationF_mr_2_s_1}
\end{subfigure}%
\begin{subfigure}{.5\textwidth}
  \centering
  \includegraphics[width=70mm]{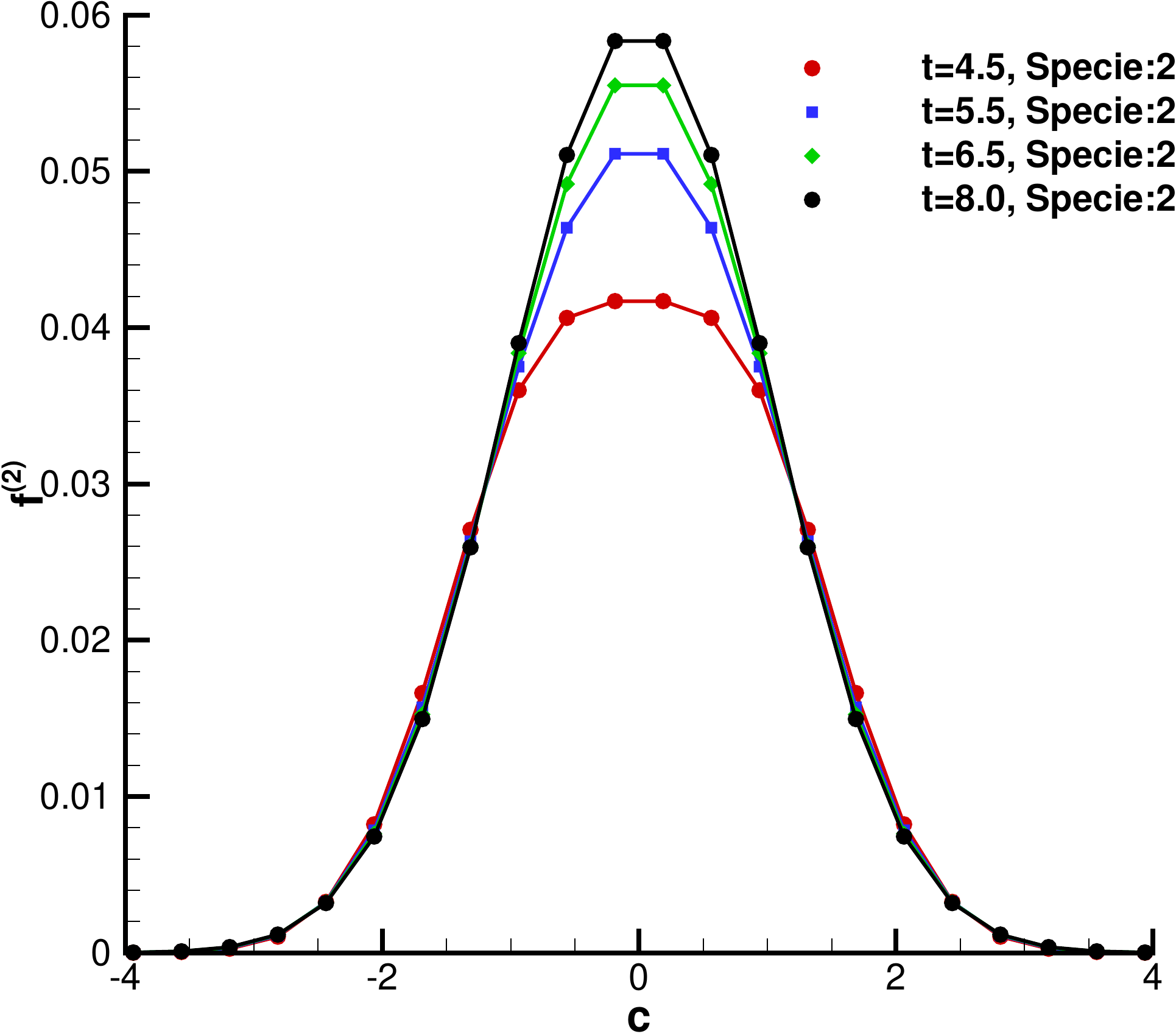}
  \caption{$m_1/m_2=2$, species 2}
  \label{fig_bkw_variationF_mr_2_s_2}
\end{subfigure}
\begin{subfigure}{.5\textwidth}
  \centering
  \includegraphics[width=70mm]{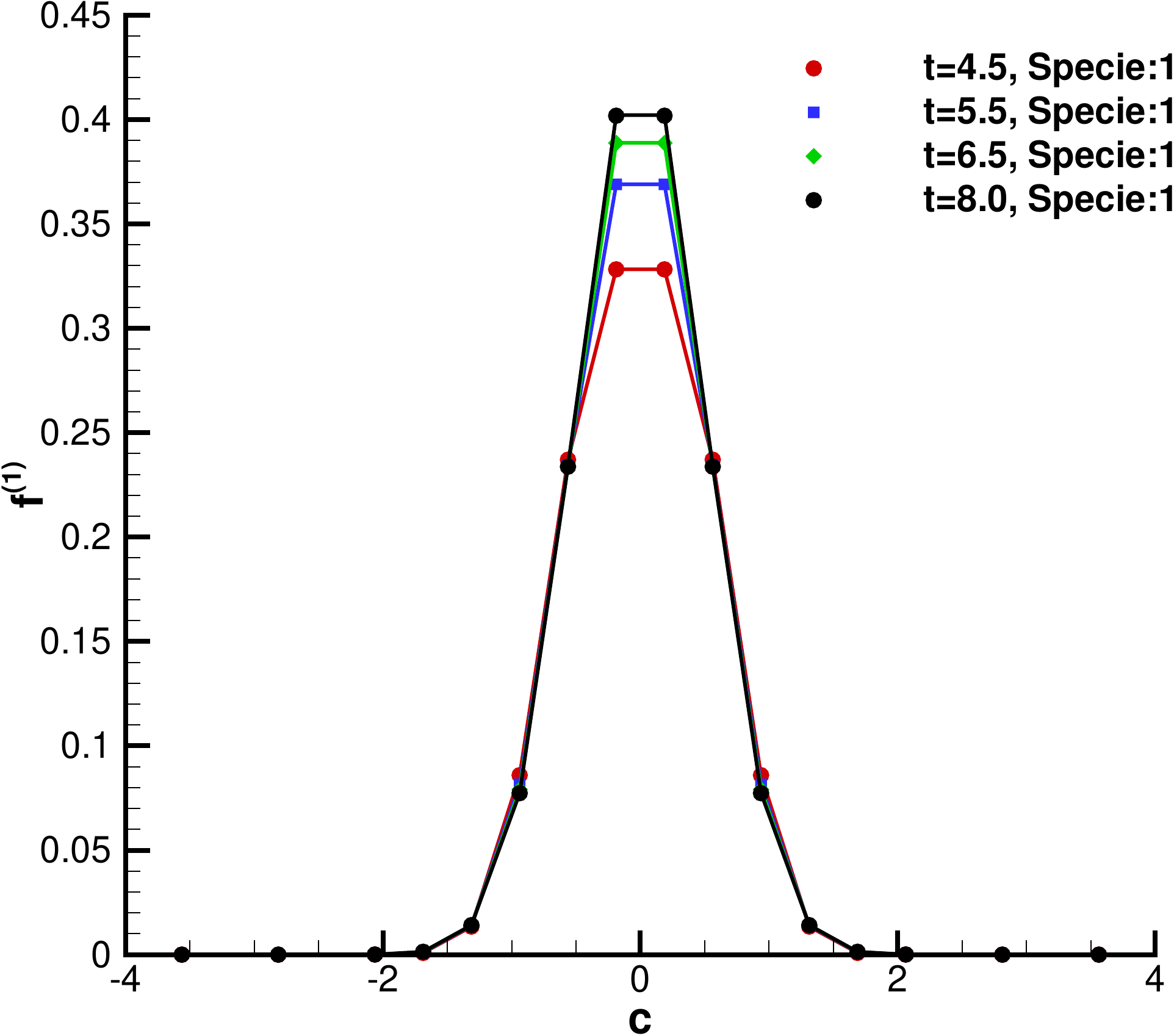}
  \caption{$m_1/m_2= 4$, species 1}
  \label{fig_bkw_variationF_mr_4_s_1}
\end{subfigure}%
\begin{subfigure}{.5\textwidth}
  \centering
  \includegraphics[width=70mm]{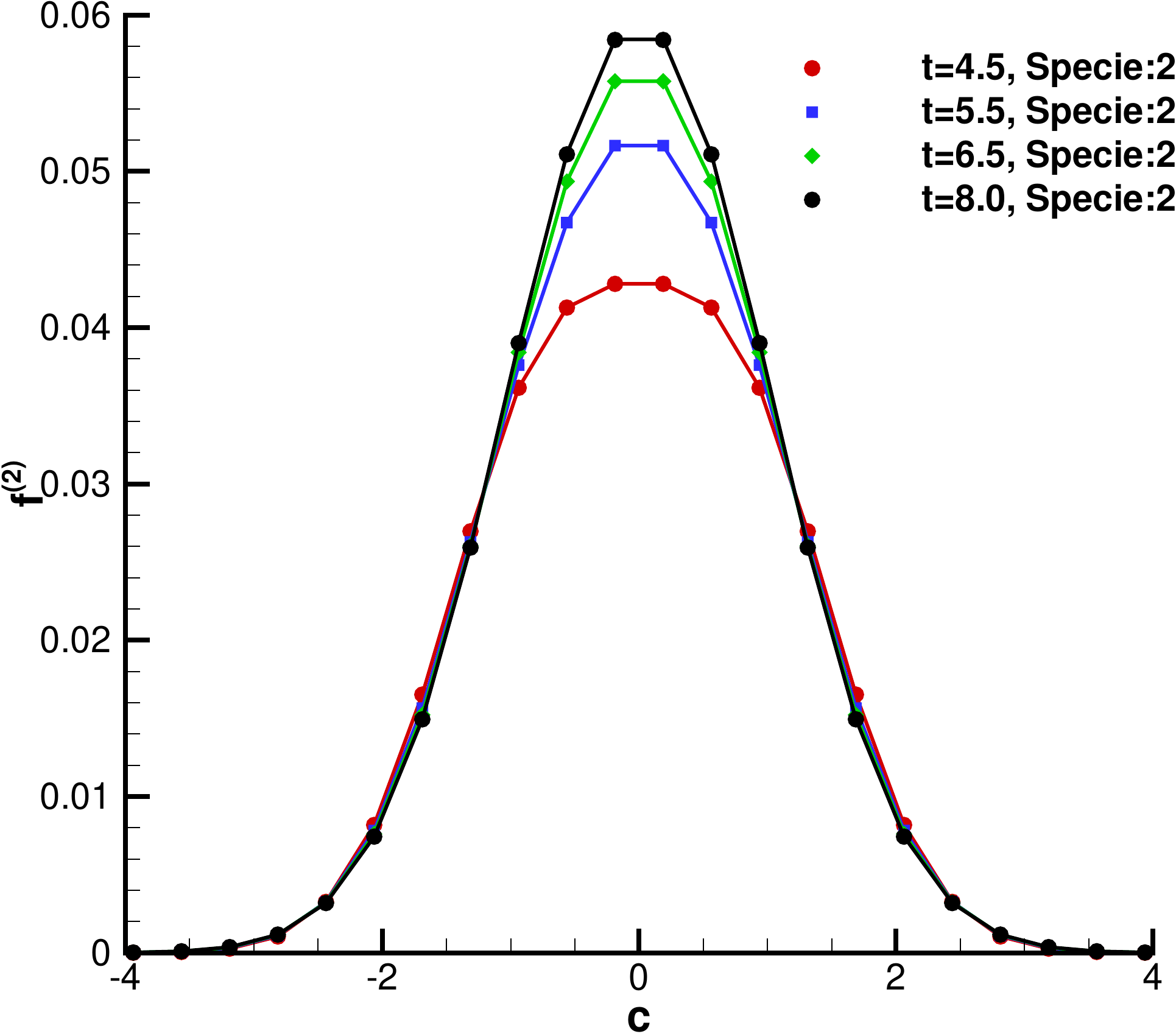}
  \caption{$m_1/m_2=4$, species 2}
  \label{fig_bkw_variationF_mr_4_s_2}
\end{subfigure}
\begin{subfigure}{.5\textwidth}
  \centering
  \includegraphics[width=70mm]{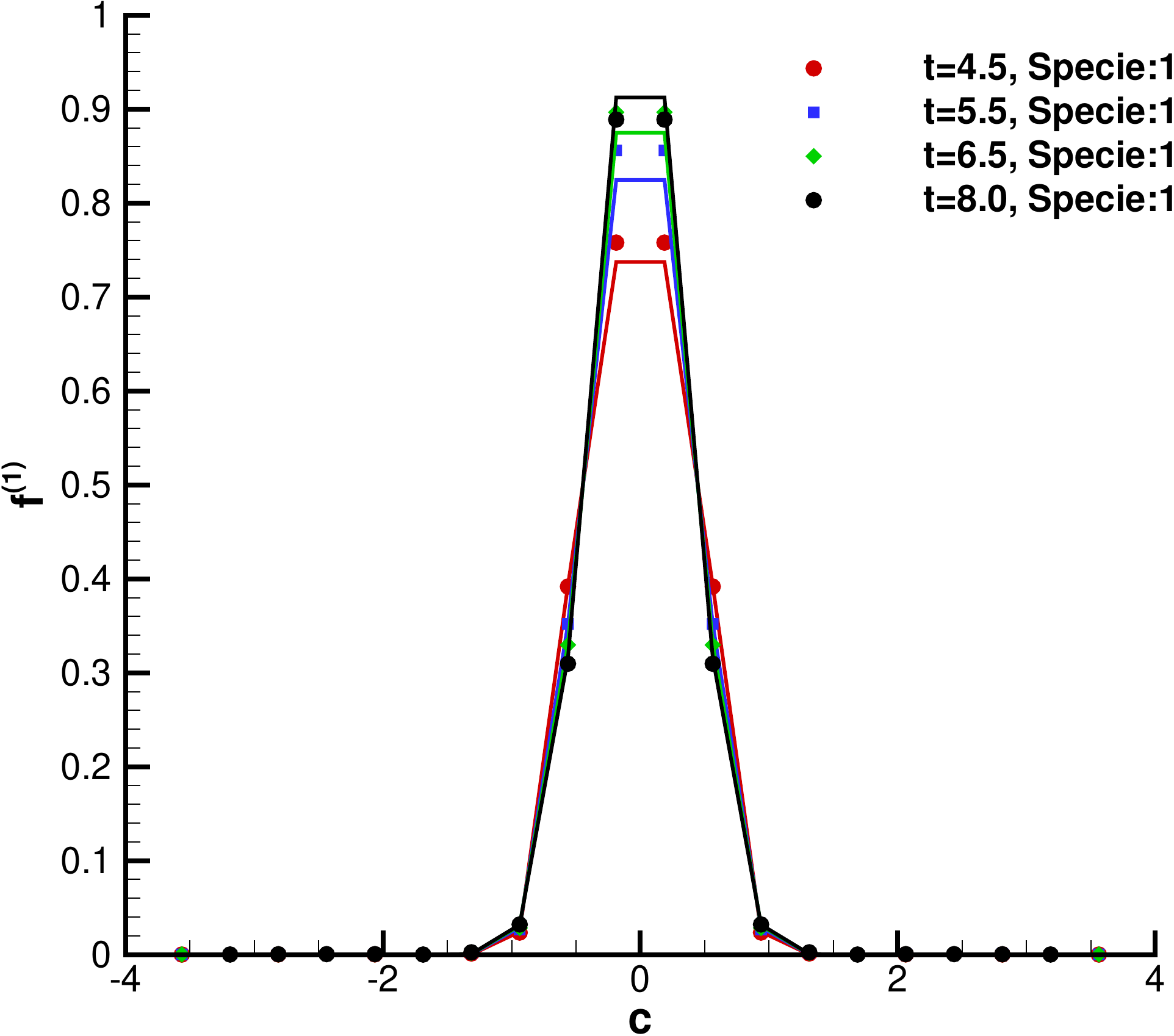}
  \caption{$m_1/m_2=8$, species 1}
  \label{fig_bkw_variationF_mr_8_s_1}
\end{subfigure}%
\begin{subfigure}{.5\textwidth}
  \centering
  \includegraphics[width=70mm]{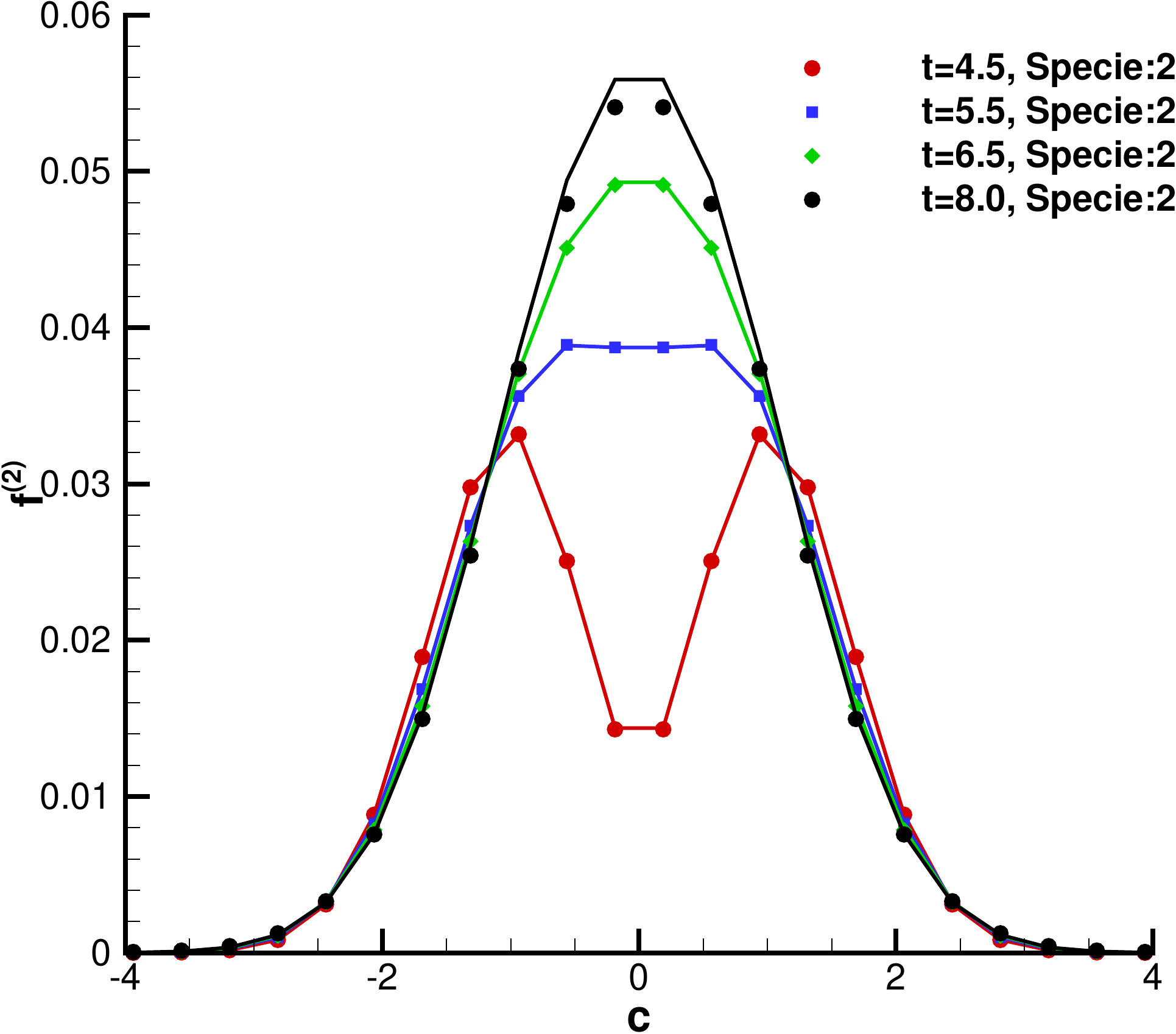}
  \caption{$m_1/m_2=8$, species 2}
  \label{fig_bkw_variationF_mr_8_s_2}
\end{subfigure}
\caption{Spatially homogeneous Krook-Wu solution. Evolution of $\displaystyle f^\I,\;i=\{1,2\}$ sliced along the velocity domain centerline, i.e., $f^\I(:, N/2, N/2)$ for different mass ratios. The exact solutions (solid lines) are plotted using $N=64$. The numerical solutions (symbols) are evaluated using $N=64$, $M=6$, $N_{\rho}=64$. A fixed velocity domain $[-12,\;12]^3$ has been used for all cases. SSP-RK2 with $\Delta t=0.01$ is used for time stepping. Note that the x-axis has been zoomed to $[-4,4]$ for better visibility. }
\label{fig_bkw_variationF}
\end{figure*}

Next we evolve the solution using the SSP-RK2 with time step $\Delta t=0.01$. Figure~\ref{fig_bkw_variationF} illustrates the time evolution of the distribution function sliced along the velocity domain centerline, i.e., $f^\I(:, N/2, N/2)$. It is observed that: a) the distribution function of the heavy particles becomes more skewed as the mass ratio increases; b) as time goes by, the distribution function tends toward the Maxwellian. 

\subsection{Spatially inhomogeneous case}

\subsubsection{Normal shock with HS collision kernel}
\label{sec:normal}

As a first example in the spatially inhomogeneous case, we consider the normal shock wave and compare our results with the finite difference solutions reported in \cite{KAT01}. Four cases are considered here whose numerical parameters are described in Table~\ref{tab_normalShock_conditions}. The boundary conditions at upstream and downstream are the in-flow equilibrium boundary (see eqn.~(\ref{eq_bc_inlet})). We solve the Boltzmann equation until the solution reaches a steady state. A convergence criterion of $(\|f^{n+1}-f^{n}\|_{L^2}/\|f^{n}\|_{L^2})$ /$(\|f^{2}-f^{1}\|_{L^2}/\|f^{1}\|_{L^2}) < 2\times10^{-5}$ has been used, where $f^{n}$ denotes the distribution function at $n^{th}$ time step. 

\begin{table}[!ht]
\centering
\begin{tabular}{@{}lcccc@{}}
\toprule
Parameter & Case NS-01 & Case NS-02 & Case NS-03 & Case NS-04 \\ 
\midrule
Molecular mass: $m_1$ ($\times 10^{27}\,kg$) & $6.63$ & $6.63$ & $6.63$ & $6.63$ \\ 
Molecular mass: $m_2$ ($\times 10^{27}\,kg$) & $3.315$ & $1.6575$ & $3.315$ & $3.315$ \\ 
Mass Ratio: $m_2/m_1$ & 0.5 & 0.25 & 0.5 & 0.5 \\ 
Mach number & 1.5 & 1.5 & 1.5 & 3.0 \\
Concentration: $n^\two_-/n_-=n^\two_+/n_+$ & 0.5 & 0.5 & 0.1 & 0.1\\
Non-dim physical space & $[-0.5,\,0.5]$ & $[-0.5,\,0.5]$ & $[-0.5,\,0.5]$ & $[-0.5,\,0.5]$ \\ 
Non-dim velocity space & $[-9,\,9]^3$ & $[-15,\,15]^3$ & $[-9,\,9]^3$ & $[-15,\,15]^3$ \\ 
$N^3$ & $32^3$ & $64^3$ & $32^3$ & $48^3$ \\
$N_\rho$ & $32$ & $16$ & $32$ & $48$ \\
$M$ & 12 & 12 & 12 & 12 \\
Spatial elements & $8,\,16$ & 16 & 16 & 16 \\
DG order & 3 & 3 & 3 & 3 \\
Time step ($s\times 10^{8}$) & $5.57,\,2.77$ & $1.64$ & $2.77$ & $1.64$ \\
Viscosity index: $\omega_{ij}$ & 0.5 & 0.5 & 0.5 & 0.5 \\
Scattering parameter: $\alpha_{ij}$ & 1 & 1 & 1 & 1 \\
Ref. diameter: $d_{\text{ref},ij}$ ($\times 10^{10} m$) & $2.17$ & $2.17$ & $2.17$ & $2.17$ \\
Ref. temperature: $T_{\text{ref},ij}$ ($K$) & 273 & 273 & 273 & 273 \\
Characteristic mass: $m_0$ ($\times 10^{27}\,kg$) & $6.63$ & $6.63$ & $6.63$ & $6.63$ \\
Characteristic length: $H_0$ ($mm$) & 30 & 30 & 30 & 30 \\
Characteristic velocity: $u_0$ ($m/s$) & 963.7 & 963.7 & 963.7 & 963.7 \\
Characteristic temperature: $T_0$ ($K$) & 223 & 223 & 223 & 223 \\
Characteristic number density: $n_0$ ($m^{-3}$) & $2.889 \times 10^{21}$ & $2.889 \times 10^{21}$ & $2.889 \times 10^{21}$ & $2.889 \times 10^{21}$ \\
\midrule
\multicolumn{4}{l}{Upstream conditions (subscript -)} \\
Velocity: $u_-$ ($m/s$) & 1523.737 & 1669.171 & 1353.876 & 2707.753 \\
Temperature: $T_-$ ($K$) & 223 & 223 & 223 & 223 \\
Mean free path: $\lambda_- = (\sqrt{2}\,\pi\,(n^\one_- + n^\two_-)\,d_{\text{ref},ij}^2)^{-1}$ ($m$) & 0.000827 & 0.000827 & 0.00148 & 0.00148 \\
Number density: $n^\one_-$ ($m^{-3}$) & $2.889 \times 10^{21}$ & $2.889 \times 10^{21}$ & $2.889 \times 10^{21}$ & $2.889 \times 10^{21}$ \\
Number density: $n^\two_-$ ($m^{-3}$) & $2.889 \times 10^{21}$ & $2.889 \times 10^{21}$ & $3.209 \times 10^{20}$ & $3.209 \times 10^{20}$ \\
\midrule
\multicolumn{4}{l}{Downstream conditions (subscript +)} \\
Velocity: $u_+$ ($m/s$) & 888.847 & 973.683 & 789.761 & 902.584 \\
Temperature: $T_+$ ($K$) & 333.338 & 333.338 & 333.338 & 817.667 \\
Number density: $n^\one_+$ ($m^{-3}$) & $4.953 \times 10^{21}$ & $4.953 \times 10^{21}$ & $4.953 \times 10^{21}$ & $8.669 \times 10^{21}$ \\
Number density: $n^\two_+$ ($m^{-3}$) & $4.953 \times 10^{21}$ & $4.953 \times 10^{21}$ & $5.502 \times 10^{20}$ & $9.633 \times 10^{20}$ \\
\midrule
\multicolumn{2}{l}{Initial conditions} \\
Velocity: $u$ ($m/s$) & \multicolumn{4}{l}{$u_- + (u_+ - u_-)\;x/H_0$} \\
Temperature: $T$ ($K$) & \multicolumn{4}{l}{$T_- + (T_+ - T_-)\;x/H_0$} \\
Number density: $n^\one$ ($m^{-3}$) & \multicolumn{4}{l}{$n^\one_- + (n^\one_+ - n^\one_-)\;x/H_0$} \\
Number density: $n^\two$ ($m^{-3}$) & \multicolumn{4}{l}{$n^\two_- + (n^\two_+ - n^\two_-)\;x/H_0$} \\
\bottomrule
\end{tabular}
\caption{Numerical parameters for normal shock wave \cite{KAT01}.}
\label{tab_normalShock_conditions}
\end{table}

\begin{figure*}[tbp]
\begin{subfigure}{0.5\textwidth}
  \centering
  \includegraphics[width=70mm]{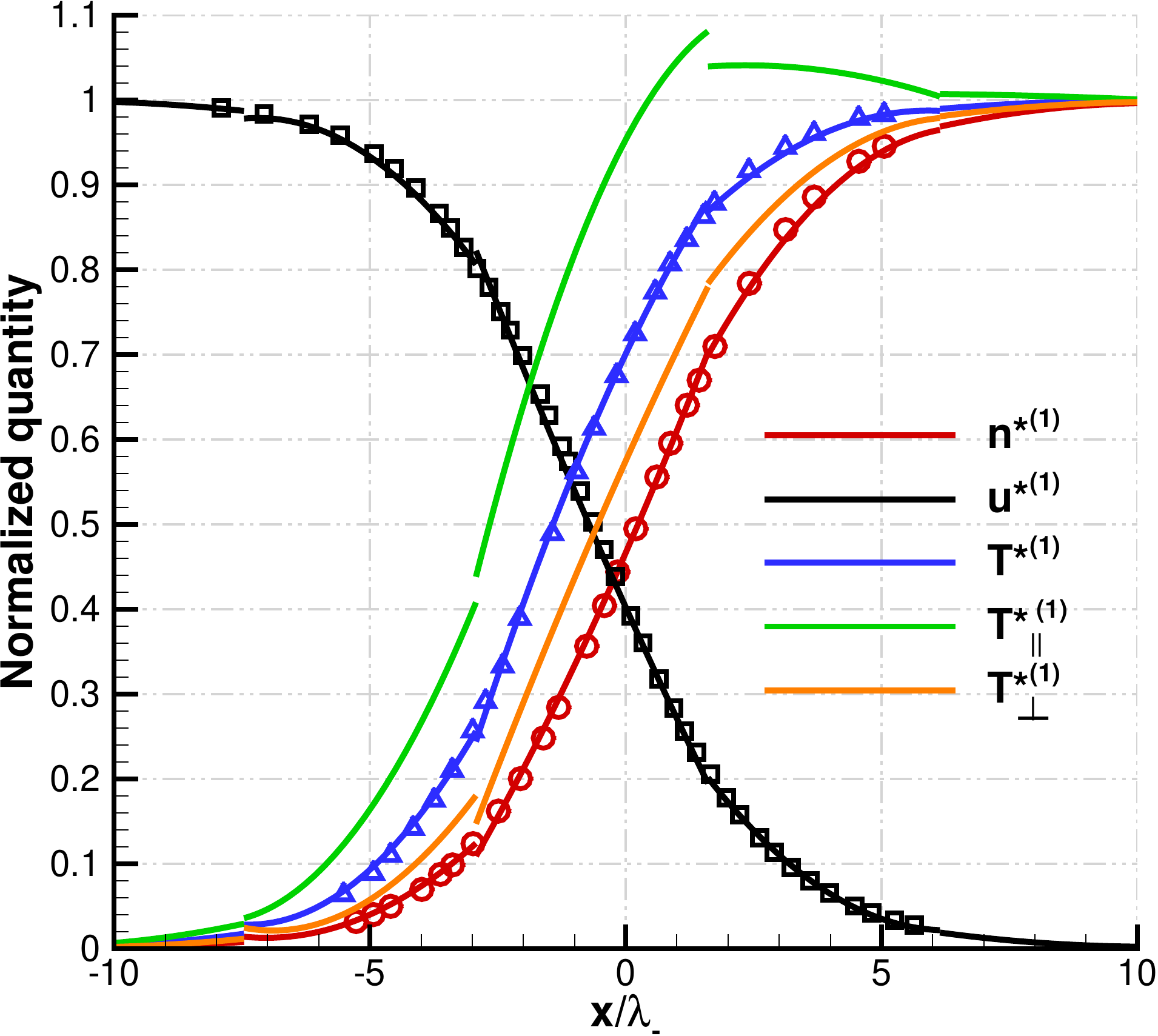}
  \caption{Case NS-01, 8 elements, species 1}
  \label{fig_shockHeKosuge_Ma_1_5_frac_0_5_mr_0_5_8elem_species1}
\end{subfigure}
\begin{subfigure}{0.5\textwidth}
  \centering
  \includegraphics[width=70mm]{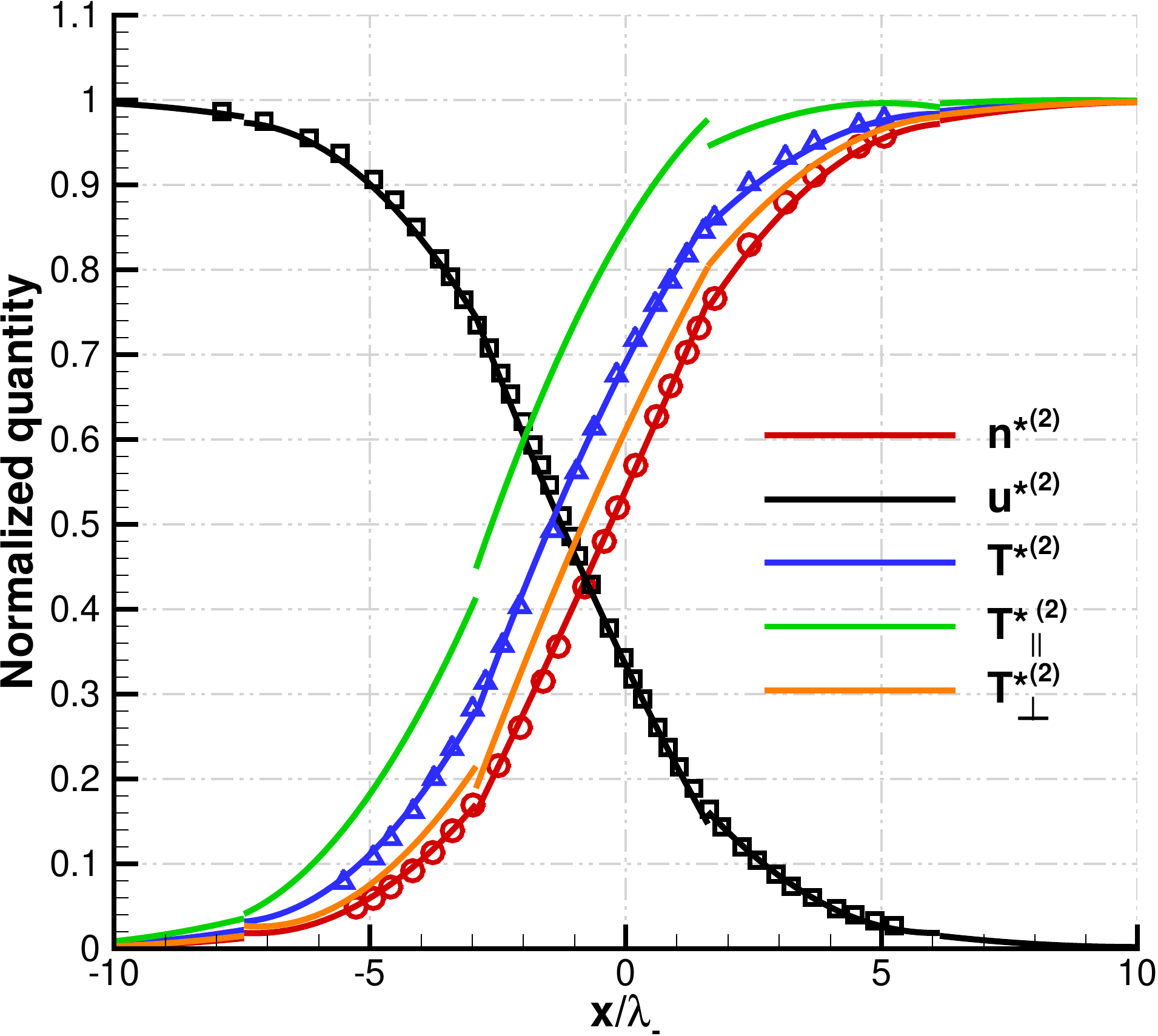}
  \caption{Case NS-01, 8 elements, species 2}
  \label{fig_shockHeKosuge_Ma_1_5_frac_0_5_mr_0_5_8elem_species2}
\end{subfigure}
\begin{subfigure}{0.5\textwidth}
  \centering
  \includegraphics[width=70mm]{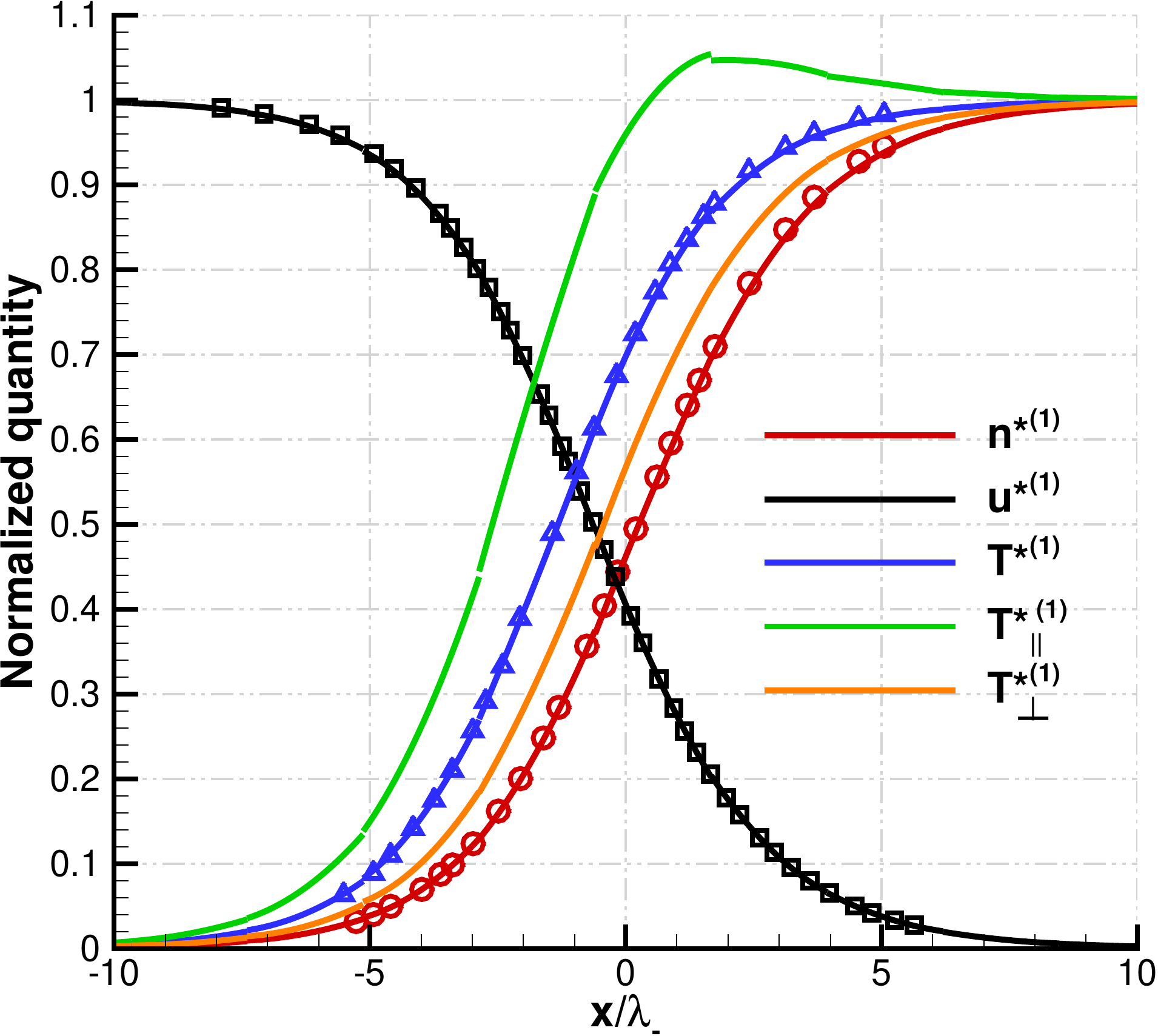}
  \caption{Case NS-01, 16 elements, species 1}
  \label{fig_shockHeKosuge_Ma_1_5_frac_0_5_mr_0_5_16elem_species1}
\end{subfigure}
\begin{subfigure}{0.5\textwidth}
  \centering
  \includegraphics[width=70mm]{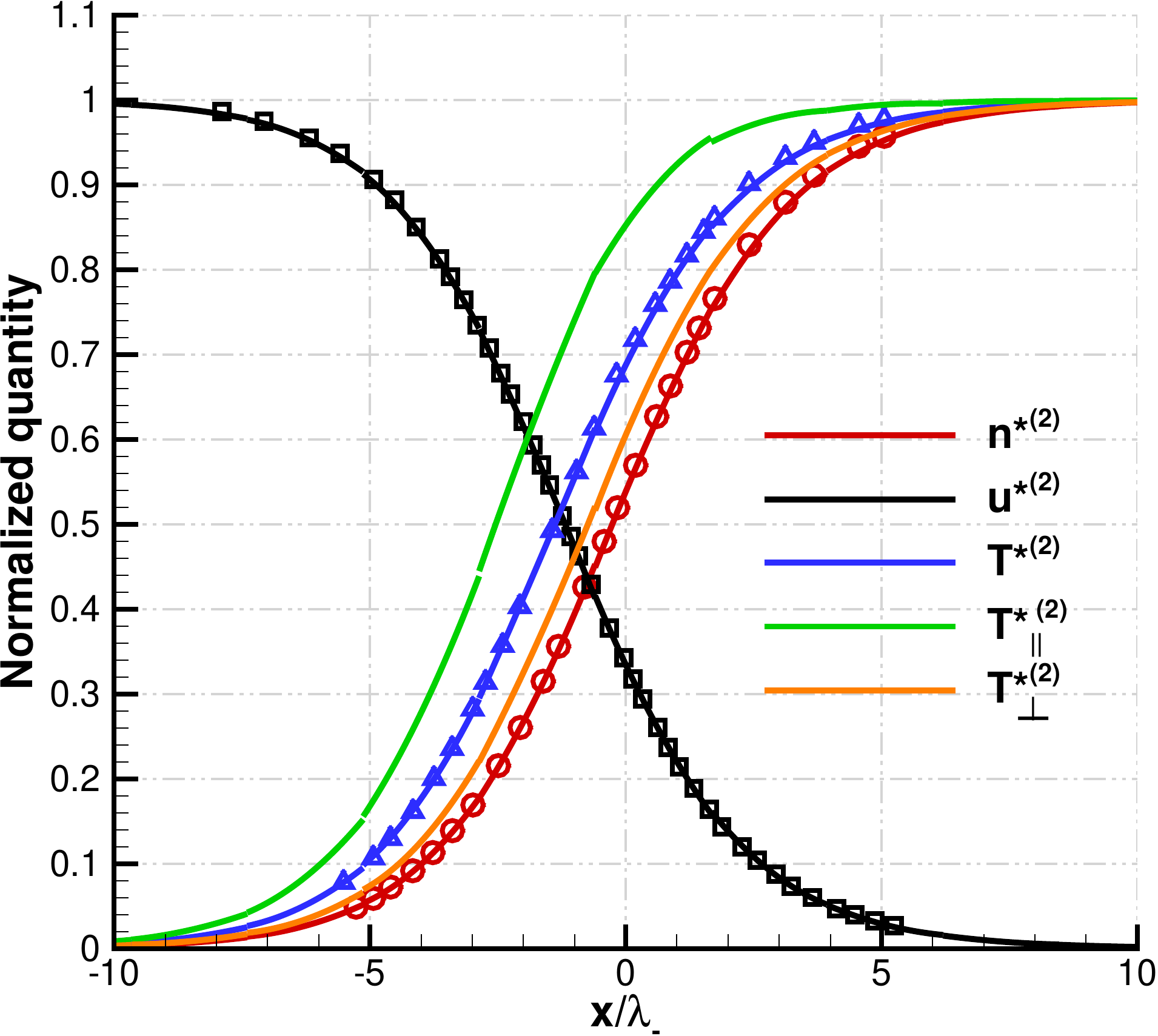}
  \caption{Case NS-01, 16 elements, species 2}
  \label{fig_shockHeKosuge_Ma_1_5_frac_0_5_mr_0_5_16elem_species2}
\end{subfigure}
\begin{subfigure}{0.5\textwidth}
  \centering
  \includegraphics[width=70mm]{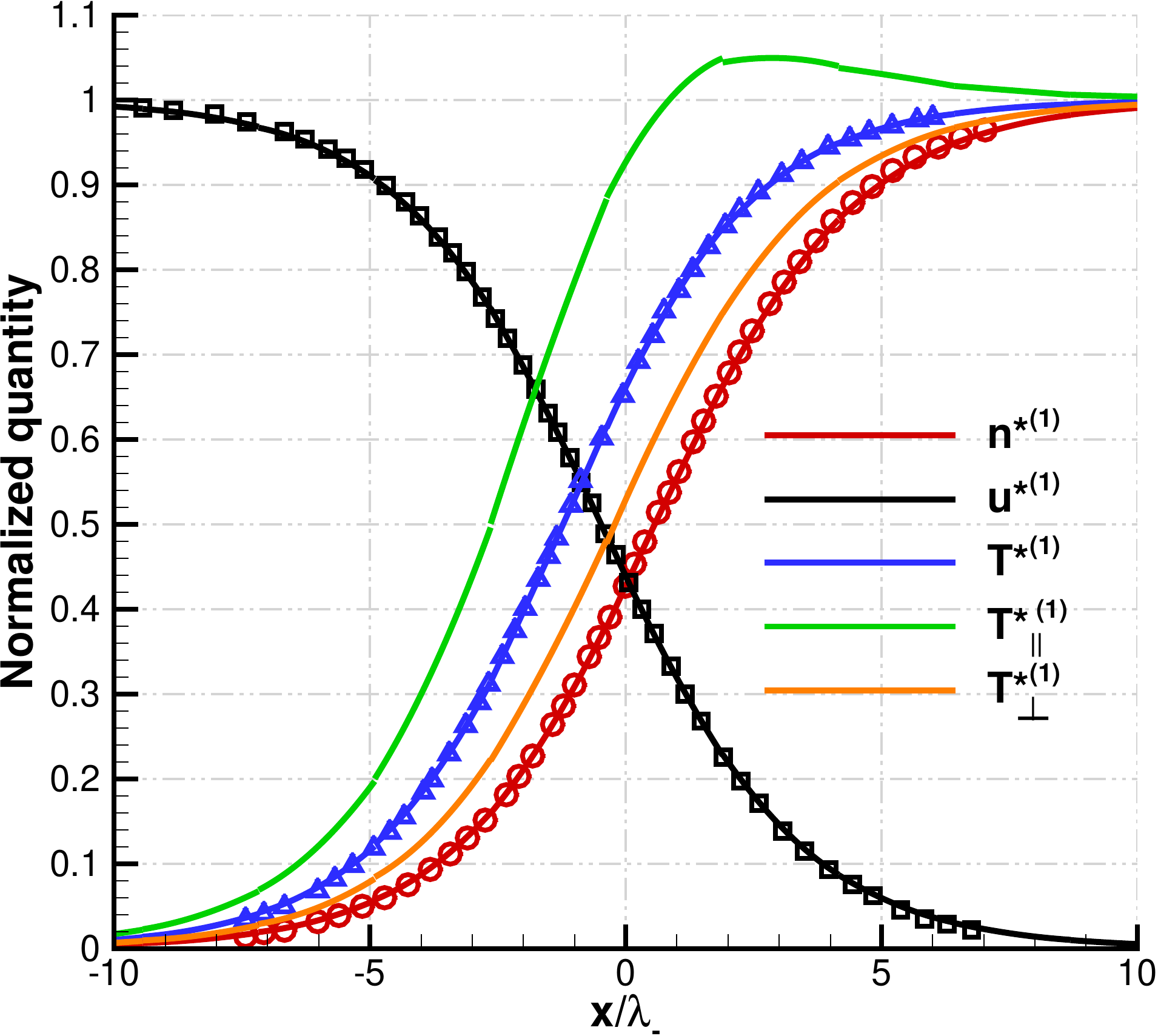}
  \caption{Case NS-02, 16 elements, species 1}
\end{subfigure}
\begin{subfigure}{0.5\textwidth}
  \centering
  \includegraphics[width=70mm]{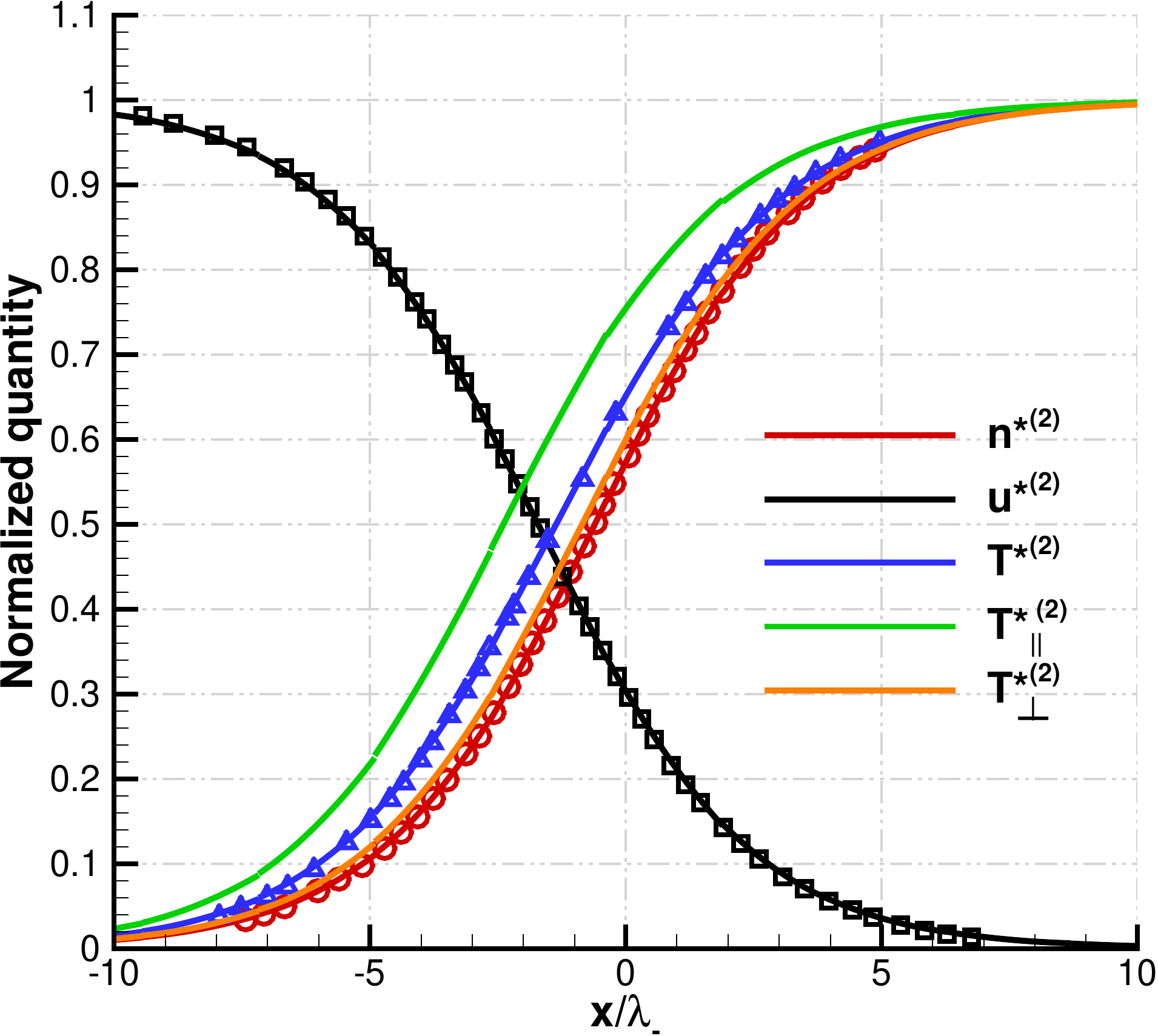}
  \caption{Case NS-02, 16 elements, species 2}
\end{subfigure}
\caption{Variation of normalized flow properties along the domain for Mach~1.5 normal shock with $n^\two_-/n_-=0.5$: (a--b) $m_2/m_1=0.5$ (Case NS-01) with 8 elements, (c--d) $m_2/m_1=0.5$ (Case NS-01) with 16 elements, and (e--f) $m_2/m_1=0.25$ (Case NS-02) with 16 elements. Symbols denote results from \cite{KAT01}, and lines denote DGFS solutions. Note that the position of the shock wave has been adjusted to the location with the average number density $(n_-+n_+)/2$ as per \cite{KAT01}. The normalized quantities are defined using: $n^{*\I}=(n^\I-n^\I_-)/(n^\I_+-n^\I_-)$, $T^{*\I}=(T^\I-T_-)/(T_+-T_-)$, $u^{*\I}=(u^\I-u_+)/(u_--u_+)$, $T_\parallel^{*\I}=(T_\parallel^\I-T_-)/(T_+-T_-)$, and $T_\perp^{*\I}=(T_\perp^\I-T_-)/(T_+-T_-)$.}
\label{fig_shockHeKosuge_Ma_1_5_frac_0_5_mr_0_5}
\end{figure*}

Figure~\ref{fig_shockHeKosuge_Ma_1_5_frac_0_5_mr_0_5} shows the bulk properties (number density, temperature, velocity, parallel/perpendicular temperature components) for Mach~1.5 normal shock with mass ratios $m_2/m_1=0.5$ and $m_2/m_1=0.25$. Based on these results, one can infer that DGFS recovers the normal shock reasonably well. In particular, from Figures~\ref{fig_shockHeKosuge_Ma_1_5_frac_0_5_mr_0_5_8elem_species1}, \ref{fig_shockHeKosuge_Ma_1_5_frac_0_5_mr_0_5_8elem_species2}, we observe that a Mach~1.5 shock can be captured with just 8 elements within \textit{engineering} $\pm 5\%$ accuracy. Note that the discontinuity in the flow profile is the characteristic of the DG method. The discontinuity expectedly vanishes upon refining the grid as in Figures~\ref{fig_shockHeKosuge_Ma_1_5_frac_0_5_mr_0_5_16elem_species1}, \ref{fig_shockHeKosuge_Ma_1_5_frac_0_5_mr_0_5_16elem_species2}.

\begin{figure}[!ht]
\begin{subfigure}{0.5\textwidth}
  \centering
  \includegraphics[width=70mm]{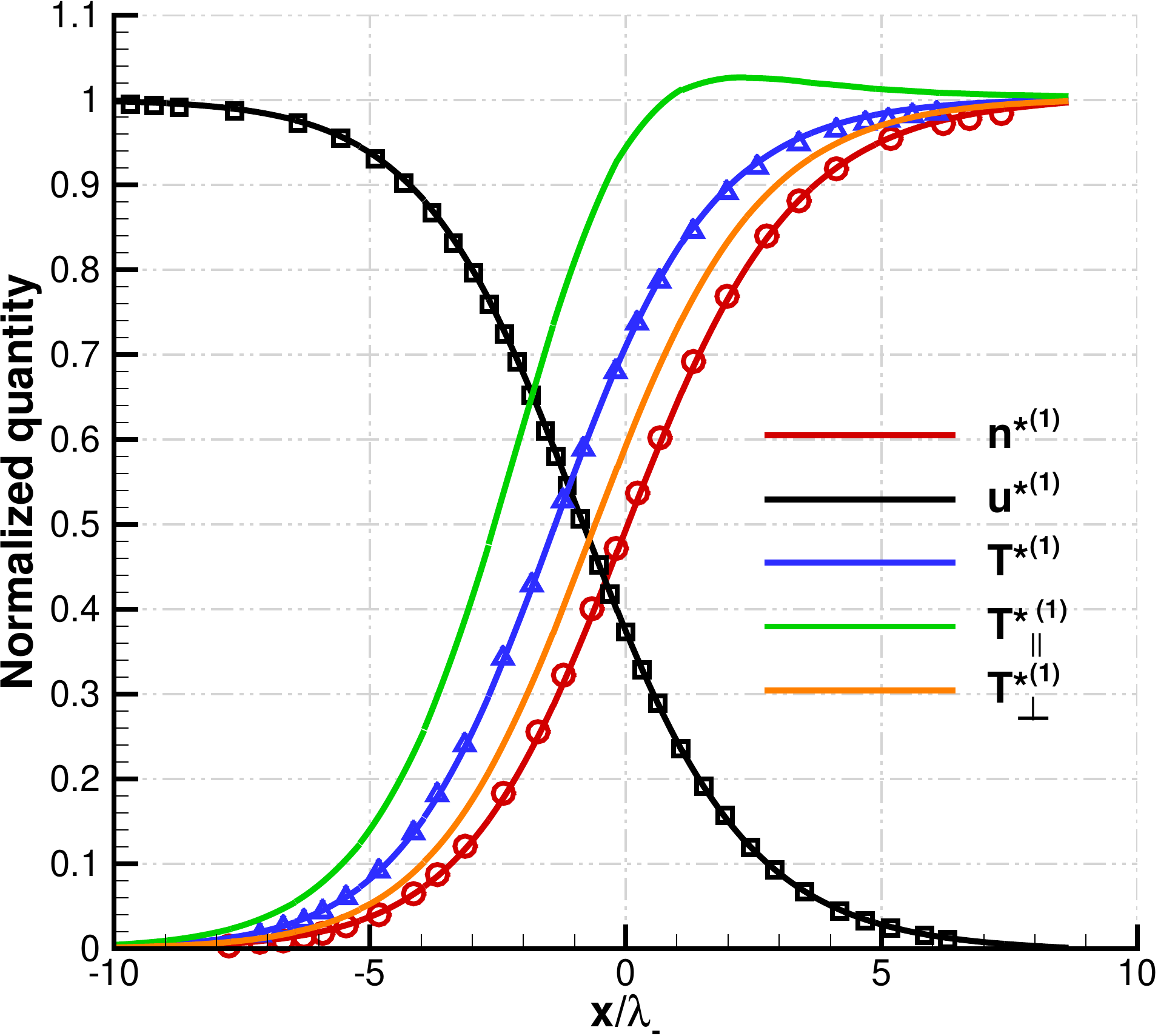}
  \caption{Case NS-03, 16 elements, species 1}
\end{subfigure}
\begin{subfigure}{0.5\textwidth}
  \centering
  \includegraphics[width=70mm]{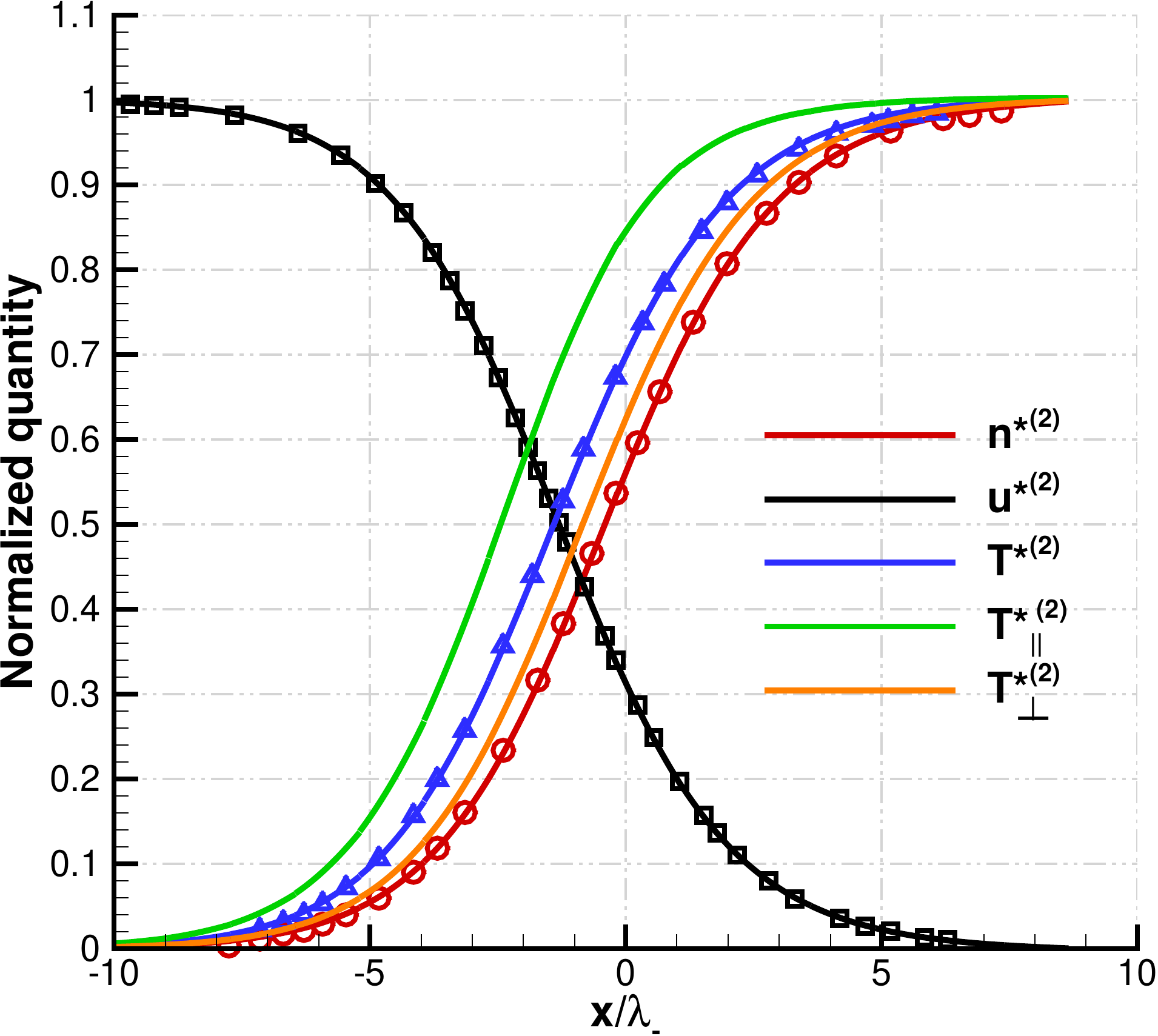}
  \caption{Case NS-03, 16 elements, species 2}
\end{subfigure}
\begin{subfigure}{0.5\textwidth}
  \centering
  \includegraphics[width=70mm]{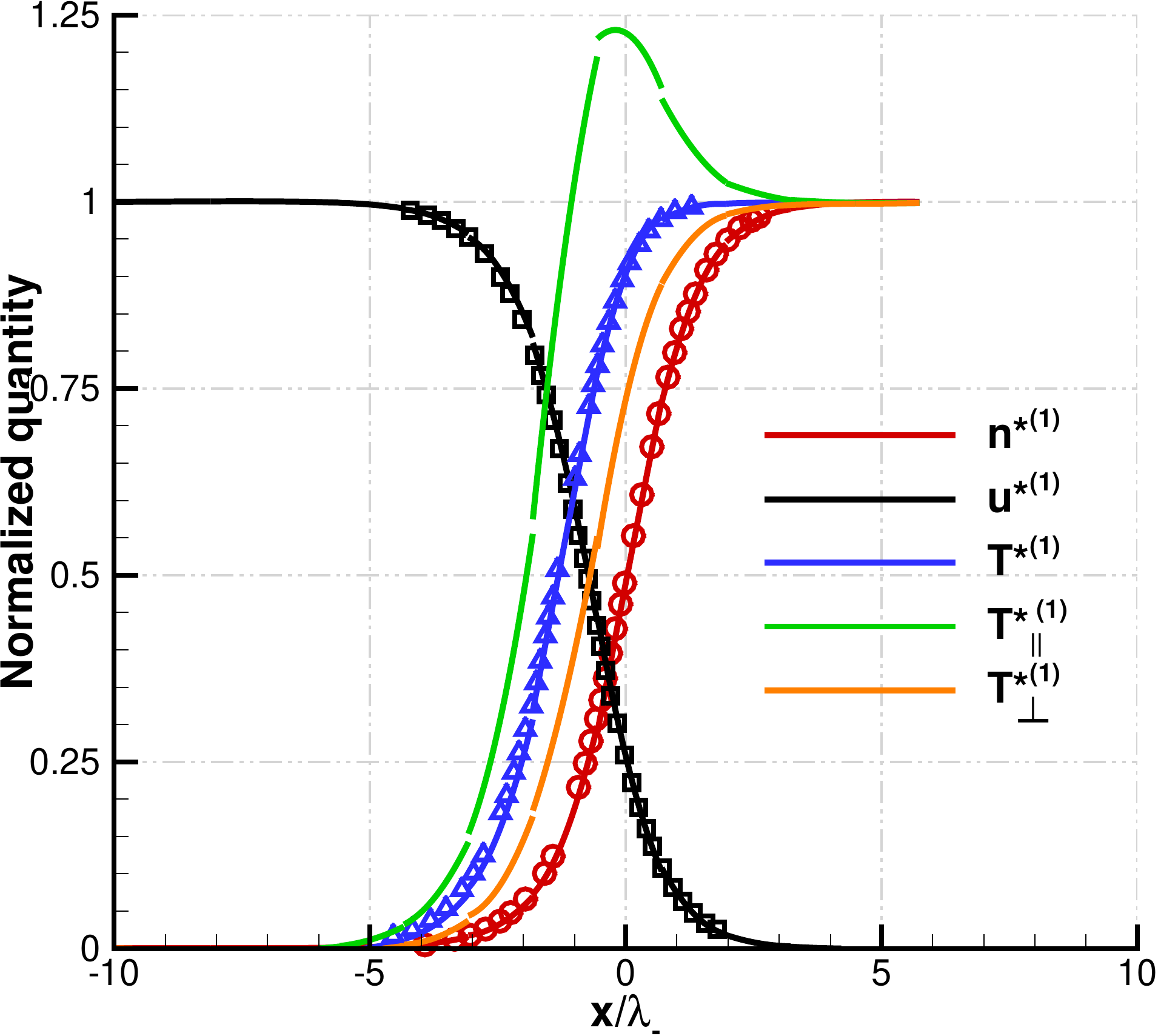}
  \caption{Case NS-04, 16 elements, species 1}
\end{subfigure}
\begin{subfigure}{0.5\textwidth}
  \centering
  \includegraphics[width=70mm]{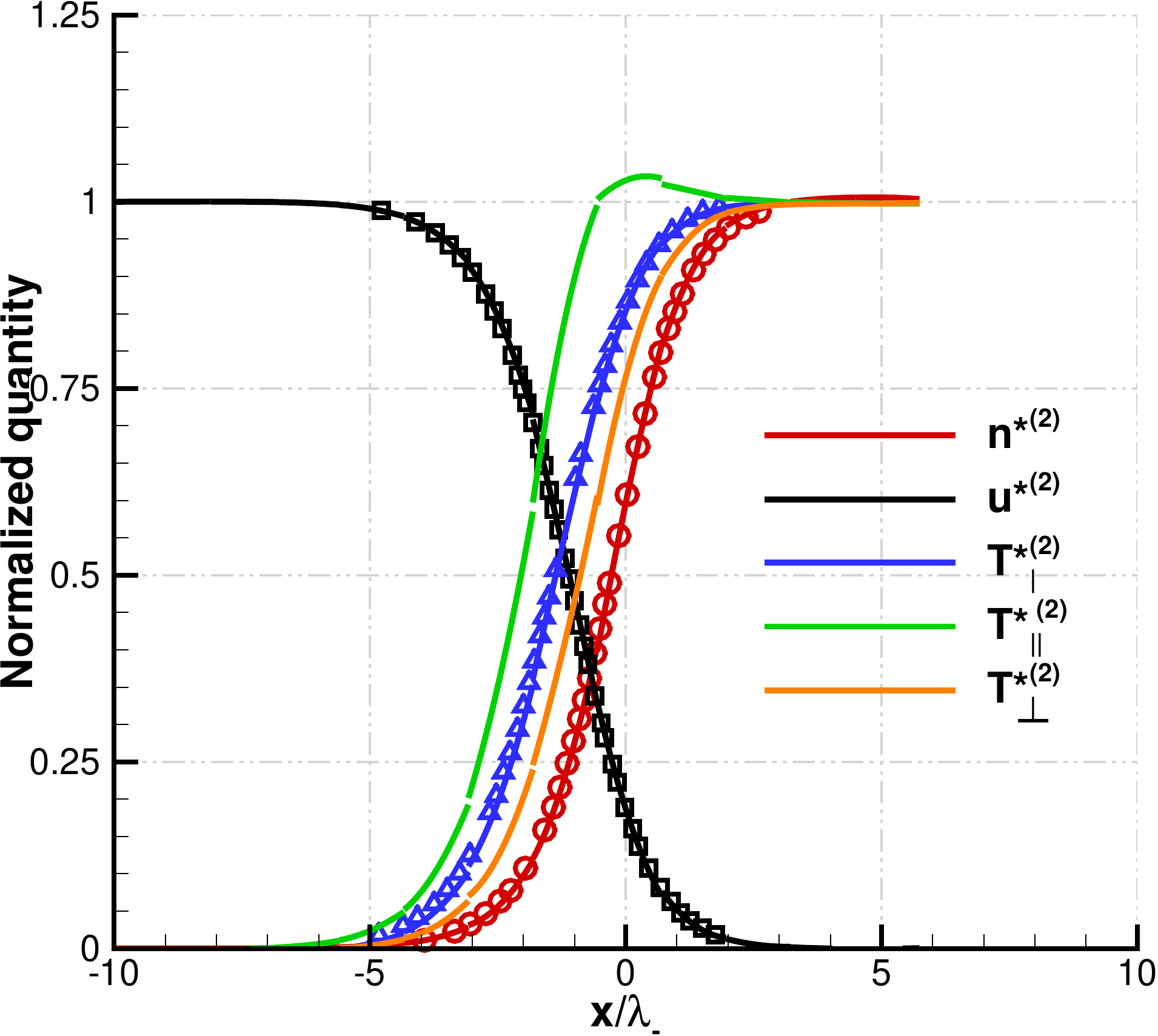}
  \caption{Case NS-04, 16 elements, species 2}
\end{subfigure}
\caption{Variation of normalized flow properties along the domain for normal shock with $m_2/m_1=0.5$, $n^\two_-/n_-=0.1$: (a--b) Mach~1.5 (Case NS-03), and (c--d) Mach~3 (Case NS-04). Symbols denote results from \cite{KAT01}, and lines denote DGFS solutions. Note that the position of the shock wave has been adjusted to the location with the average number density $(n_-+n_+)/2$ as per \cite{KAT01}. Definition of the normalized quantities is the same as in caption of Figure~\ref{fig_shockHeKosuge_Ma_1_5_frac_0_5_mr_0_5}.}
\label{fig_shockHeKosuge_frac_0_1_mr_0_5}
\end{figure}

Figure~\ref{fig_shockHeKosuge_frac_0_1_mr_0_5} shows the bulk properties (number density, temperature, and velocity, parallel/perpendicular temperature components) for Mach~1.5 and Mach~3 normal shock for mass ratio $m_2/m_1=0.5$ at low concentration $n_-^\two/n_-=0.1$. Again, we observe a fair agreement with the reference solutions.

\subsubsection{Solver configurations}

In the sections that follow, we consider the standard benchmark cases of Fourier heat transfer, oscillatory Couette flow, Couette flow, and Fick's diffusion problem at different Knudsen numbers for different collision kernels including VHS and VSS kernels. The results are compared with those obtained from DSMC with equivalent molecular collision models.

All the cases, unless otherwise noted, employ Argon-Krypton mixture. The collision model parameters are tabulated in Table~\ref{tab_collision} (as provided in \cite{Bird}). The reference diameters are selected so as to maintain the reference viscosity (cf. eqn. (4.62) in \cite{Bird}). Note that the viscosity index ($\omega_{ij}$) and scattering index ($\alpha_{ij}$) are empirical parameters, which are calibrated against experiments so that DSMC simulations reproduce experimental observations. The values of these parameters need to be recalibrated for different temperature ranges and different molecules. It is worth noting that there are hundreds of works on recalibration of transport coefficients. In the present DGFS formulation, no recalibration is needed (since the fast collision solver works for general collision kernels), i.e., one can directly use the HS/VHS/VSS model parameters from DSMC literature. %

\begin{table}[!ht]
\centering
\begin{tabular}{@{}lcc@{}}
\toprule
Mixture & Ar-Kr & Ar-Kr \\ 
Collision kernel & VHS & VSS \\ 
Molecular mass: $m_1$ ($\times 10^{27}\,kg$) & $66.3$ & $66.3$ \\ 
Molecular mass: $m_2$ ($\times 10^{27}\,kg$) & $139.1$ & $139.1$ \\ 
Reference viscosity: $\mu_{\text{ref},1}$ ($\times 10^{5}\,\text{Pa}\cdot s$) & $2.117$ & $2.117$ \\ 
Reference viscosity: $\mu_{\text{ref},2}$ ($\times 10^{5}\,\text{Pa}\cdot s$) & $2.328$ & $2.328$ \\ 
Viscosity index: $(\omega_{11},\,\omega_{22})$ & $(0.81,\,0.8)$ & $(0.81,\,0.8)$ \\
Viscosity index: $(\omega_{12},\,\omega_{21})$ & $(0.805,\,0.805)$ & $(0.805,\,0.805)$ \\
Scattering parameter: $(\alpha_{11},\,\alpha_{22})$ & $(1,\,1)$ & $(1.4,\,1.32)$ \\
Scattering parameter: $(\alpha_{12},\,\alpha_{21})$ & $(1,\,1)$ & $(1.36,\,1.36)$ \\
Ref. diameter: $(d_{\text{ref},11},\,d_{\text{ref},22})$ ($\times 10^{10} m$) & $(4.17,\,4.76)$ & $(4.11,\,4.7)$ \\
Ref. diameter: $(d_{\text{ref},11},\,d_{\text{ref},22})$ ($\times 10^{10} m$) & $(4.465,\,4.465)$ & $(4.405,\,4.405)$ \\
Ref. temperature: $(T_{\text{ref},11},\,T_{\text{ref},22})$ ($K$) & $(273,\,273)$ & $(273,\,273)$ \\
Ref. temperature: $(T_{\text{ref},12},\,T_{\text{ref},21})$ ($K$) & $(273,\,273)$ & $(273,\,273)$ \\
\bottomrule
\end{tabular}
\caption{VHS and VSS model parameters for different mixture systems \cite{Bird}.}
\label{tab_collision}
\end{table}

\textit{SPARTA} \cite{gallis2014direct} has been employed for carrying out DSMC simulations in the present work. It implements the DSMC method as proposed by Bird \cite{Bird}. The solver has been benchmarked \cite{gallis2014direct} and widely used for studying hypersonic, subsonic and thermal gas flow problems \cite{gallis2017molecular, gallis2016direct, sebastiao2018direct,jaiswal2018femta,jaiswal2018dsmc}. In this work, cell size less than $\lambda/3$ has been ensured in all test cases. A minimum of 30 DSMC simulator particles per species per cell are used in conjunction with the no-time collision (NTC) algorithm. Each steady-state simulation has been averaged for a minimum 100,000 steps so as to minimize the statistical noise. 

More specifically, for all the cases except oscillatory Couette flow, DSMC-SPARTA simulations employ 500 cells, $>$100 particles per cell, a time step of $2 \times 10^{-9}$ sec, 1 million unsteady time steps, and 100 million steady time steps. These DSMC parameters have been in part taken from \cite{JAH19} where the authors investigated the single-species rarefied gas flow problems. The parameters have been selected partially to minimize the statistical fluctuations and linear time-stepping errors inherent to DSMC simulations. We, however, note that these parameters are very conservative from a numerical simulation perspective.

\subsubsection{Fourier heat transfer of Argon-Krypton mixture using VHS collision kernel}

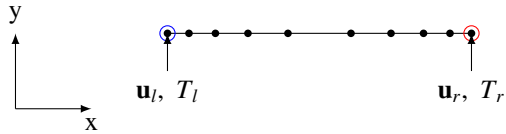
\begin{figure}[!ht]
	\centering
    \begin{tikzpicture}
       \def\pr{1.23};
       \foreach \x in {1,...,5}
         \fill ({-\pr + (\pr)^\x},0cm) circle (0.05cm);

       \foreach \x in {1,...,5}
         \fill ({ 4 + \pr - (\pr)^(\x)},0cm) circle (0.05cm);

       \draw(0,0) -- (4,0) ;
       \draw[blue] (0.0,0.0) circle (0.1cm);
       \draw[red] (4.0,0.0) circle (0.1cm);

       \draw[-latex] (0.0,-0.5) -- (0.0, 0.0) node[below, yshift=-0.5cm] {$\mathbf{u}_l,\;T_l$};
       \draw[-latex] (4.0,-0.5) -- (4.0, 0.0) node[below, yshift=-0.5cm] {$\mathbf{u}_r,\;T_r$};

       \draw[-latex] (-2.0,-1) -- (-1.0, -1) node[anchor=north] {x};
       \draw[-latex] (-2.0,-1) -- (-2.0, 0) node[anchor=south] {y};
    \end{tikzpicture}
	\caption{Numerical setup for 1D Fourier and Couette flows. Distance between the walls is fixed as $H_0=10^{-3}$ m. Note that the cells are finer in the near-wall region.}
	\label{fig_fourierHeatTransferSchematic}
\end{figure}

In the current test case, we consider the effect of temperature gradient on the solution. The coordinates are chosen such that the walls are parallel to the $y$ direction and $x$ is the direction perpendicular to the walls. The geometry as well as boundary conditions are shown in Figure \ref{fig_fourierHeatTransferSchematic}. We consider six cases for a range of temperature gradients and rarefaction levels. The numerical parameters for these six cases are given in Table~\ref{tab_fourier_conditions}.

\begin{table}[!ht]
\centering
\setlength{\tabcolsep}{0.3em}
\begin{tabular}{@{}lcccccc@{}}
\toprule
Parameter & Case F-01 & Case F-02 & Case F-03 & Case F-04 & Case F-05 & Case F-06 \\ 
\midrule
Mixture & Ar-Kr & Ar-Kr & Ar-Kr & Ar-Kr & Ar-Kr & Ar-Kr \\
Collision kernel & VHS & VHS & VHS & VHS & VHS & VHS \\
Non-dim physical space & $[0,\,1]$ & $[0,\,1]$  & $[0,\,1]$ & $[0,\,1]$  & $[0,\,1]$ & $[0,\,1]$ \\ 
Non-dim velocity space & $[-5,\,5]^3$ & $[-5,\,5]^3$ & $[-5,\,5]^3$ & $[-9,\,9]^3$ & $[-9,\,9]^3$ & $[-9,\,9]^3$ \\ 
$N^3$ & $32^3$ & $32^3$ & $32^3$ & $64^3$ & $64^3$ & $64^3$ \\
$N_\rho$ & $32$ & $32$ & $32$ & $64$ & $64$ & $64$ \\
$M$ & 12 & 12 & 12 & 12 & 12 & 12 \\
Spatial elements & 4 & 4 & 4 & 4 & 4 & 4 \\
DG order & 3 & 3 & 3 & 3 & 3 & 3 \\
Time step (s) & $2\times 10^{-8}$ & $2\times 10^{-8}$ & $2\times 10^{-8}$ & $2\times 10^{-8}$ & $2\times 10^{-8}$ & $2\times 10^{-8}$ \\
Mass: $m_0$ & $m_\text{Ar}=m_1$ & $m_\text{Ar}=m_1$ & $m_\text{Ar}=m_1$ & $m_\text{Ar}=m_1$ & $m_\text{Ar}=m_1$ & $m_\text{Ar}=m_1$ \\
Length: $H_0$ ($mm$) & 1 & 1 & 1 & 1 & 1 & 1 \\
Velocity: $u_0$ ($m/s$) & 337.2 & 337.2 & 337.2 & 337.2 & 337.2 & 337.2 \\
Temperature: $T_0$ ($K$) & 273 & 273 & 273 & 273 & 273 & 273 \\
Number density: $n_0$ ($m^{-3}$) & $1.680\times10^{21}$ & $8.401\times10^{20}$ & $1.680\times10^{20}$ & $1.680\times10^{21}$ & $8.401\times10^{20}$ & $1.680\times10^{20}$ \\
\midrule
\multicolumn{4}{l}{Left wall (purely diffuse) boundary conditions (subscript $l$)} \\
Velocity: $u_l$ ($m/s$) & $0$ & $0$ & $0$ & $0$ & $0$ & $0$ \\
Temperature: $T_l$ ($K$) & 263 & 263 & 263 & 223 & 223 & 223  \\
\midrule
\multicolumn{4}{l}{Right wall (purely diffuse) boundary conditions (subscript $r$)} \\
Velocity: $u_r$ ($m/s$) & $0$ & $0$ & $0$ & $0$ & $0$ & $0$ \\
Temperature: $T_r$ ($K$) & 283 & 283 & 283 & 323 & 323 & 323  \\
\midrule
\multicolumn{2}{l}{Initial conditions} \\
Velocity: $u$ ($m/s$) & 0 & 0 & 0 & 0 & 0 & 0 \\
Temperature: $T$ ($K$) & 273 & 273 & 273 & 273 & 273 & 273 \\
Number density: $n^\one$ ($m^{-3}$) & $1.680\times10^{21}$ & $8.401\times10^{20}$ & $1.680\times10^{20}$ & $1.680\times10^{21}$ & $8.401\times10^{20}$ & $1.680\times10^{20}$ \\
Number density: $n^\two$ ($m^{-3}$) & $8.009\times10^{20}$ & $4.004\times10^{20}$ & $8.009\times10^{19}$ & $8.009\times10^{20}$ & $4.004\times10^{20}$ & $8.009\times10^{19}$ \\
Knudsen: $(\Kn_{11},\,\Kn_{22})$ & $(0.770,\,0.591)$ & $(1.541,\,1.182)$ & $(7.703,\,5.912)$ & $(0.770,\,0.591)$ & $(1.541,\,1.182)$ & $(7.703,\,5.912)$ \\
Knudsen: $(\Kn_{12},\,\Kn_{21})$ & $(0.782,\,0.540)$ & $(1.564,\,1.080)$ & $(7.820,\,5.399)$ & $(0.782,\,0.540)$ & $(1.564,\,1.080)$ & $(7.820,\,5.399)$ \\
\bottomrule
\end{tabular}
\caption{Numerical parameters for Fourier heat transfer. The molecular collision parameters for Ar-Kr system are provided in Table~\ref{tab_collision}.}
\label{tab_fourier_conditions}
\end{table}

\begin{figure}[!ht]
\centering
\begin{subfigure}{.5\textwidth}
  \centering
  \includegraphics[width=80mm]{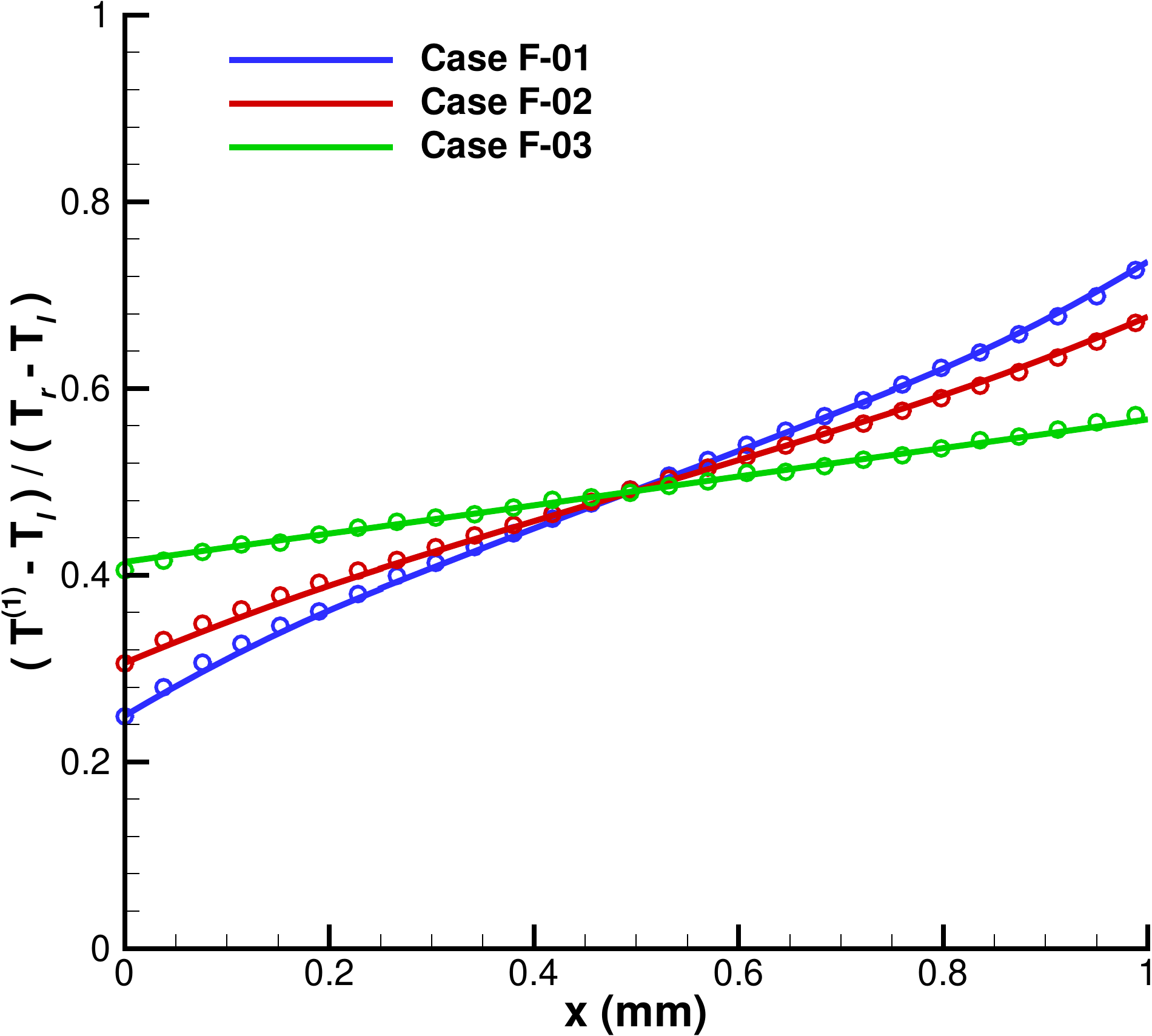}
  \caption{species 1: Argon, $\Delta T=20K$}
\end{subfigure}%
\begin{subfigure}{.5\textwidth}
  \centering
  \includegraphics[width=80mm]{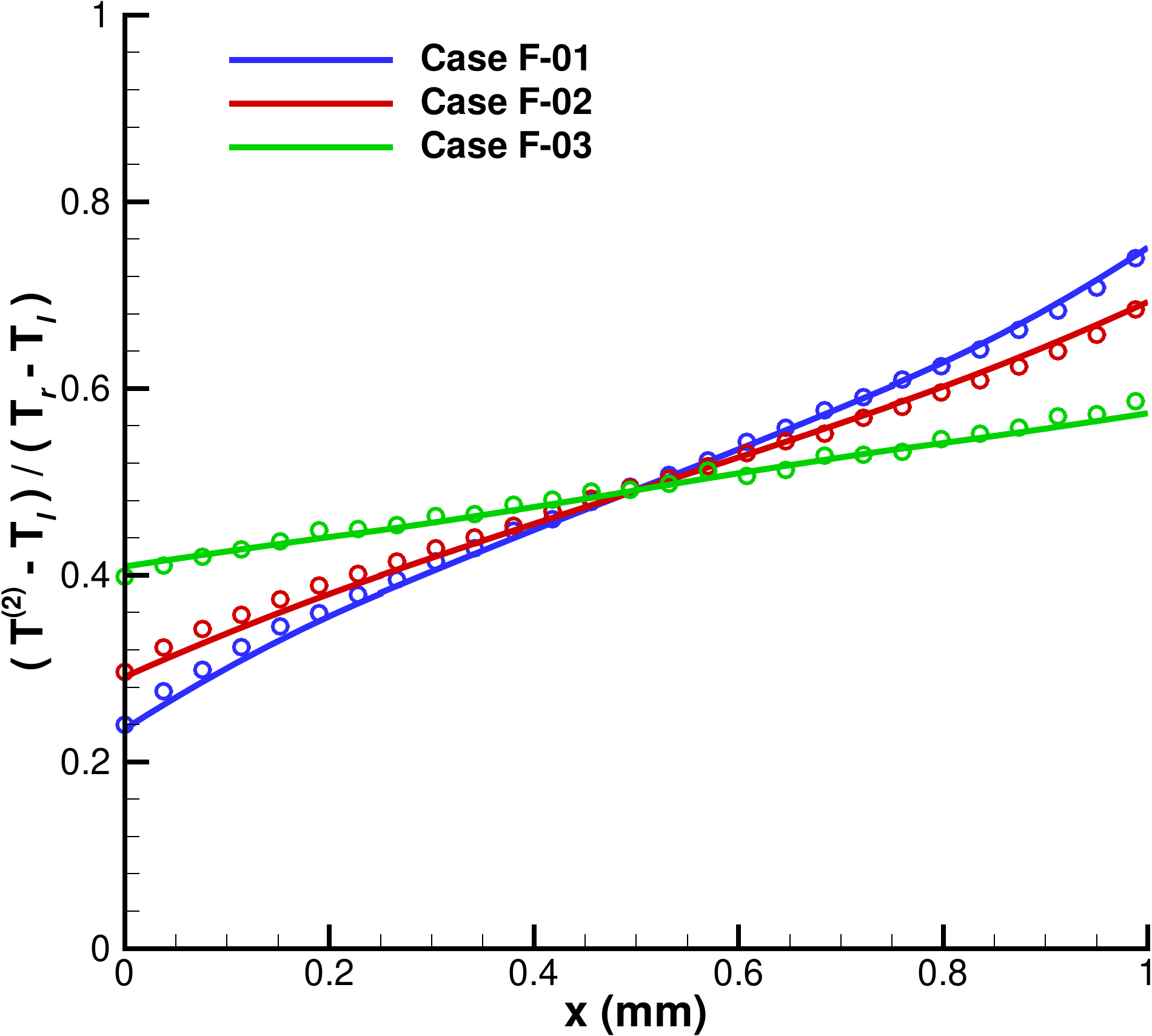}
  \caption{species 2: Krypton, $\Delta T=20K$}
\end{subfigure}
\begin{subfigure}{.5\textwidth}
  \centering
  \includegraphics[width=80mm]{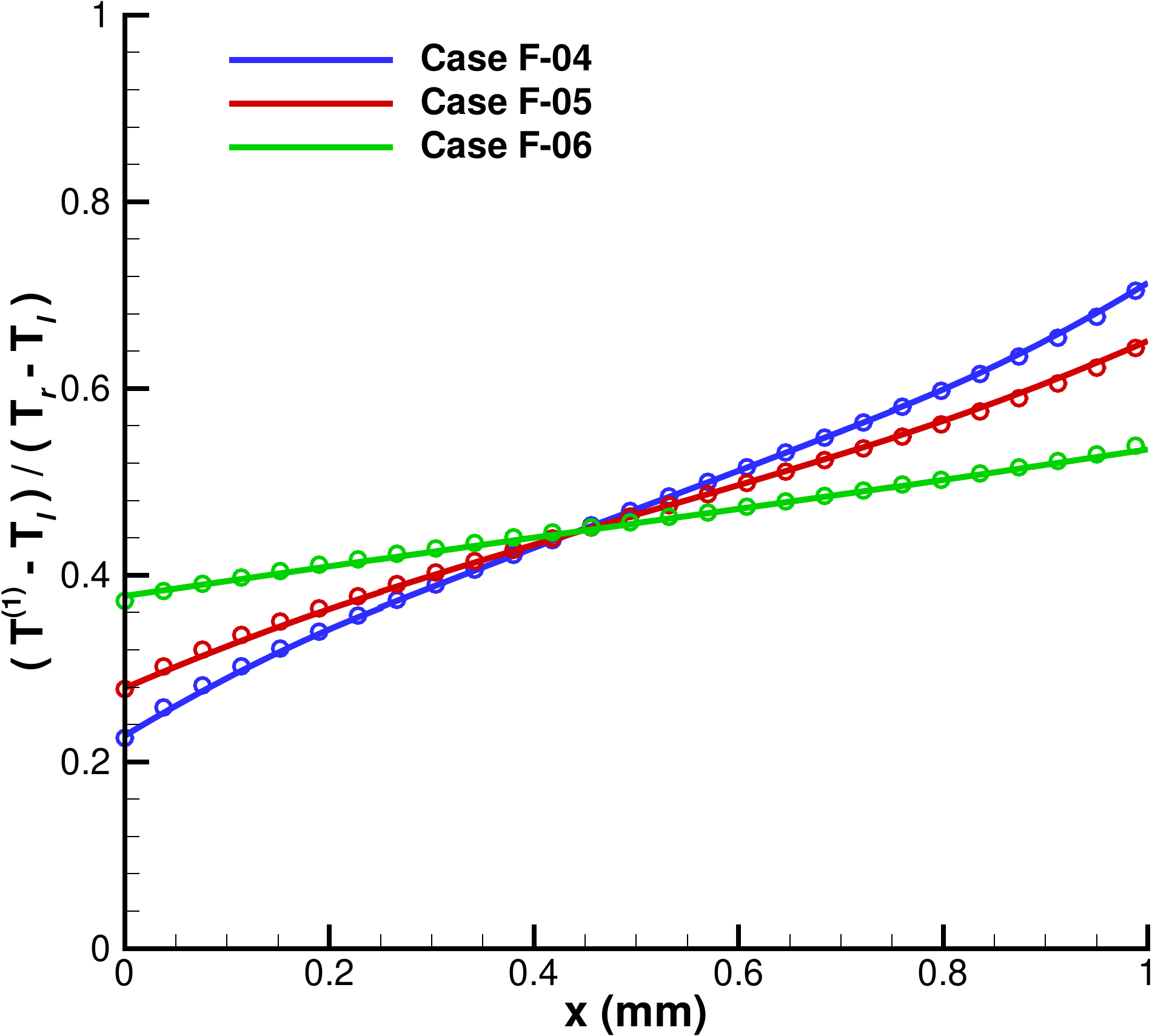}
  \caption{species 1: Argon, $\Delta T=100K$}
\end{subfigure}%
\begin{subfigure}{.5\textwidth}
  \centering
  \includegraphics[width=80mm]{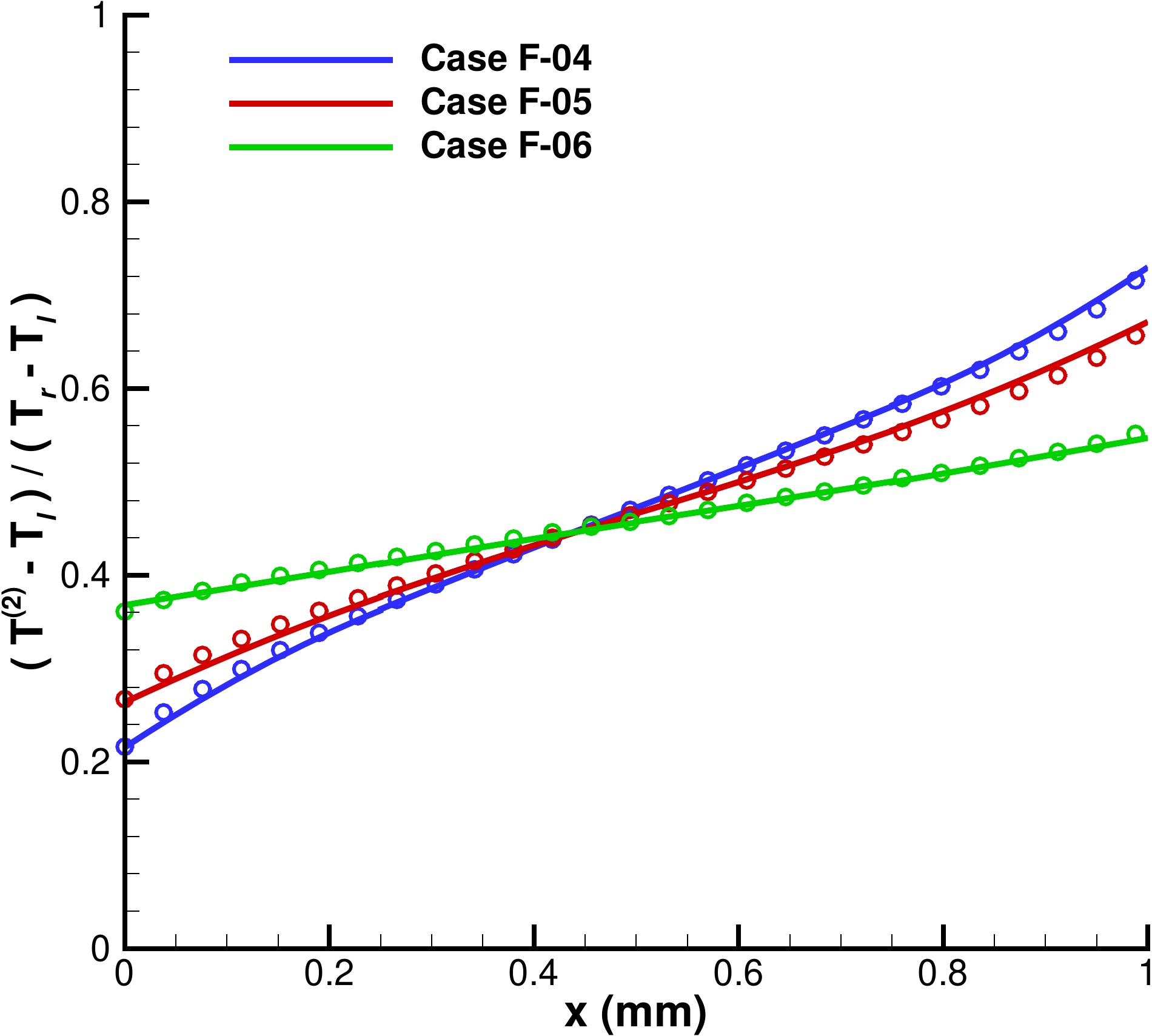}
  \caption{species 2: Krypton, $\Delta T=100K$}
\end{subfigure}
\caption{Variation of normalized temperature $(T^\I-T_l)/(T_r-T_l),\;i=\{1,2\}$ along the domain length for Fourier heat transfer obtained with DSMC and DGFS using VHS collision kernel for Argon-Krypton mixture. Symbols denote DSMC solutions, and lines denote DGFS solutions. Numerical parameters are provided in Table~\ref{tab_fourier_conditions}.}
\label{fig_fourier_T}
\end{figure}

Figure~\ref{fig_fourier_T} shows the variation of normalized temperature along the domain length for different initial mixture densities: a--b) $\Delta T=20$ (Case F-01, F-02, F-03), and c--d) $\Delta T=100$ (Case F-05, F-06, F-07). The results are compared against DSMC. We note minor ($1-2\%$) discrepancy between DGFS and DSMC for Krypton in the bulk-region away from the walls. Note however that the amount of predicted temperature jump is consistent between DSMC and DGFS for both species. %
   
\subsubsection{Oscillatory Couette flow of Argon-Krypton mixture using VHS collision kernel}

In the current test case, we consider the effect of \textit{transient} momentum transport for verifying the temporal accuracy of the DGFS. The schematic remains the same as in the previous test case. The left wall is at rest, and the right wall moves with a velocity of $u=(0, v_a\sin{(\zeta t)}, 0) \;m/s$, where $v_a$ is the amplitude of oscillation. The simulation parameters are given in Table~\ref{tab_oscCouette_conditions}. The present case is run for two different wall velocities: a) $v_a=50\;m/s$, and b) $v_a=500\;m/s$. Argon-Krypton mixture with VHS collision model is taken as the working gas. Specifically for DSMC simulations, the domain is discretized into 50 cells with 100000 particles per cell (PPC). For $v_a=50\;m/s$ case, a time step of $2\times 10^{-10}$ sec is employed. For $v_a=500\;m/s$ case, a time step of $2\times 10^{-11}$ sec is employed. The results are averaged for every 1000 (Navg) time steps. These DSMC simulation parameters have been taken from \cite{JAH19}. Note that such low DSMC time steps are particularly needed for obtaining time accurate results since the time stepping is inherently linear in traditional DSMC method \cite{Bird}.

\begin{table}[!ht]
\centering
\begin{tabular}{@{}lcc@{}}
\toprule
Parameter & Case OC-01 & Case OC-02 \\ 
\midrule
Mixture & Ar-Kr & Ar-Kr \\
Collision kernel & VHS & VHS \\
Non-dim physical space & $[0,\,1]$ & $[0,\,1]$ \\ 
Non-dim velocity space & $[-5,\,5]^3$ & $[-9,\,9]^3$ \\ 
$N^3$ & $24^3$ & $48^3$ \\
$N_\rho$ & $24$ & $48$ \\
$M$ & 6 & 6 \\
Spatial elements & 4 & 4 \\
DG order & 3 & 3 \\
Time step (s) & $2\times 10^{-8}$ & $2\times 10^{-8}$ \\
Characteristic mass: $m_0$ & $m_\text{Ar}=m_1$ & $m_\text{Ar}=m_1$ \\
Characteristic length: $H_0$ ($mm$) & 1 & 1 \\
Characteristic velocity: $u_0$ ($m/s$) & 337.2 & 337.2 \\
Characteristic temperature: $T_0$ ($K$) & 273 & 273 \\
Characteristic number density: $n_0$ ($m^{-3}$) & $8.401\times10^{20}$ & $8.401\times10^{20}$ \\
\midrule
\multicolumn{2}{l}{Initial conditions} \\
Velocity: $u$ ($m/s$) & 0 & 0 \\
Temperature: $T$ ($K$) & 273 & 273  \\
Number density: $n^\one$ ($m^{-3}$) & $8.401\times10^{20}$ & $8.401\times10^{20}$ \\
Number density: $n^\two$ ($m^{-3}$) & $4.004\times10^{20}$ & $4.004\times10^{20}$ \\
Knudsen number: $(\Kn_{11},\,\Kn_{22})$ & $(1.541,\,1.182)$ & $(1.541,\,1.182)$ \\
Knudsen number: $(\Kn_{12},\,\Kn_{21})$ & $(1.564,\,1.080)$ & $(1.564,\,1.080)$ \\
\midrule
\multicolumn{2}{l}{Left wall (purely diffuse) boundary conditions (subscript $l$)} \\
Velocity: $u_l$ ($m/s$) & $(0,\,0,\,0)$ & $(0,\,0,\,0)$ \\
Temperature: $T_l$ ($K$) & 273 & 273  \\
\midrule
\multicolumn{2}{l}{Right wall (purely diffuse) boundary conditions (subscript $r$)} \\
Velocity: $u_r$ ($m/s$) & $(0,\,50 sin(\zeta t),\,0)$ & $(0,\,500 sin(\zeta t),\,0)$ \\
Temperature: $T_r$ ($K$) & 273 & 273 \\
Period of oscillation: $\zeta$ ($s^{-1}$) & $2\pi/(5 \times 10^{-5})$ & $2\pi/(5 \times 10^{-5})$ \\
Velocity amplitude: $v_a$ ($m/s$) & $50$ & $500$ \\
\bottomrule
\end{tabular}
\caption{Numerical parameters for oscillatory Couette flow. The molecular collision parameters for Ar-Kr system are provided in Table~\ref{tab_collision}.}
\label{tab_oscCouette_conditions}
\end{table}

\begin{figure}[!ht]
\begin{subfigure}{.5\textwidth}
  \centering
  \includegraphics[width=75mm]{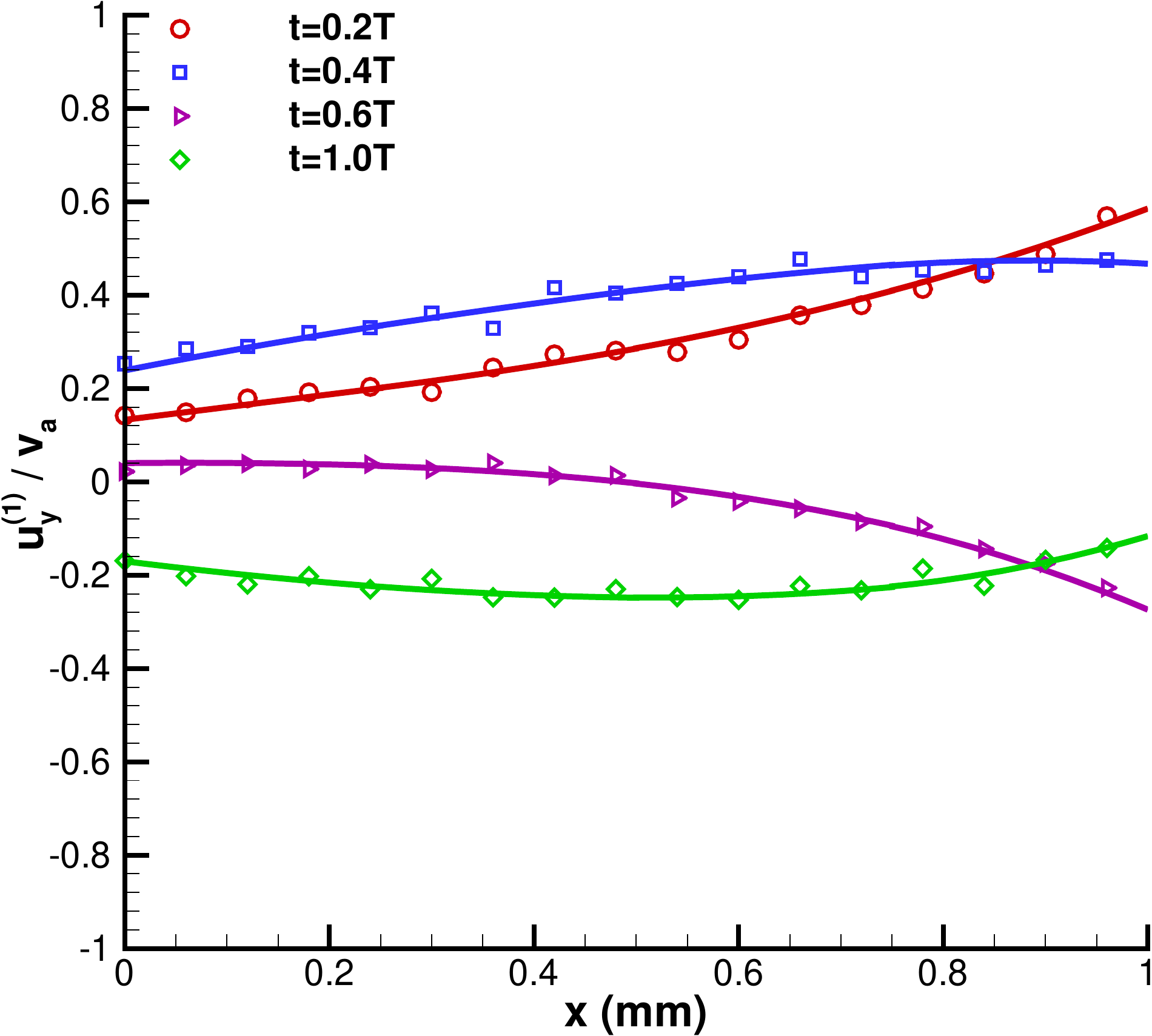}
  \caption{Case OC-01, $v_a=50\;m/s$, species 1: Argon}
  \label{fig_oscCouette_U_Ar}
\end{subfigure}%
\begin{subfigure}{.5\textwidth}
  \centering
  \includegraphics[width=75mm]{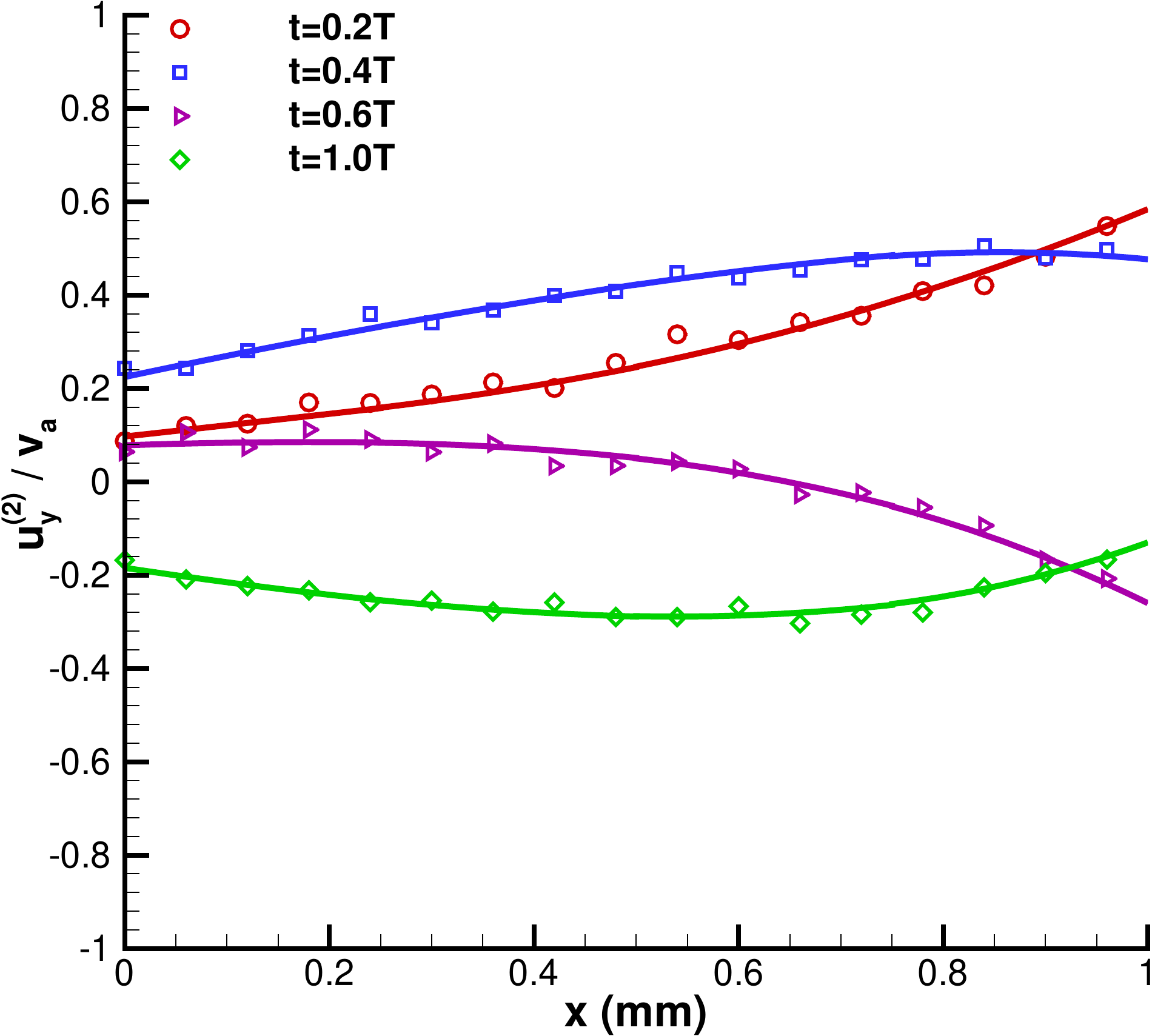}
  \caption{Case OC-01, $v_a=50\;m/s$, species 2: Krypton}
  \label{fig_oscCouette_U_Kr}
\end{subfigure}
\begin{subfigure}{.5\textwidth}
  \centering
  \includegraphics[width=75mm]{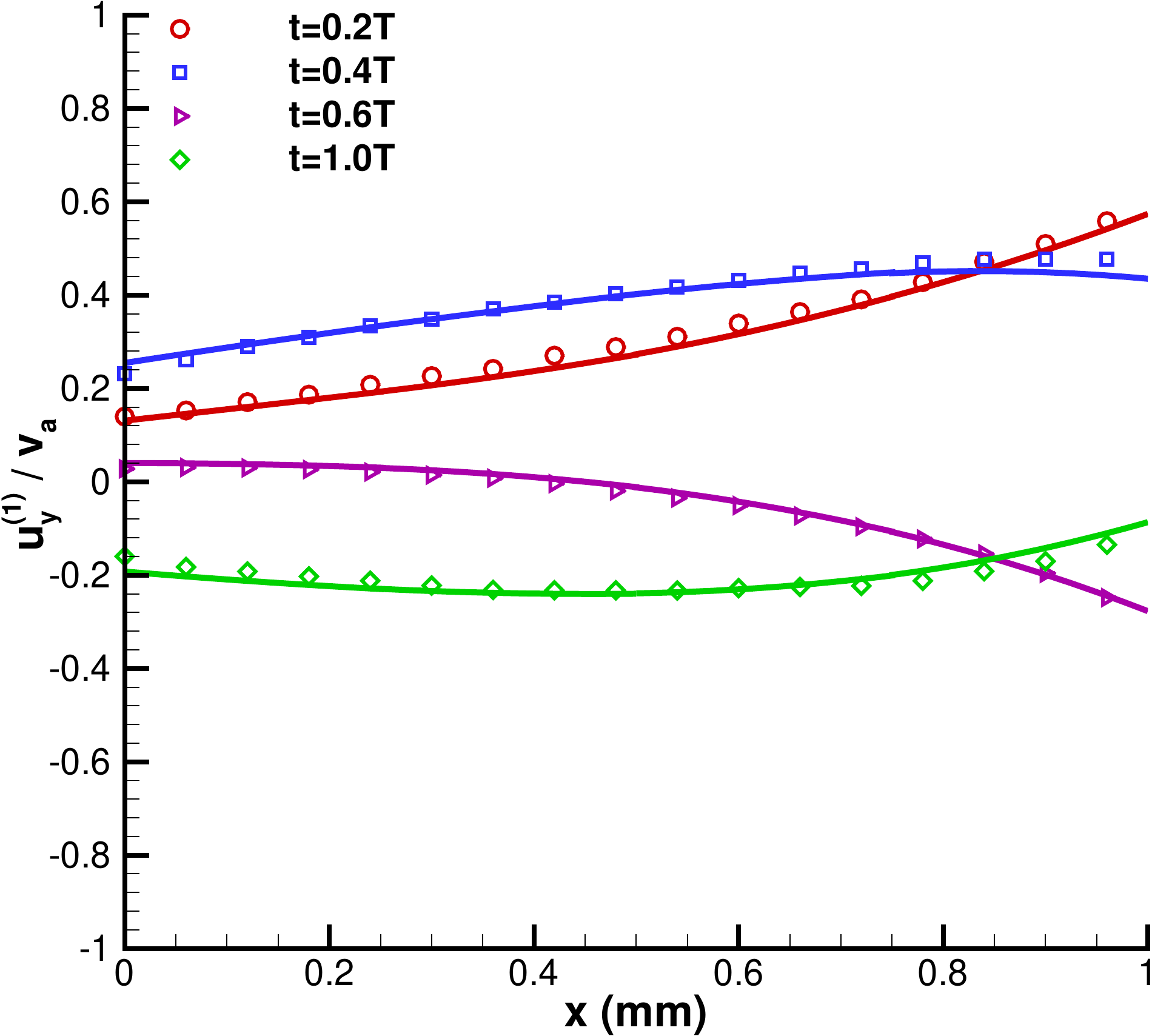}
  \caption{Case OC-02, $v_a=500\;m/s$, species 1: Argon}
  \label{fig_oscCouette_U_Ar_U500}
\end{subfigure}%
\begin{subfigure}{.5\textwidth}
  \centering
  \includegraphics[width=75mm]{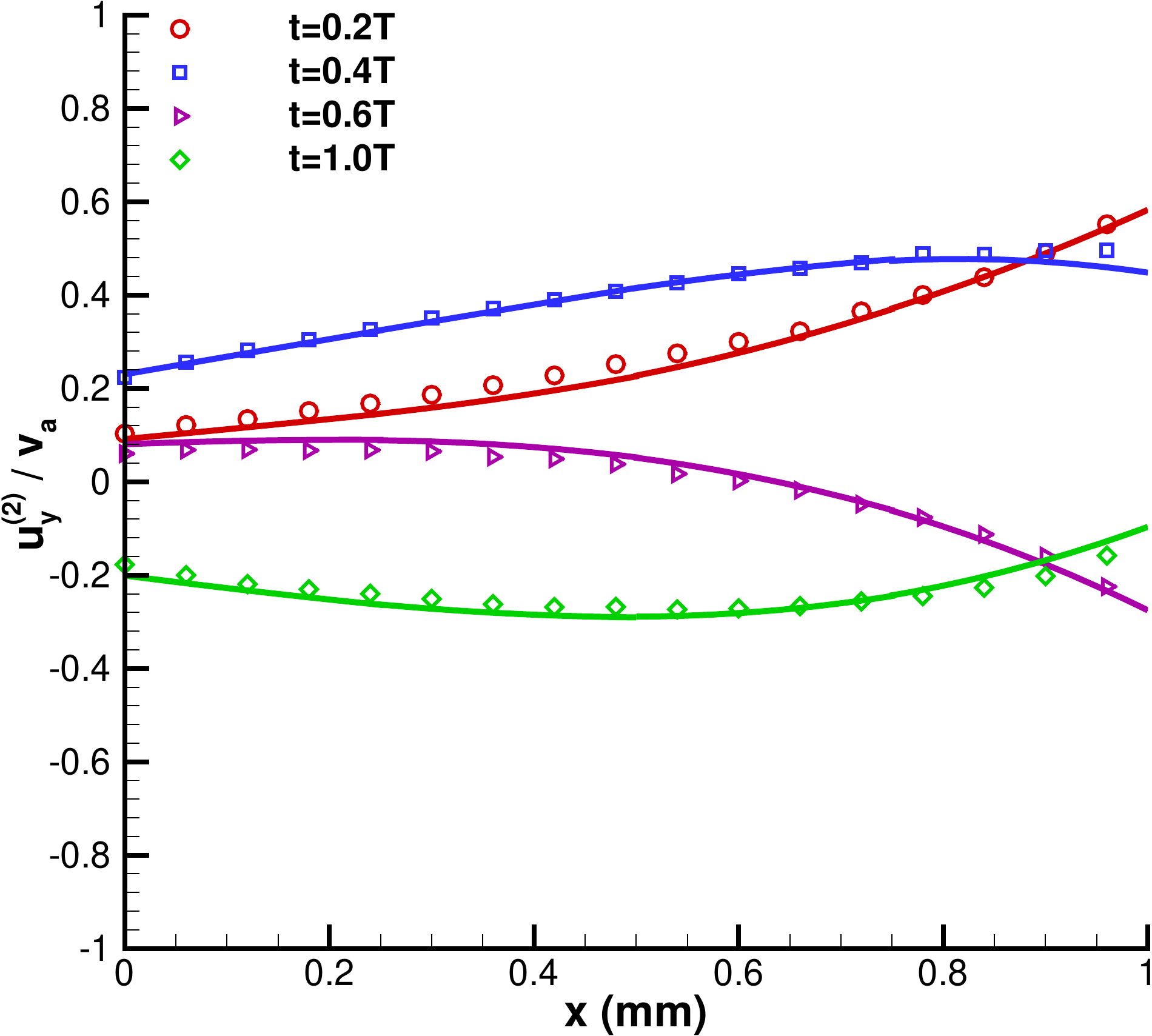}
  \caption{Case OC-02, $v_a=500\;m/s$, species 2: Krypton}
  \label{fig_oscCouette_U_Kr_U500}
\end{subfigure}
\caption{Variation of normalized velocity $u^\I_y/v_a,\;i=\{1,2\}$ along the domain length for oscillatory Couette flow obtained with DSMC and DGFS using VHS collision model for Argon-Krypton mixture. Symbols denote DSMC solutions, and lines denote DGFS solutions.}
\label{fig_oscCouette_U}
\end{figure}

Figure~\ref{fig_oscCouette_U} illustrates the results for the oscillatory Couette flow along the domain length for different $v_a$. Ignoring the statistical noise, we observe a good agreement between DGFS and DSMC. Note in particular that for both species, the amount of slip at the left wall are different -- which is in accordance with the conservation principles. Moreover, the amount of slip is consistent between DSMC and DGFS.

\subsubsection{Couette flow of Argon-Krypton mixture using VSS collision kernel}

Phenomenological scattering models are designed and calibrated (against experiments) so as to recover the correct transport properties. VSS model, in particular, recovers two transport properties: a) viscosity and b) diffusion \cite{Bird}. Couette flow serves as a test case for reproducing the correct viscosity coefficient (the test case for reproducing the correct diffusion coefficient is provided in the later sections). In the current test case, the schematic remains the same as in the previous test case. The left and right parallel walls move with a velocity of $u_w=(0, \mp50, 0) \;m/s$. The simulation parameters are given in Table~\ref{tab_oscCouette_conditions}. Argon-Krypton mixture with VSS collision kernel is taken as the working gas.

\begin{table}[!ht]
\centering
\begin{tabular}{@{}lcc@{}}
\toprule
Parameter & Case C-01  \\ 
\midrule
Mixture & Ar-Kr \\
Collision kernel & VSS \\
Non-dim physical space & $[0,\,1]$ \\ 
Non-dim velocity space & $[-7,\,7]^3$ \\ 
$N^3$ & $32^3$ \\
$N_\rho$ & $32$ \\
$M$ & 12 \\
Spatial elements & 4 \\
DG order & 3 \\
Time step (s) & $2\times 10^{-8}$ \\
Characteristic mass: $m_0$ & $m_\text{Ar}=m_1$ \\
Characteristic length: $H_0$ ($mm$) & 1 \\
Characteristic velocity: $u_0$ ($m/s$) & 337.2 \\
Characteristic temperature: $T_0$ ($K$) & 273 \\
Characteristic number density: $n_0$ ($m^{-3}$) & $1.680\times10^{21}$ \\
\midrule
\multicolumn{2}{l}{Initial conditions} \\
Velocity: $u$ ($m/s$) & 0 \\
Temperature: $T$ ($K$) & 273 \\
Number density: $n^\one$ ($m^{-3}$) & $1.680\times10^{21}$ \\
Number density: $n^\two$ ($m^{-3}$) & $8.009\times10^{20}$ \\
Knudsen number: $(\Kn_{11},\,\Kn_{22})$ & $(0.793,\,0.606)$  \\
Knudsen number: $(\Kn_{12},\,\Kn_{21})$ & $(0.803,\,0.555)$ \\
\midrule
\multicolumn{2}{l}{Left wall (purely diffuse) boundary conditions (subscript $l$)} \\
Velocity: $u_l$ ($m/s$) & $(0,\,-50,\,0)$ \\
Temperature: $T_l$ ($K$) & 273 \\
\midrule
\multicolumn{2}{l}{Right wall (purely diffuse) boundary conditions (subscript $r$)} \\
Velocity: $u_r$ ($m/s$) & $(0,\,+50,\,0)$ \\
Temperature: $T_r$ ($K$) & 273 \\
\bottomrule
\end{tabular}
\caption{Numerical parameters for Couette flow. The molecular collision parameters for Ar-Kr system are provided in Table~\ref{tab_collision}.}
\label{tab_oscCouette_conditions}
\end{table}

\begin{figure}[!ht]
\begin{subfigure}{.5\textwidth}
  \centering
  \includegraphics[width=80mm]{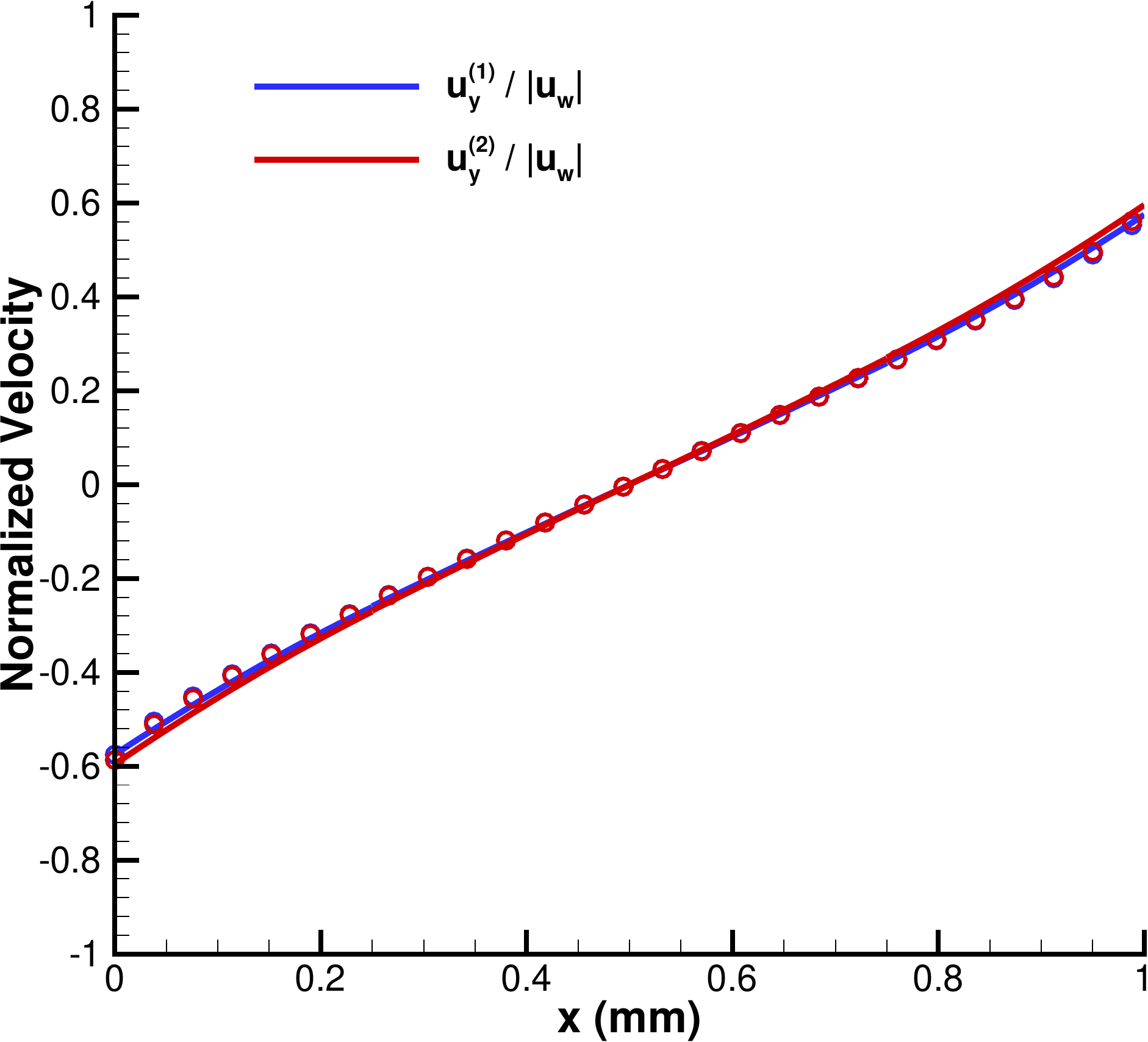}
  \caption{normalized $y$-component of velocity}
  \label{fig_couetteVSS_U}
\end{subfigure}%
\begin{subfigure}{.5\textwidth}
  \centering
  \includegraphics[width=80mm]{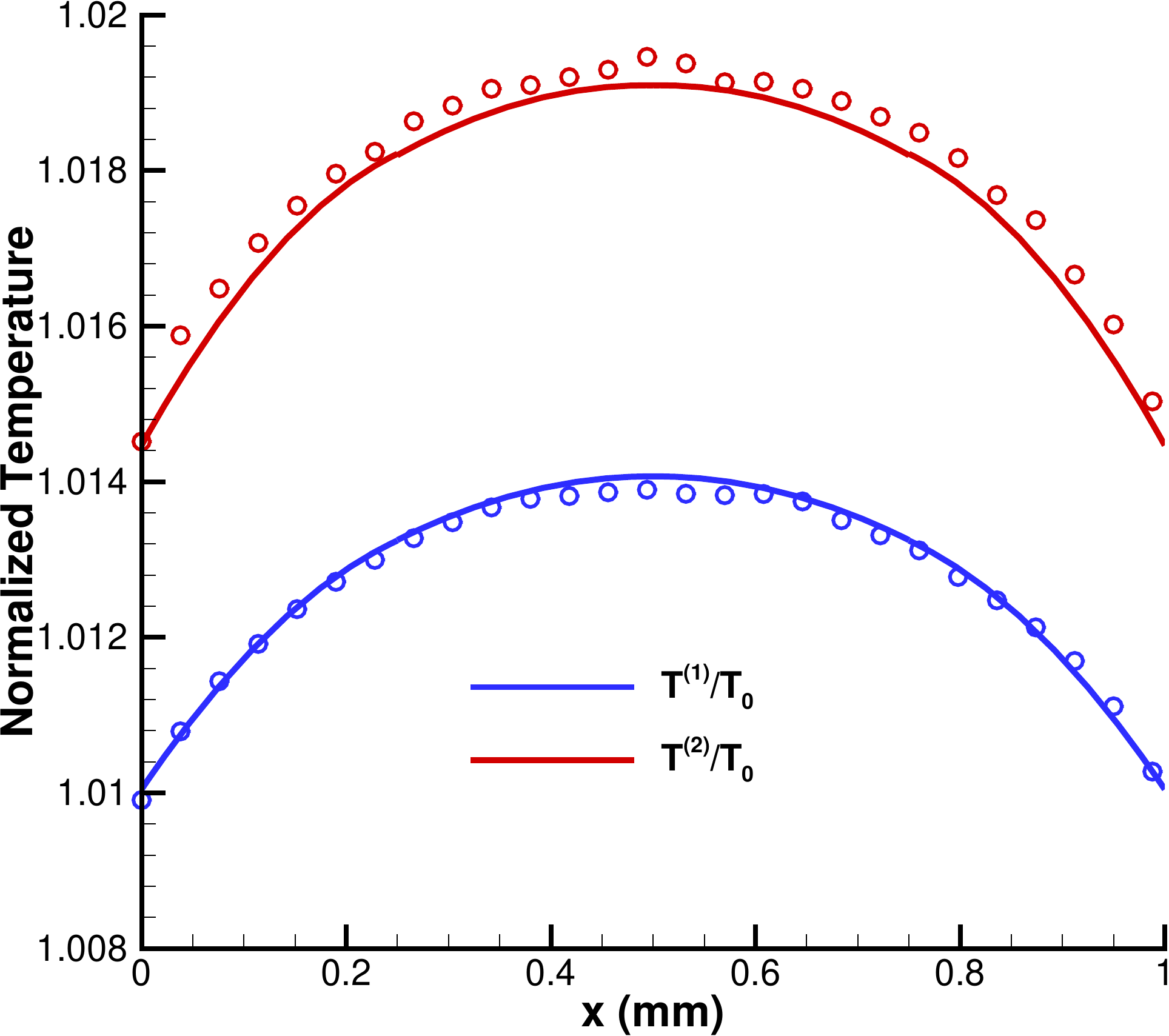}
  \caption{normalized temperature}
  \label{fig_couetteVSS_T}
\end{subfigure}
\caption{Variation of normalized $y$-velocity, and temperature along the domain for Couette flow (Case C-01) obtained with DSMC and DGFS using VSS collision kernel for Argon-Krypton mixture. Symbols denote DSMC solutions, and lines denote DGFS solutions.}
\label{fig_couetteVSS_U_T}
\end{figure}

Figure~\ref{fig_couetteVSS_U_T} illustrates the velocity and temperature along the domain length for both species. Ignoring the statistical noise, we observe an excellent agreement between DGFS and DSMC. 

\begin{figure}[!ht]
\begin{subfigure}{.5\textwidth}
  \centering
  \includegraphics[width=80mm]{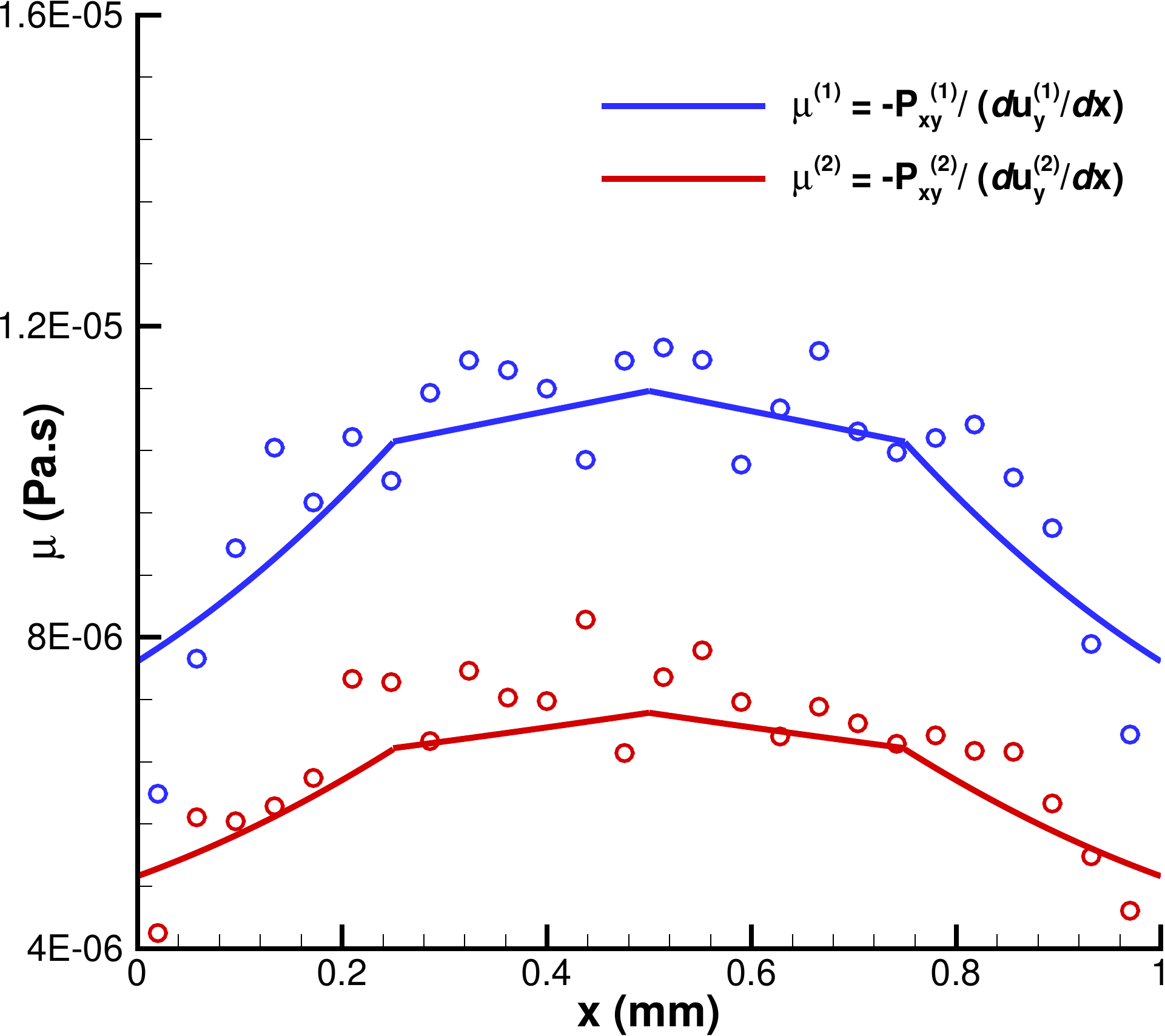}
  \caption{viscosity, 4 elements and $K=3$}
  \label{fig_couetteVSS_mu_s4k3v32m12}
\end{subfigure}
\begin{subfigure}{.5\textwidth}
  \centering
  \includegraphics[width=80mm]{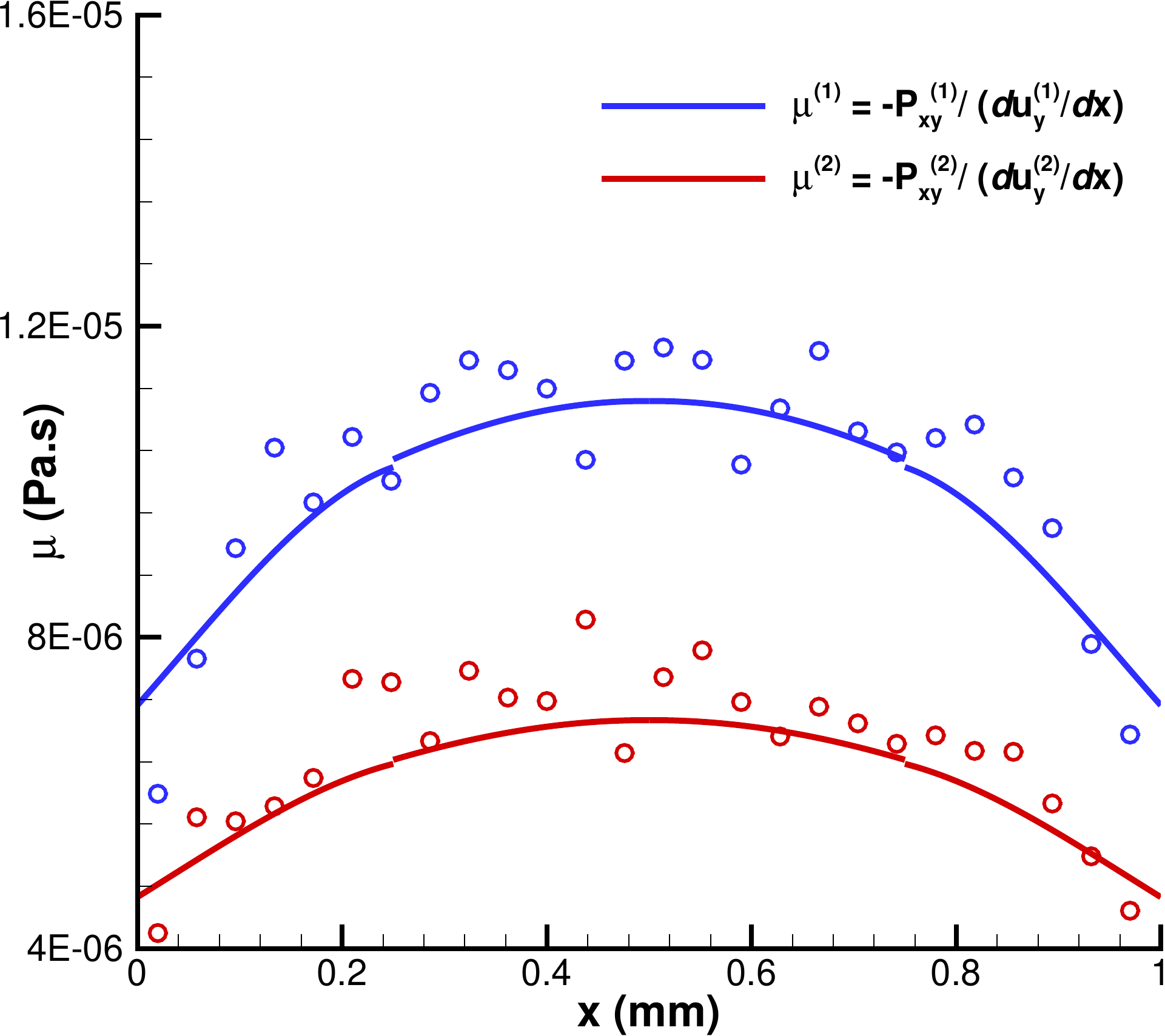}
  \caption{viscosity, 4 elements and $K=4$}
  \label{fig_couetteVSS_mu_s4k4v32m12}
\end{subfigure}
\caption{Variation of viscosity along the domain for Couette flow (Case C-01) obtained with DSMC and DGFS using VSS collision model for Argon-Krypton mixture. The physical space is discretized using 4 elements: a) $K=3$, b) $K=4$. Symbols denote DSMC solutions, and lines denote DGFS solutions.}
\label{fig_couetteVSS_mu}
\end{figure}

The viscosity $\mu^\I$ can be recovered from the 1-D Couette flow simulations using the relation between shear-stress and velocity-gradient \cite{weaver2014effect,jaiswal2018dsmc}:
\begin{align} \label{eq_viscosity}
\mu^\I = -\frac{\mathbb{P}^\I_{xy}}{\partial u^\I_y/\partial x}.
\end{align}
For consistency, we use eqn.~(\ref{eq_viscosity}) for both DSMC and DGFS. For computing the derivative in (\ref{eq_viscosity}), we use centered finite difference for DSMC, and the polynomial derivative for DGFS. Figure~\ref{fig_couetteVSS_mu} illustrates the variation of viscosity along the domain for both species. It is observed that: (a) the viscosity is lower for the heavier (Kr) species since the mixture contains $\sim32\%$ Kr, and $\sim68\%$ Ar; and (b) both DSMC and DGFS match well within the expected statistical scatter inherent to DSMC simulations. Note that, in the present simulation, we use DG scheme with $K=3$ which implies that the underlying polynomial is quadratic. Hence all the bulk properties including velocity should be a quadratic polynomial. Since the viscosity (\ref{eq_viscosity}) contains the derivative of the velocity, the overall reconstructed viscosity should be linear, as we observe in Figure~\ref{fig_couetteVSS_mu_s4k3v32m12}. Upon increasing $K$, we recover the smooth high order polynomial for viscosity as illustrated in Figure~\ref{fig_couetteVSS_mu_s4k4v32m12}.

From a computation viewpoint, DSMC-SPARTA simulations with 500 cells, 900 particles per cell, a time step of 2e-9 sec, 1 million unsteady time steps, and 20 million steady time steps, on 24 (Intel-Xeon Gold) processors took 26086.45 sec. The parameters have been selected to minimize the statistical fluctuations and linear time-stepping errors inherent to DSMC simulations. On the other hand, DGFS simulations on a single (Titan X Pascal) GPU with 4 elements, $K=3$, $N^3=32^3$, $M=12$ took 6020.19 sec to achieve  $(\|f^{n+1}-f^{n}\|_{L^2}/\|f^{n}\|_{L^2})/(\|f^{2}-f^{1}\|_{L^2}/\|f^{1}\|_{L^2}) < 5\times10^{-6}$. Note that these are representative simulation times for indicating the computational efforts required in DGFS and DSMC for 1-D simulations. Our experience shows that even heavily tuned codes can be further improved. A detailed comparison between CPU and GPU performance is subject of future study.

\subsubsection{Self diffusion of Argon-Argon mixture using VSS collision kernel}

In the current test case, we consider the effect of diffusive transport. The schematic remains the same as in the previous test case. Argon-Argon mixture with VSS collision kernel is taken as the working gas. To differentiate between two types of Argon, we tag the molecules as $\text{Ar}_1$ and $\text{Ar}_2$. At the left boundary, $\text{Ar}_1$ enters and exits at the right boundary. At the right boundary, $\text{Ar}_2$ enters and exits at the left boundary. The molecules enter the domain with zero mean velocity. The simulation parameters are provided in Table~\ref{tab_selfDiffusion_conditions}.

\begin{table}[!ht]
\centering
\begin{tabular}{@{}lcc@{}}
\toprule
Parameter & Case SD-01 & Case SD-02 \\ 
\midrule
Mixture & Ar-Ar & Ar-Ar \\
Collision kernel & VSS & VSS \\
Non-dim physical space & $[0,\,1]$ & $[0,\,1]$ \\ 
Non-dim velocity space & $[-5.09,\,5.09]^3$ & $[-5.09,\,5.09]^3$ \\ 
$N^3$ & $32^3$ & $32^3$ \\
$N_\rho$ & $32$ & $32$ \\
$M$ & 12 & 12 \\
Spatial elements & 4 & 4 \\
DG order & 3 & 3 \\
Time step (s) & $2\times 10^{-8}$ & $2\times 10^{-8}$ \\
Viscosity index: ($\omega_{ij}$) & 0.81 & 0.81 \\
Scattering index: ($\alpha_{ij}$) & 1.4 & 1.4 \\
Characteristic mass: $m_0$ & $m_\text{Ar}=m_1$ & $m_\text{Ar}=m_1$ \\
Characteristic length: $H_0$ ($mm$) & 1 & 1 \\
Characteristic velocity: $u_0$ ($m/s$) & 337.2 & 337.2 \\
Characteristic temperature: $T_0$ ($K$) & 273 & 273 \\
Characteristic number density: $n_0$ ($m^{-3}$) & $1.680\times10^{21}$ & $8.401\times10^{21}$ \\
\midrule
\multicolumn{2}{l}{Initial conditions} \\
Velocity: $u$ ($m/s$) & 0 & 0 \\
Temperature: $T$ ($K$) & 273 & 273  \\
Number density: $n^\one$ ($m^{-3}$) & $1.680\times10^{21}$ & $8.401\times10^{21}$ \\
Number density: $n^\two$ ($m^{-3}$) & $1.680\times10^{21}$ & $8.401\times10^{21}$ \\
Knudsen number: $(\Kn_{11},\,\Kn_{22})$ & $(0.793,\,0.793)$ & $(0.159,\,0.159)$ \\
Knudsen number: $(\Kn_{12},\,\Kn_{21})$ & $(0.793,\,0.793)$ & $(0.159,\,0.159)$ \\
\midrule
\multicolumn{2}{l}{Left boundary conditions (subscript $l$)} \\
\multicolumn{2}{l}{Ar$_1$ enters: inlet boundary condition for Ar$_1$} \\
Velocity: $u_l$ ($m/s$) & $(0,\,0,\,0)$ & $(0,\,0,\,0)$ \\
Temperature: $T_l$ ($K$) & 273 & 273 \\
Number density: $n^\one$ ($m^{-3}$) & $1.680\times10^{21}$ & $8.401\times10^{21}$ \\
\multicolumn{2}{l}{Ar$_2$ freely exits} \\
\midrule
\multicolumn{2}{l}{Right wall (purely diffuse) boundary conditions (subscript $r$)} \\
\multicolumn{2}{l}{Ar$_2$ enters: inlet boundary condition for Ar$_2$} \\
Velocity: $u_r$ ($m/s$) & $(0,\,0,\,0)$ & $(0,\,0,\,0)$ \\
Temperature: $T_r$ ($K$) & 273 & 273 \\
Number density: $n^\two$ ($m^{-3}$) & $1.680\times10^{21}$ & $8.401\times10^{21}$ \\
\multicolumn{2}{l}{Ar$_1$ freely exits} \\
\bottomrule
\end{tabular}
\caption{Numerical parameters for Ar-Ar self diffusion. The molecular collision parameters for Ar are provided in Table~\ref{tab_collision}.}
\label{tab_selfDiffusion_conditions}
\end{table}

\begin{figure}[!ht]
\begin{subfigure}{.5\textwidth}
  \centering
  \includegraphics[width=80mm]{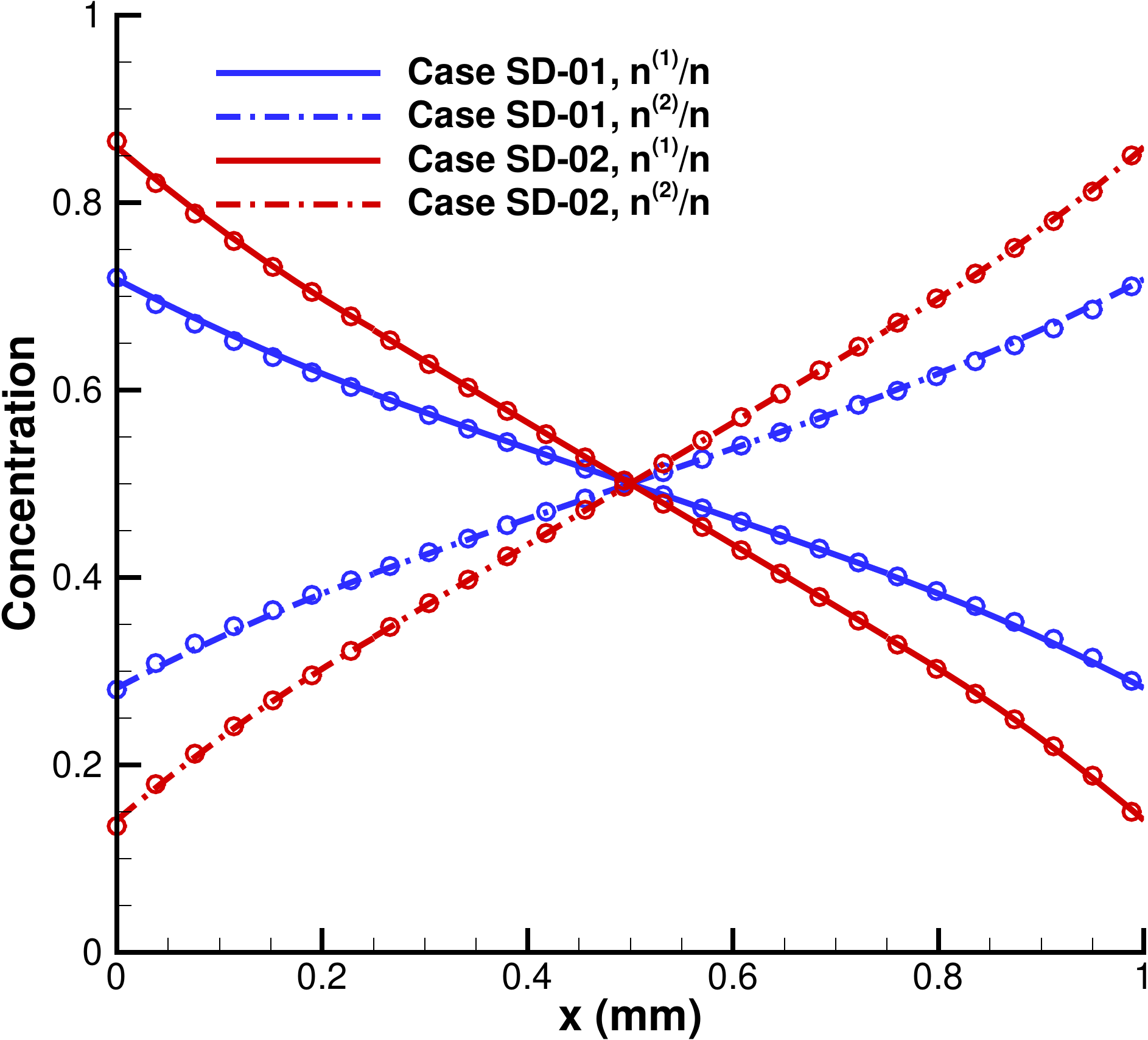}
  \caption{concentration}
  \label{fig_selfDiffusion_nden}
\end{subfigure}%
\begin{subfigure}{.5\textwidth}
  \centering
  \includegraphics[width=80mm]{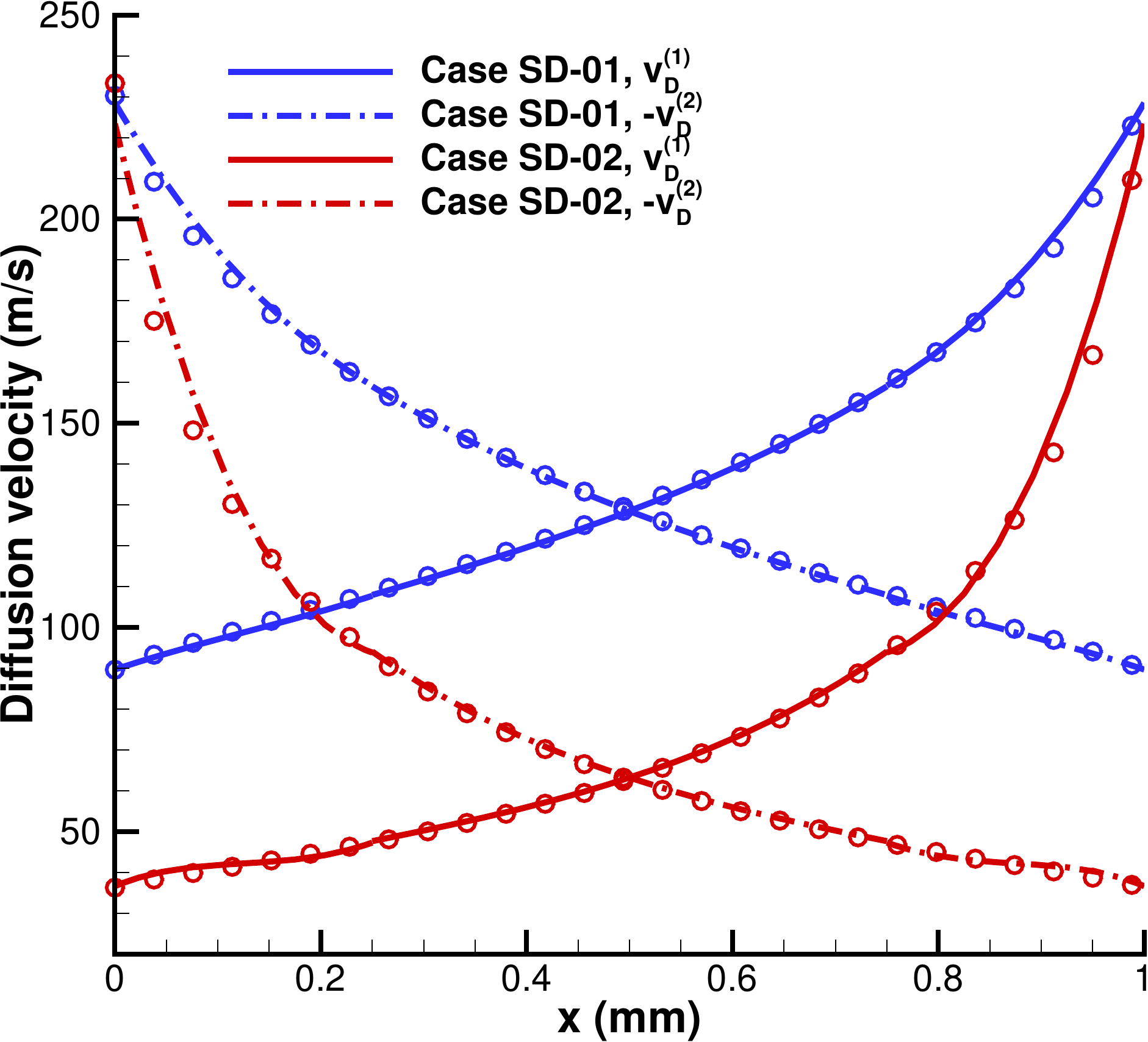}
  \caption{diffusion velocity}
  \label{fig_selfDiffusion_diffusionVelocity}
\end{subfigure}
\caption{Variation of number density and diffusion velocity along the domain for self-diffusion cases obtained with DSMC and DGFS using VSS collision kernel for Argon-Argon mixture. Symbols denote DSMC results, and lines denote DGFS results.}
\label{fig_selfDiffusion_nfrac_vD}
\end{figure}

Figure~\ref{fig_selfDiffusion_nden} shows the variation of concentration $n^\I/n$ along the domain. Since the species-1 enters from the left boundary and exits at right, we observe a drop in species-1 concentration as we move towards the right boundary. Conversely for species-2, since the species-2 enters from the right boundary and exits at left, we observe a drop in species-2 concentration as we move towards the left boundary. The non-linearity of the concentration profile and the associated slip at the boundaries can be explained through the low mixture density, and the fact that the flow is in slip regime. It is also worth noting that throughout the domain at any given $x$ location, the sum of the concentrations of two species is unity which asserts that the numerical formulation is conservative.

Figure~\ref{fig_selfDiffusion_diffusionVelocity} shows the variation of diffusion velocity along the domain. Since the species-1 enters from the left boundary and exits at right, we observe a low net diffusion speed (magnitude of the diffusion velocity) for the first species and a high diffusion speed for the second species. Conversely at the right boundary, since the species-2 enters from the right boundary and exits at left, we observe a low diffusion speed for the second species and high diffusion speed for the second species. 

\begin{figure}[!ht]
\begin{subfigure}{.5\textwidth}
  \centering
  \includegraphics[width=80mm]{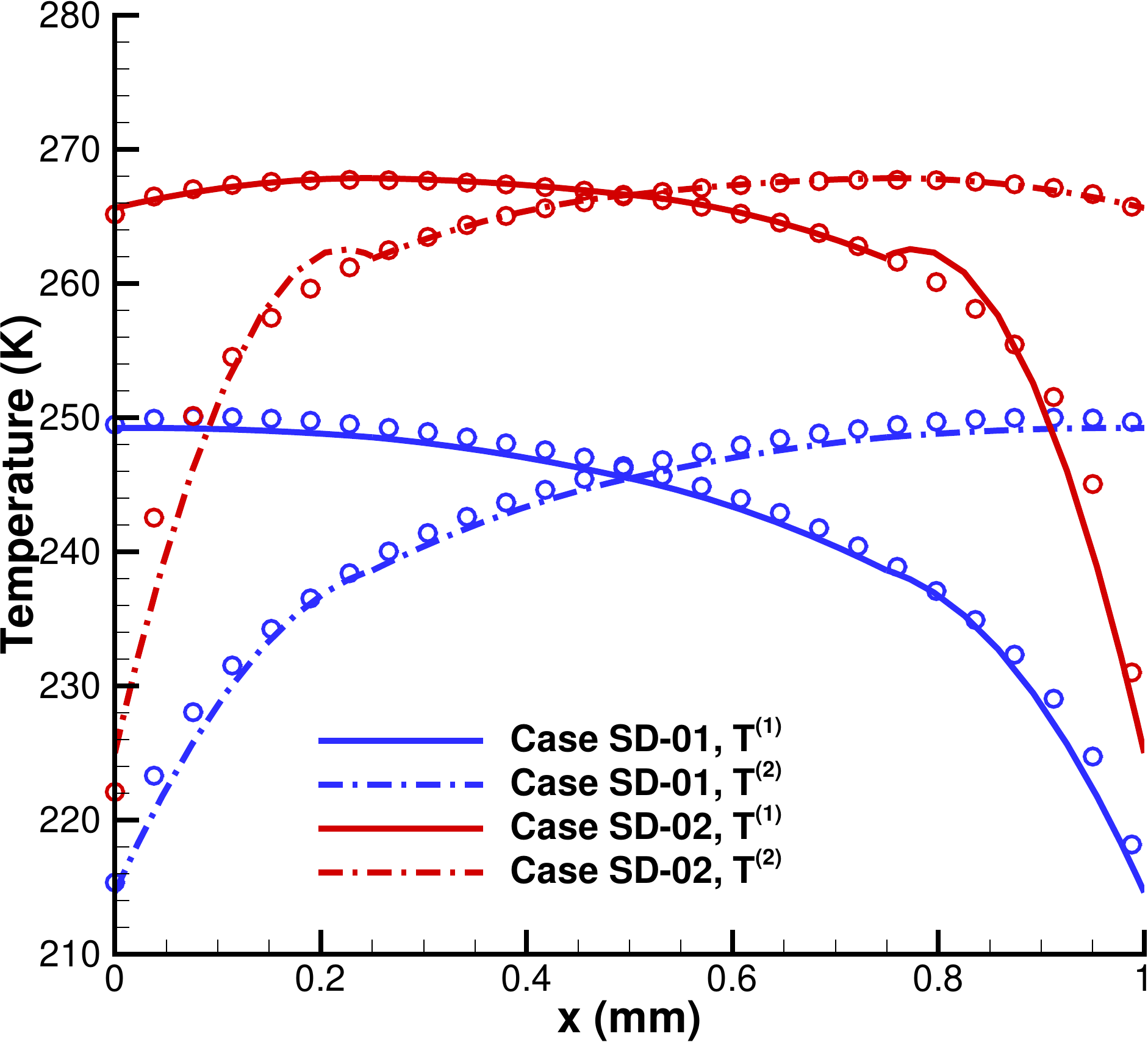}
  \caption{temperature, 4 elements and $K=3$}
  \label{fig_selfDiffusion_temperature_s4k3v32m12}
\end{subfigure}
\begin{subfigure}{.5\textwidth}
  \centering
  \includegraphics[width=80mm]{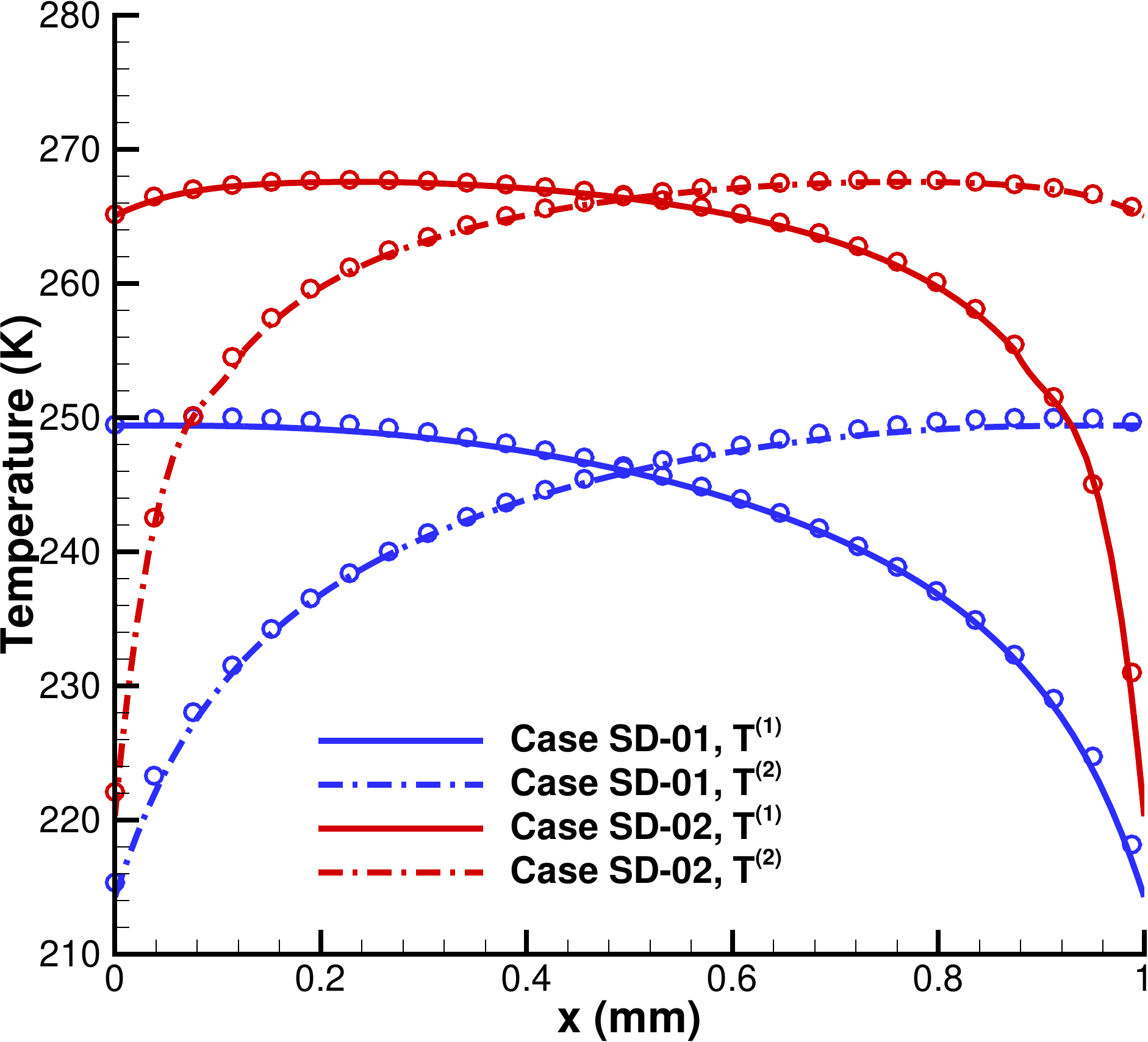}
  \caption{temperature, 8 elements and $K=4$}
  \label{fig_selfDiffusion_temperature_s8k4v32m12}
\end{subfigure}
\caption{Variation of temperature along the domain for self-diffusion cases obtained with DSMC and DGFS using VSS collision kernel for Argon-Argon mixture ($\alpha_{ij}=1.4$). The physical space is discretized using: a) 4 elements and $K=3$, b) 8 elements and $K=4$. Symbols denote DSMC results, and lines denote DGFS results.}
\label{fig_selfDiffusion_T}
\end{figure}

Figure~\ref{fig_selfDiffusion_T} illustrates the temperature profile along the domain, where we observe a drop in temperatures of the two species. Based upon these results, it can be inferred that DGFS can resolve the strong gradients in temperature and diffusion velocity with just 4 elements and $K=3$ within \textit{engineering} accuracy. 

For this test case, the self-diffusion coefficient is given as (cf. eqn.~(12.18) in \cite{Bird})
\begin{align}
D^{(11)} = D^{(12)} = -(u_x^{(1)} - u_x^{(2)}) \; \frac{n^{(1)} \; n^{(2)}}{n^2} \frac{\Delta x}{\Delta(n^{(1)}/n)}.
\label{eq_diffusion}
\end{align}
While writing this equation, it is also assumed that the coefficient of thermal diffusion is low, and therefore the effect of temperature gradient is negligible (see eqn.~(8.4.7) in \cite{chapman1990mathematical}). Note that this equation is an approximation to the diffusion equation, derived from leading order Chapman expansion (see section 8.4 in \cite{chapman1990mathematical}), and therefore, strictly speaking, the values computed from this equation might not be fully accurate especially for the rarefied flows, since the higher order terms have not been accounted for. It is worth noting that the full diffusion equation based on moment of the distribution function is highly non-trivial from a computation perspective, and is therefore rarely implemented in DSMC codes. For consistency, we use (\ref{eq_diffusion}) for both DSMC and DGFS. For computing the derivatives in (\ref{eq_diffusion}), we use centered finite difference for DSMC, and the polynomial derivative for DGFS. In particular, for DSMC simulations, we used 500 cells, 2000 particles per cell, a time step of $1 \times 10^{-8}$ sec, and averaged the results for 1 million time steps to minimize the statistical scatter in diffusion coefficients.

\begin{figure*}[tbp]
\centering
\begin{subfigure}{.5\textwidth}
  \centering
  \includegraphics[width=70mm]{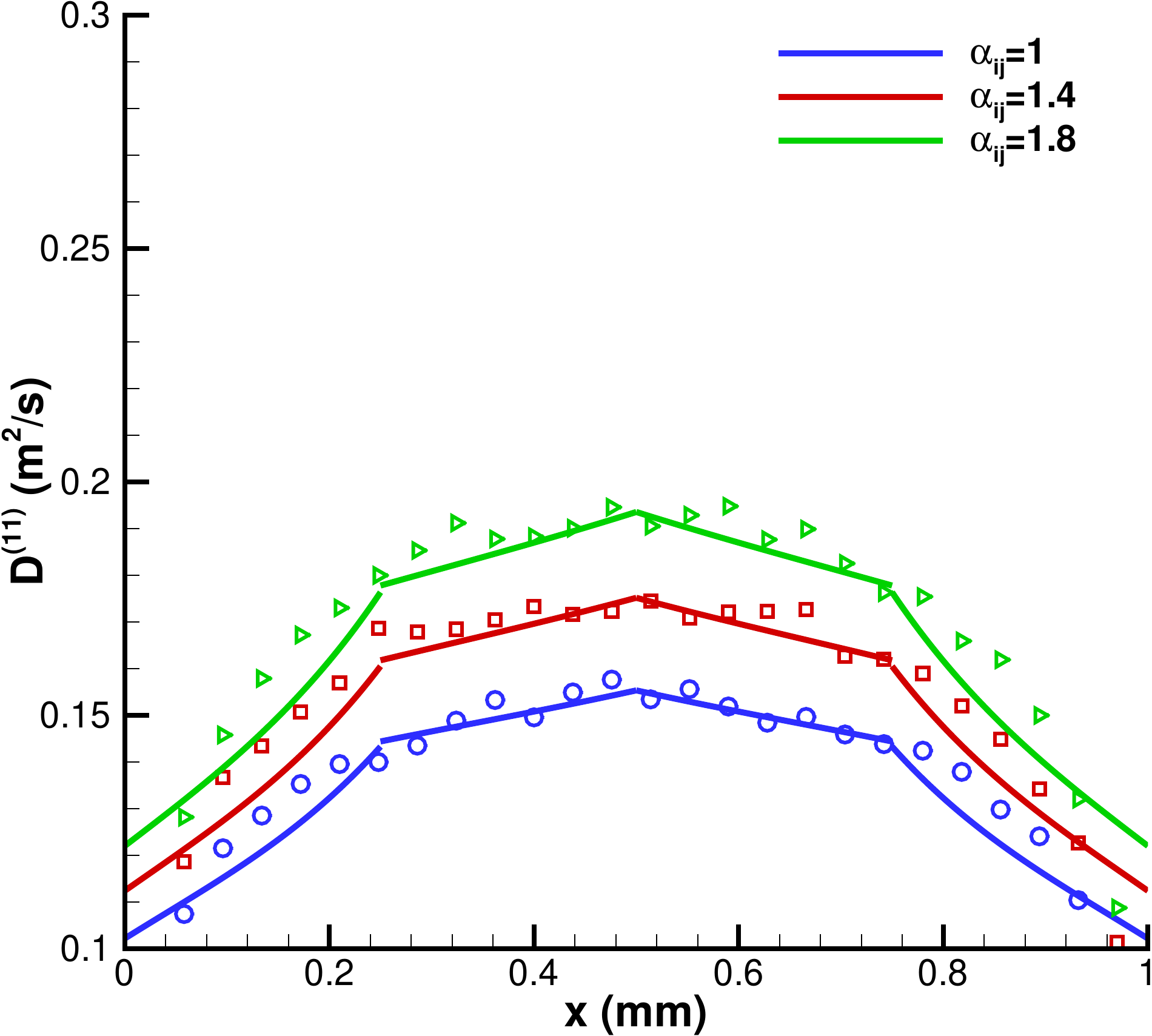}
  \caption{Case SD-01, 4 elements and $K=3$}
  \label{fig_selfDiffusion_diffusionCoefficient_case01_s4k3v32m12}
\end{subfigure}%
\begin{subfigure}{.5\textwidth}
  \centering
  \includegraphics[width=70mm]{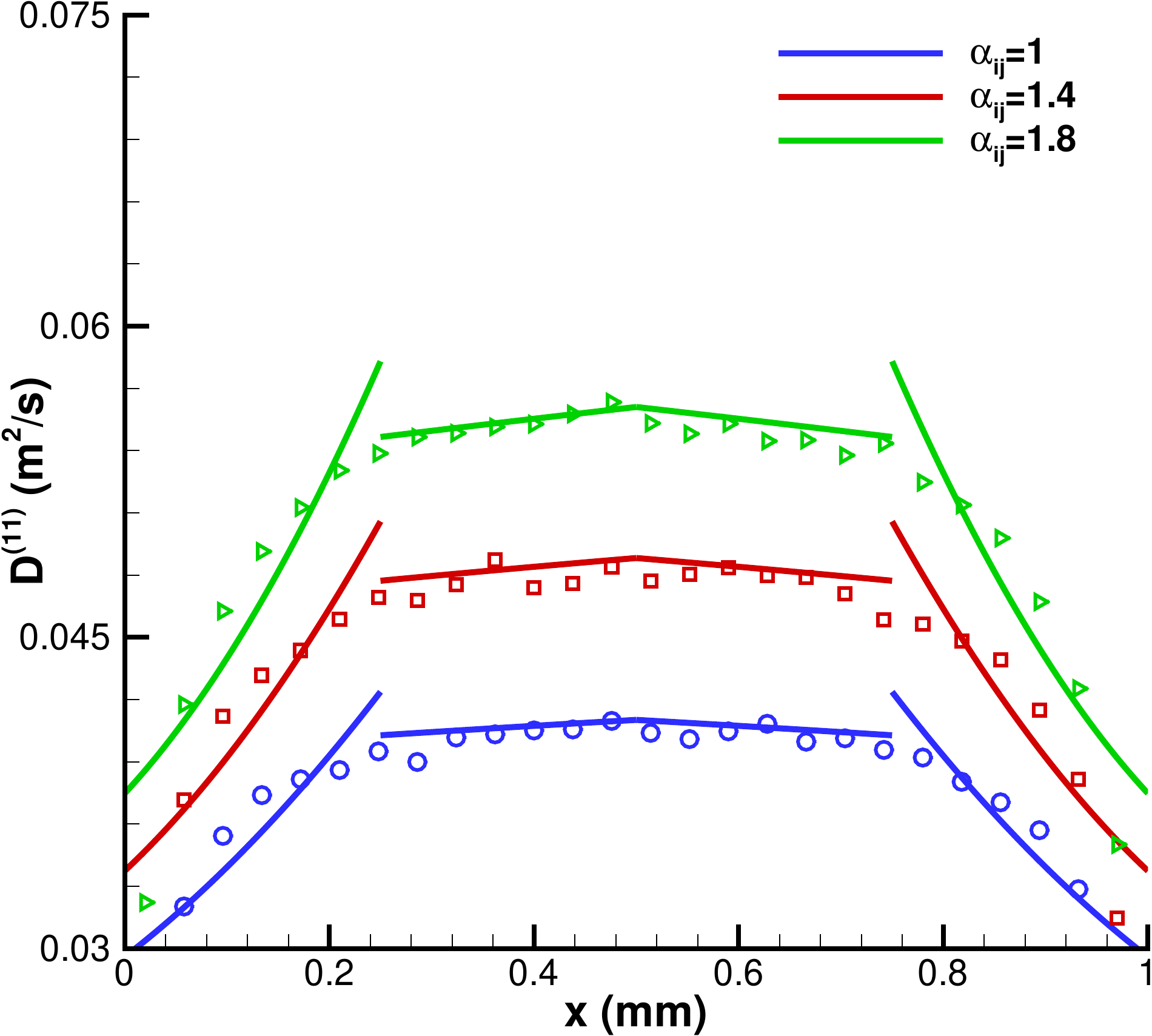}
  \caption{Case SD-02, 4 elements and $K=3$}
  \label{fig_selfDiffusion_diffusionCoefficient_case02_s4k3v32m12}
\end{subfigure}
\begin{subfigure}{.5\textwidth}
  \centering
  \includegraphics[width=70mm]{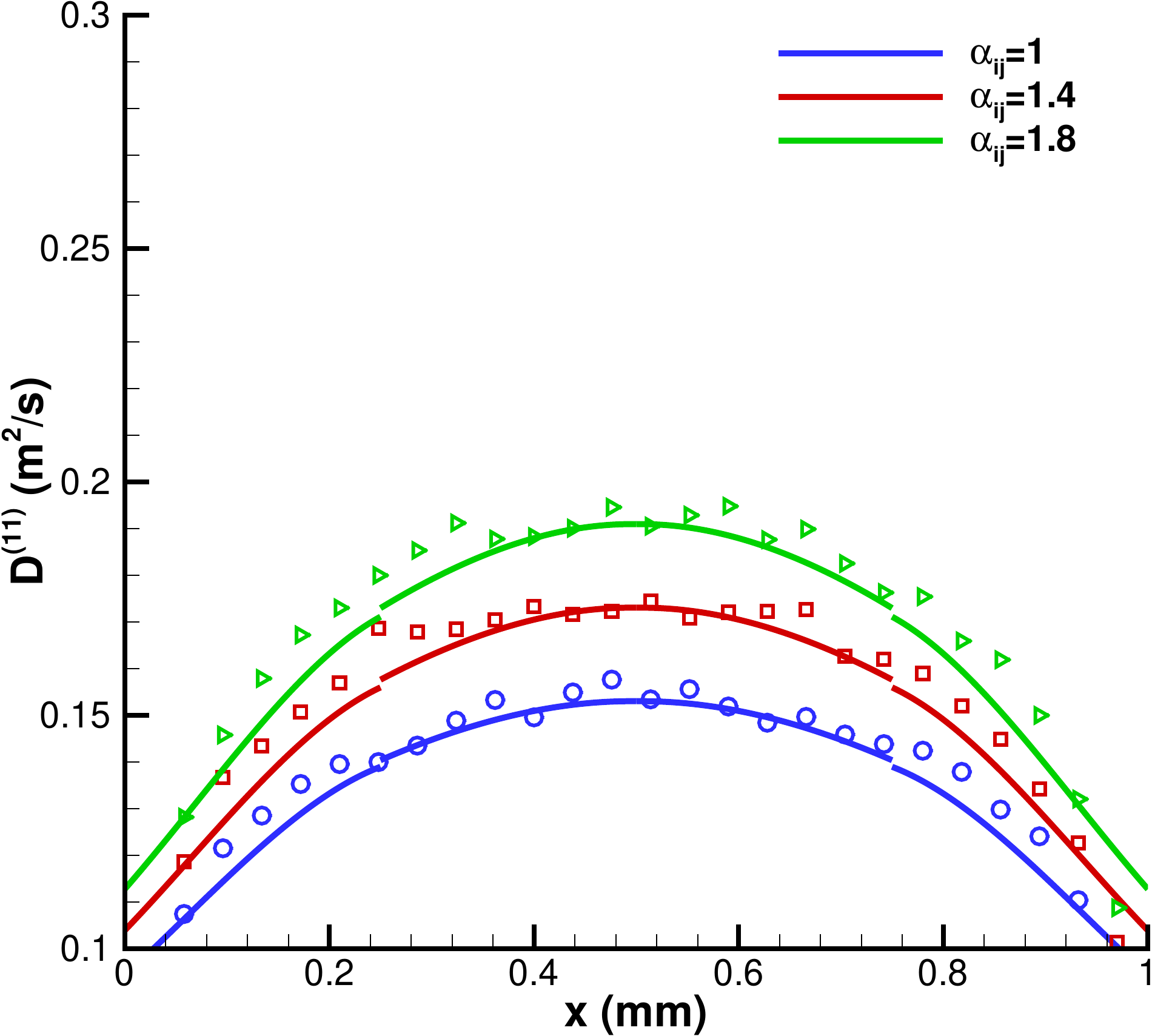}
  \caption{Case SD-01, 4 elements and $K=4$}
  \label{fig_selfDiffusion_diffusionCoefficient_case01_s4k4v32m12}
\end{subfigure}%
\begin{subfigure}{.5\textwidth}
  \centering
  \includegraphics[width=70mm]{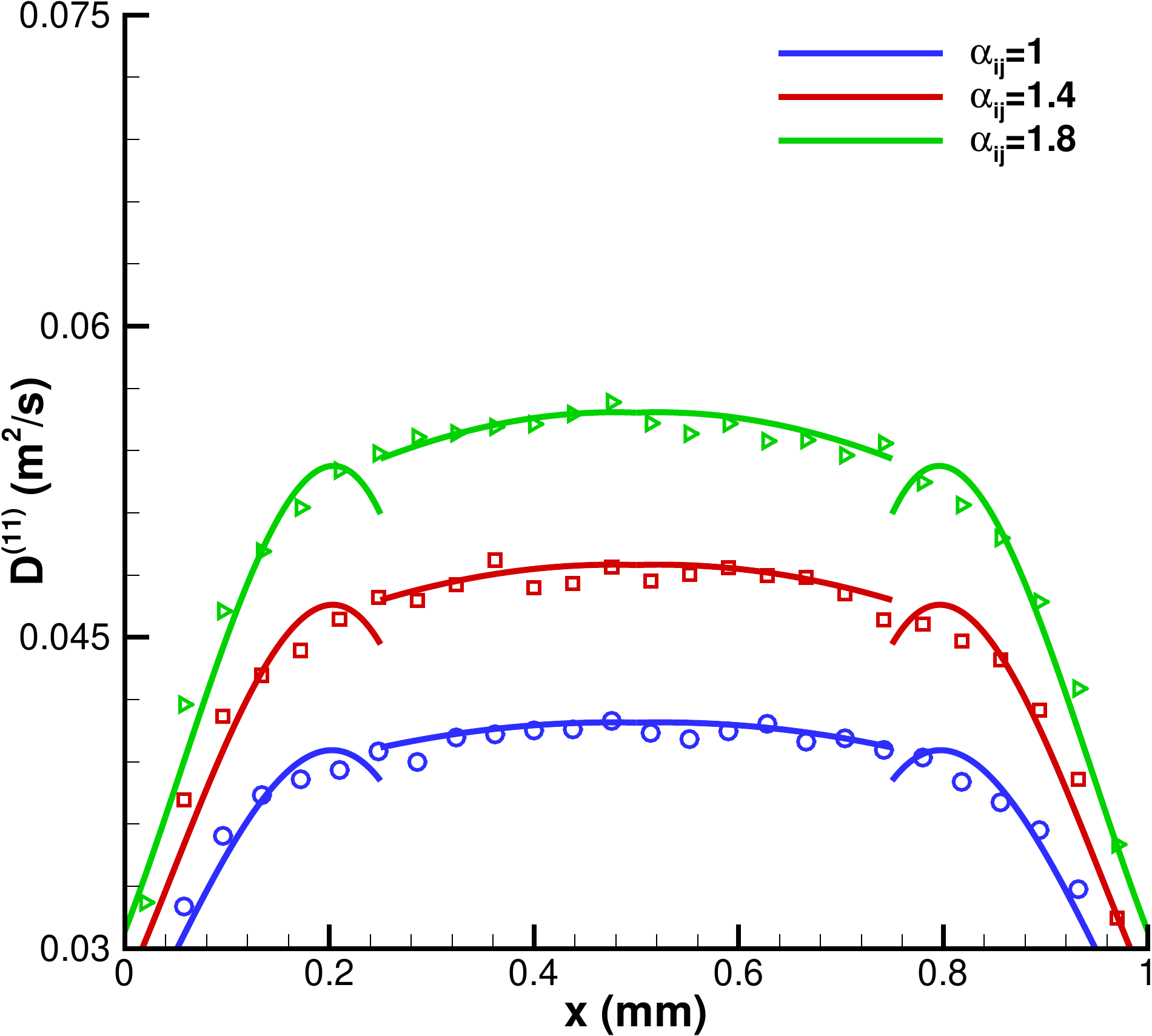}
  \caption{Case SD-02, 4 elements and $K=4$}
  \label{fig_selfDiffusion_diffusionCoefficient_case02_s4k4v32m12}
\end{subfigure}
\begin{subfigure}{.5\textwidth}
  \centering
  \includegraphics[width=70mm]{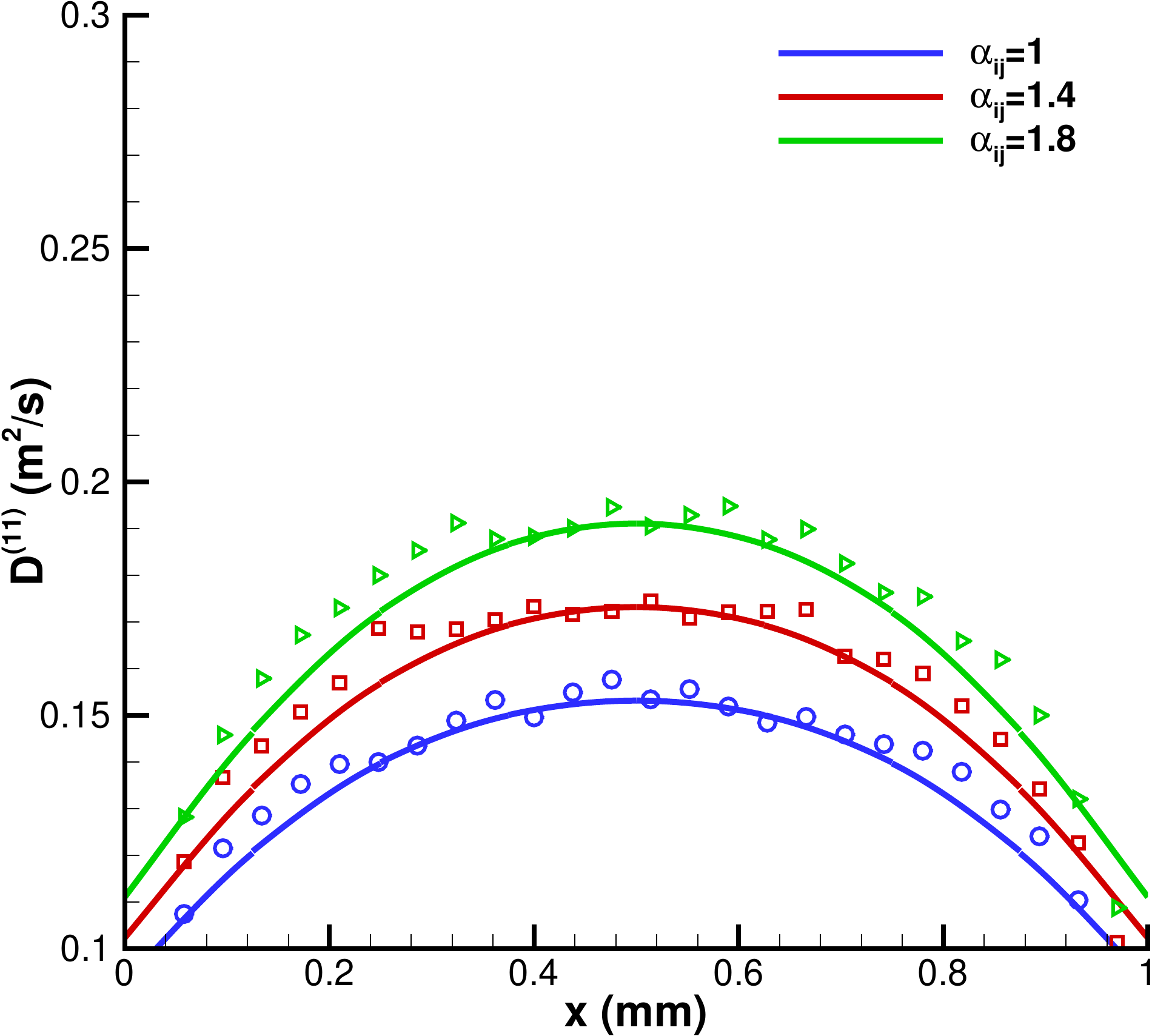}
  \caption{Case SD-01, 8 elements and $K=4$}
  \label{fig_selfDiffusion_diffusionCoefficient_case01_s8k4v32m12}
\end{subfigure}%
\begin{subfigure}{.5\textwidth}
  \centering
  \includegraphics[width=70mm]{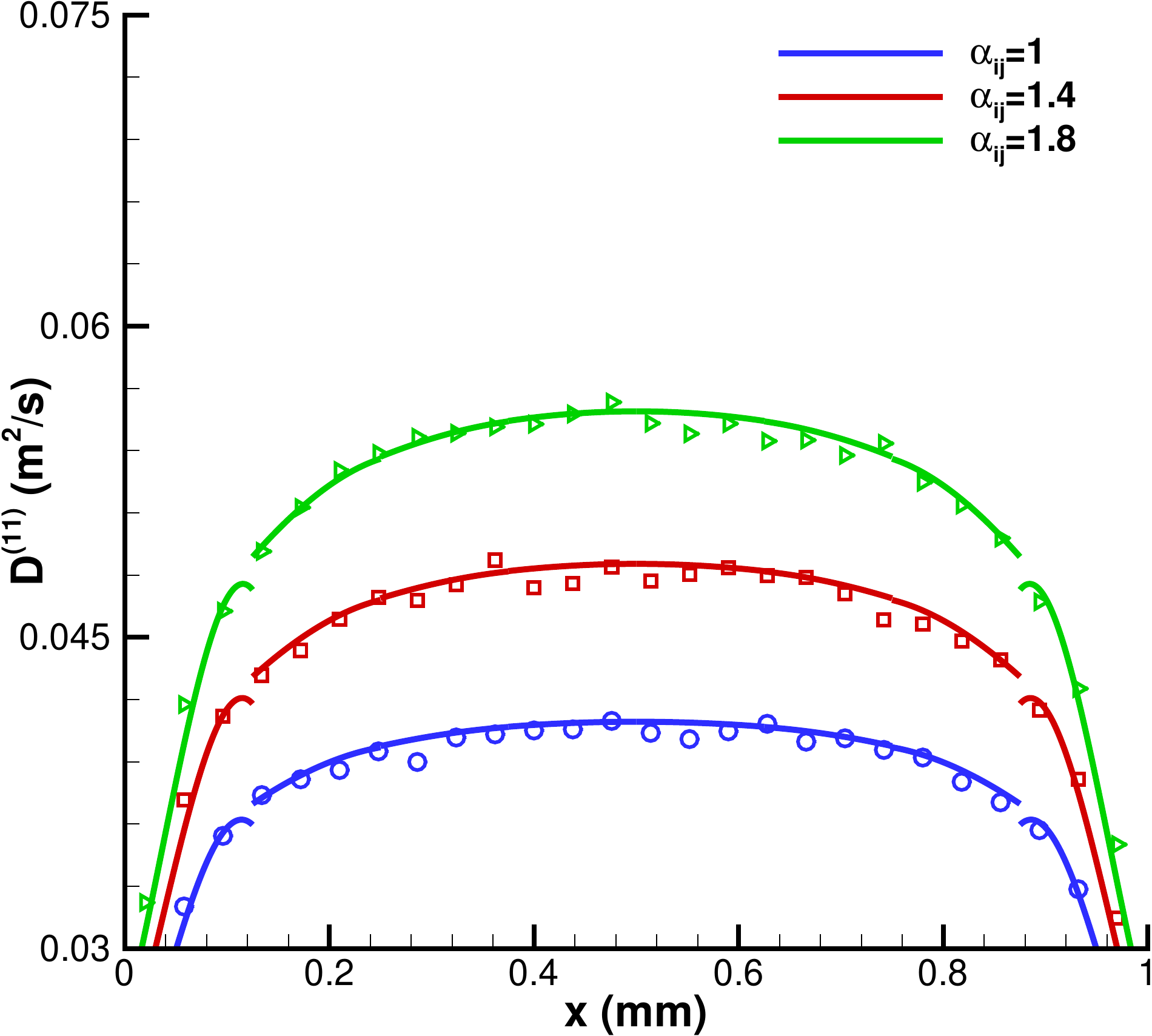}
  \caption{Case SD-02, 8 elements and $K=4$}
  \label{fig_selfDiffusion_diffusionCoefficient_case02_s8k4v32m12}
\end{subfigure}
\caption{Variation of diffusion coefficient along the domain for self-diffusion cases obtained with DSMC and DGFS using VSS collision model for Argon-Argon mixture. Note that only $\alpha_{ij}$ is varied by keeping all other parameters fixed as in Table~\ref{tab_selfDiffusion_conditions}. Symbols denote DSMC results, and lines denote DGFS results.}
\label{fig_selfDiffusion_diffusionCoefficient}
\end{figure*}

Figure~\ref{fig_selfDiffusion_diffusionCoefficient} illustrates the variation of self-diffusion coefficient along the domain as a function of scattering parameter $\alpha_{ij}$. It is observed that: (a) the diffusion coefficient increases with increase in $\alpha_{ij}$ in accordance with the VSS model (cf. eqn.~(3.75) in \cite{Bird}), (b) both DSMC and DGFS match well within the expected statistical scatter inherent to DSMC simulations, and (c) with increase in number density from Case 01 to Case 02, the diffusion coefficient decreases in accordance with (\ref{eq_diffusion}). Note that, in the present simulation, we use DG scheme with $K=3$ which implies that the underlying polynomial is quadratic. Hence all the bulk properties including number density should be a quadratic polynomial. Recall that the diffusion (\ref{eq_diffusion}) contains the derivative of the number density, and hence the overall reconstructed diffusion coefficient should be linear, which is what we observe in Figures~\ref{fig_selfDiffusion_diffusionCoefficient_case01_s4k3v32m12},\ref{fig_selfDiffusion_diffusionCoefficient_case02_s4k3v32m12}. Upon increasing $K$, we recover the smooth high-order polynomial for diffusion coefficient.

\subsubsection{Mass diffusion of Argon-Krypton mixture using VSS collision kernel}

In the current test case, we consider the effect of mass diffusion. The conditions remain the same as in previous case, except that Argon-Krypton mixture with VSS collision model is taken as the working gas. More specifically, Argon enters the left boundary and exits at the right boundary; and Krypton enters through the right and exits at left. The molecules enter the domain with zero mean velocity. We consider two cases with different initial number density. The numerical parameters for both the cases are given in Table~\ref{tab_massDiffusion_conditions}.

\begin{table}[!ht]
\centering
\begin{tabular}{@{}lcc@{}}
\toprule
Parameter & Case MD-01 & Case MD-02 \\ 
\midrule
Mixture & Ar-Kr & Ar-Kr \\
Collision kernel & VSS & VSS \\
Non-dim physical space & $[0,\,1]$ & $[0,\,1]$ \\ 
Non-dim velocity space & $[-5.09,\,5.09]^3$ & $[-5.09,\,5.09]^3$ \\ 
$N^3$ & $32^3$ & $32^3$ \\
$N_\rho$ & $32$ & $32$ \\
$M$ & 12 & 12 \\
Spatial elements & 4 & 4 \\
DG order & 3 & 3 \\
Time step (s) & $2\times 10^{-8}$ & $2\times 10^{-8}$ \\
Characteristic mass: $m_0$ & $m_\text{Ar}=m_1$ & $m_\text{Ar}=m_1$ \\
Characteristic length: $H_0$ ($mm$) & 1 & 1 \\
Characteristic velocity: $u_0$ ($m/s$) & 337.2 & 337.2 \\
Characteristic temperature: $T_0$ ($K$) & 273 & 273 \\
Characteristic number density: $n_0$ ($m^{-3}$) & $1.680\times10^{21}$ & $8.401\times10^{21}$ \\
\midrule
\multicolumn{2}{l}{Initial conditions} \\
Velocity: $u$ ($m/s$) & 0 & 0 \\
Temperature: $T$ ($K$) & 273 & 273  \\
Number density: $n^\one$ ($m^{-3}$) & $1.680\times10^{21}$ & $8.401\times10^{21}$ \\
Number density: $n^\two$ ($m^{-3}$) & $8.009\times10^{20}$ & $4.004\times10^{21}$ \\
Knudsen number: $(\Kn_{11},\,\Kn_{22})$ & $(0.793,\,0.606)$ & $(0.159,\,0.121)$ \\
Knudsen number: $(\Kn_{12},\,\Kn_{21})$ & $(0.803,\,0.555)$ & $(0.161,\,0.111)$ \\
\midrule
\multicolumn{2}{l}{Left boundary conditions (subscript $l$)} \\
\multicolumn{2}{l}{Ar enters: inlet boundary condition for Ar} \\
Velocity: $u_l$ ($m/s$) & $(0,\,0,\,0)$ & $(0,\,0,\,0)$ \\
Temperature: $T_l$ ($K$) & 273 & 273 \\
Number density: $n^\one$ ($m^{-3}$) & $1.680\times10^{21}$ & $8.401\times10^{21}$ \\
\multicolumn{2}{l}{Kr freely exits} \\
\midrule
\multicolumn{2}{l}{Right wall (purely diffuse) boundary conditions (subscript $r$)} \\
\multicolumn{2}{l}{Kr enters: inlet boundary condition for Kr} \\
Velocity: $u_r$ ($m/s$) & $(0,\,0,\,0)$ & $(0,\,0,\,0)$ \\
Temperature: $T_r$ ($K$) & 273 & 273 \\
Number density: $n^\two$ ($m^{-3}$) & $8.009\times10^{20}$ & $4.004\times10^{21}$ \\
\multicolumn{2}{l}{Ar freely exits} \\
\bottomrule
\end{tabular}
\caption{Numerical parameters for Ar-Kr mass diffusion. The molecular collision parameters for Ar-Kr system are provided in Table~\ref{tab_collision}.}
\label{tab_massDiffusion_conditions}
\end{table}

\begin{figure}[!ht]
\centering
\begin{subfigure}{.5\textwidth}
  \centering
  \includegraphics[width=80mm]{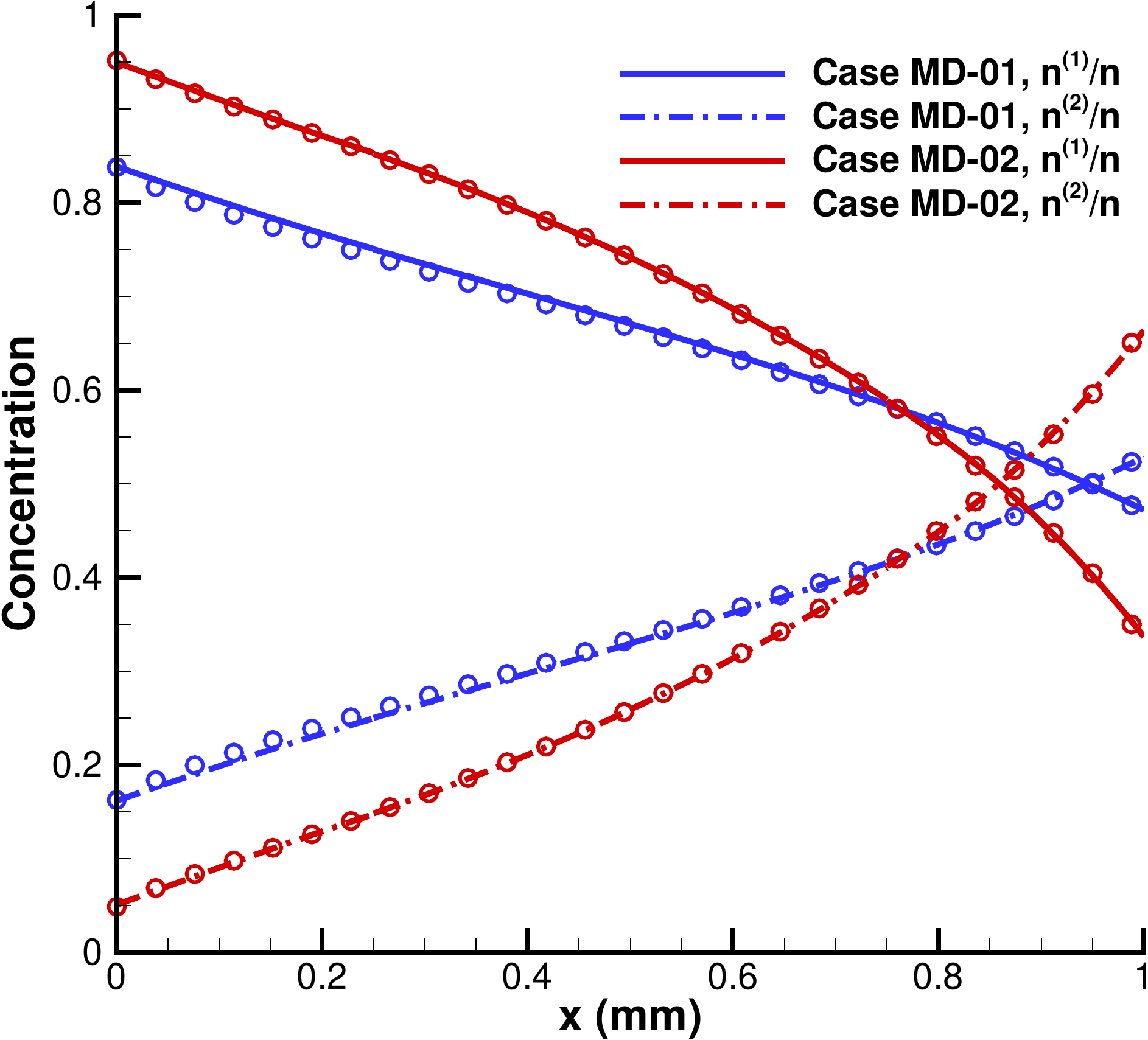}
  \caption{concentration}
  \label{fig_massDiffusion_nden}
\end{subfigure}%
\begin{subfigure}{.5\textwidth}
  \centering
  \includegraphics[width=80mm]{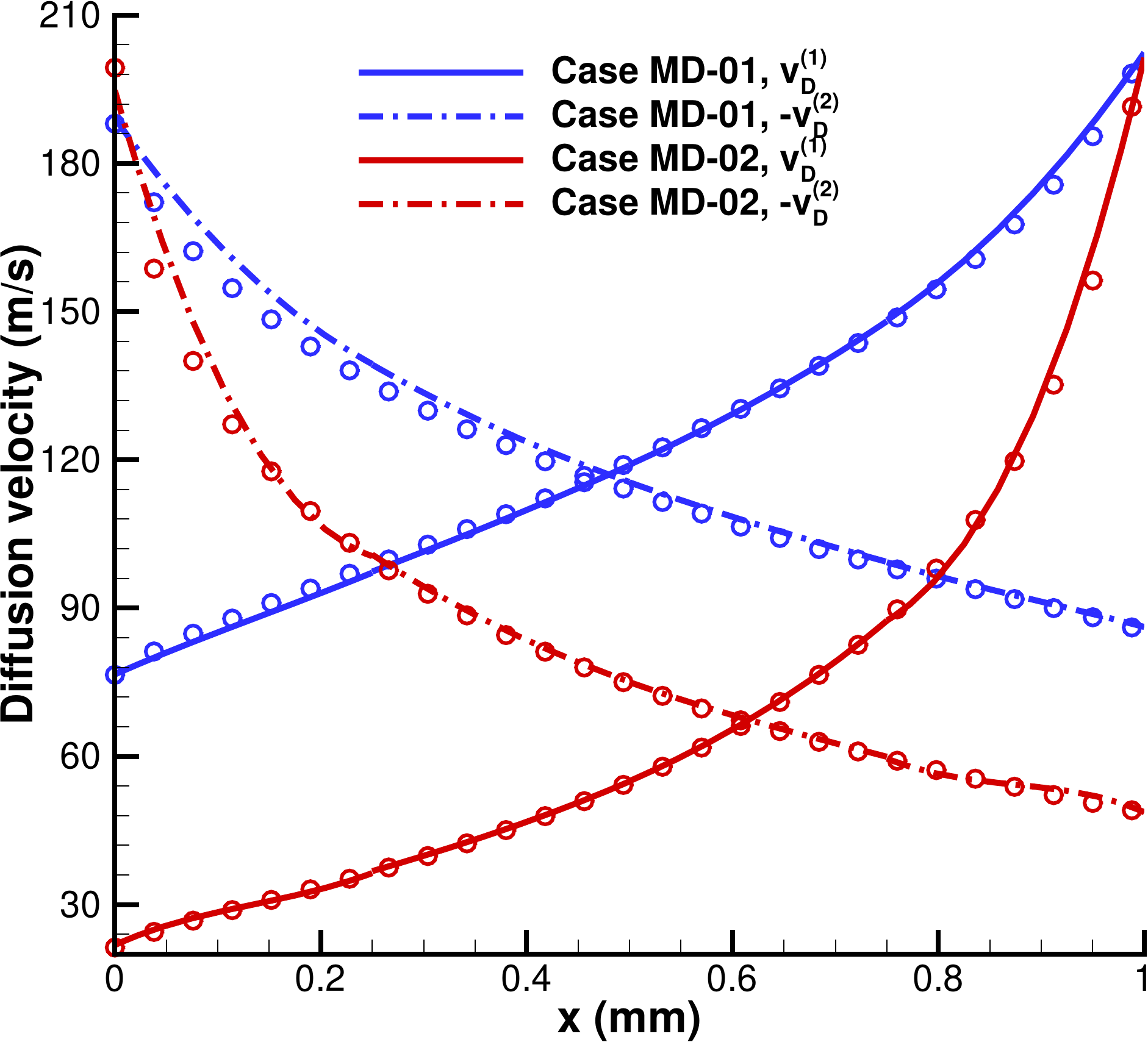}
  \caption{diffusion velocity}
  \label{fig_massDiffusion_diffusionVelocity}
\end{subfigure}
\begin{subfigure}{.5\textwidth}
  \centering
  \includegraphics[width=80mm]{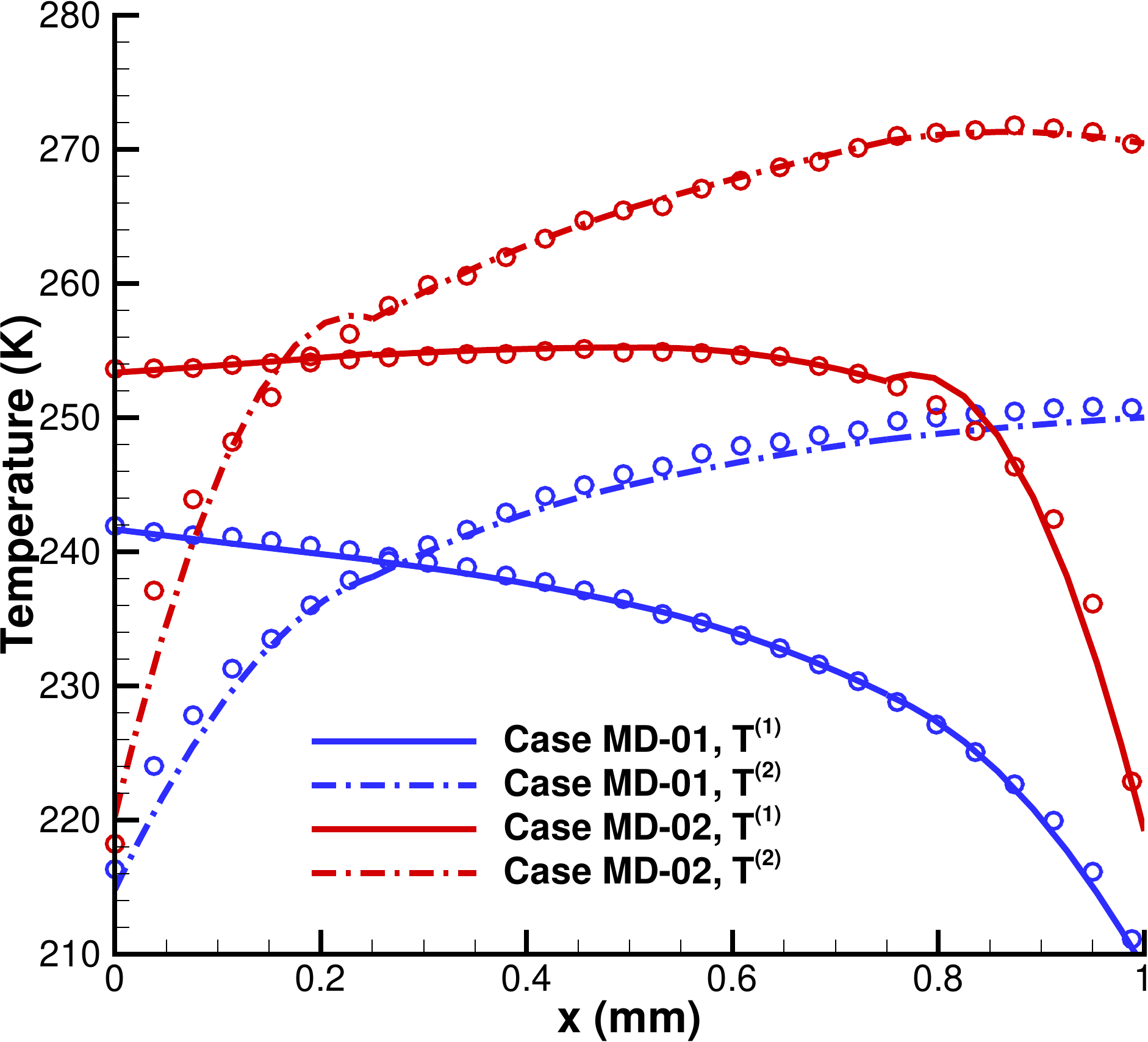}
  \caption{temperature}
  \label{fig_massDiffusion_temperature}
\end{subfigure}
\caption{Variation of number density, diffusion velocity, and temperature along the domain for mass-diffusion cases obtained with DSMC and DGFS using VSS collision model for Argon-Krypton mixture. Symbols denote DSMC results, and lines denote DGFS results.}
\label{fig_massDiffusion}
\end{figure}

Figure~\ref{fig_massDiffusion_nden} shows the variation of concentration profile for the two species. We observe that the concentration of Argon remains greater than Krypton throughout the domain, except for a small portion near the right boundary. This can be directly inferred from the mass/momentum conservation principle i.e., the heavier species diffuses slower and the lighter species diffuses faster. Therefore, after a sufficiently long time, the concentration of lighter species will be greater than that of heavier species in the major part of the domain. As in the self-diffusion case, the sum of the concentrations of both species is unity throughout the domain at any given $x$ location. The effect of the momentum conservation is more pronounced in the Figure~\ref{fig_massDiffusion_diffusionVelocity} where we observe a higher diffusion speed for the lighter species and a lower diffusion speed for the heavier species.

\section{Conclusions}
\label{sec_conclusions}

A fast spectral method for the multi-species Boltzmann collision operator has been proposed in this work. The method is designed to handle the cross-molecular interactions between dissimilar species with moderate mass ratios. In particular, it is applicable to general collision kernels which allows us to directly compare our results against the well-known stochastic DSMC solutions. The fast collision algorithm in the velocity space was then coupled with the discontinuous Galerkin discretization in the physical space to yield highly accurate numerical solutions for the full spatially inhomogeneous Boltzmann equation. The DG-type formulation employed in the present work has advantage of having high order accuracy at the element-level, and its element-local compact nature (and that of our collision algorithm) enables effective parallelization on massively parallel architectures. 

To validate our solver, extensive numerical tests were performed, including the spatially homogeneous Krook-Wu solution for Maxwell molecules where the exact solution is known; normal shock wave for HS where finite difference solutions are available for comparison, and Fourier, oscillatory Couette, Couette, self diffusion, mass diffusion problems where different collision kernels (VHS and VSS), Knudsen numbers, and mass ratios were considered and the results were compared well with DSMC solutions.

\section*{Acknowledgement}

This work was supported by NSF grant DMS-1620250 and NSF CAREER grant DMS-1654152.

\section*{References}

\bibliographystyle{elsarticle-num-names}
\bibliography{notes.bib}

\end{document}